%% file: ms.tex
\documentclass{emulateapj}
\usepackage{natbib}
\usepackage{color}
\usepackage{verbatim}
\usepackage{rotating}

\def\2pr{^{\prime \prime}}

\def\greatsim{\mathrel{\raise.3ex\hbox{$>$\kern-.75em\lower1ex\hbox{$\sim$}}}}
\def\lesssim{\mathrel{\raise.3ex\hbox{$<$\kern-.75em\lower1ex\hbox{$\sim$}}}}
\def\gs{\mathrel{\raise0.27ex\hbox{$>$}\kern-0.70em % Greater/squiggles
\lower0.71ex\hbox{{$\scriptstyle \sim$}}}}
\def\ls{\mathrel{\raise0.27ex\hbox{$<$}\kern-0.70em % Less than/squiggles
\lower0.71ex\hbox{{$\scriptstyle \sim$}}}}

\shorttitle{BOSS SN Spectroscopy}
\shortauthors{Olmstead et al.}

\input{astromacro}
\input{outlinemacro}

\newboolean{draft}
\setboolean{draft}{false}

\begin{document}

\title{
Host Galaxy Spectra and Consequences for SN Typing From The SDSS SN Survey}

\author{
Matthew~D.~Olmstead\altaffilmark{1},
Peter~J.~Brown\altaffilmark{1,2},
Masao~Sako\altaffilmark{3},
Bruce~Bassett\altaffilmark{4,5,6},
Dmitry~Bizyaev\altaffilmark{7},
J.~Brinkmann\altaffilmark{7},
Joel~R.~Brownstein\altaffilmark{1},
Howard~Brewington\altaffilmark{7},
Heather~Campbell\altaffilmark{8},
Chris~B.~D'Andrea\altaffilmark{9},
Kyle ~S.~Dawson\altaffilmark{1},
Garrett~L.~Ebelke\altaffilmark{7},
Joshua~A.~Frieman\altaffilmark{10,11},
Llu\'is~Galbany\altaffilmark{12,13},
Peter~Garnavich\altaffilmark{14},
Ravi~R.~Gupta\altaffilmark{3},
Renee~Hlozek\altaffilmark{15},
Saurabh~W.~Jha\altaffilmark{16},
Martin~Kunz\altaffilmark{17,4},
Hubert~Lampeitl\altaffilmark{9},
Elena~Malanushenko\altaffilmark{7},
Viktor~Malanushenko\altaffilmark{7},
John~Marriner\altaffilmark{10},
Ramon~Miquel\altaffilmark{12,18},
Antonio~D.~Montero-Dorta\altaffilmark{1},
Robert~C.~Nichol\altaffilmark{9},
Daniel~J.~Oravetz\altaffilmark{7}, 
Kaike~Pan\altaffilmark{7},
Donald~P.~Schneider\altaffilmark{19,20},
Audrey~E.~Simmons\altaffilmark{7},
Mathew~Smith\altaffilmark{21},
Stephanie~A.~Snedden\altaffilmark{7}
}

\altaffiltext{1}{Department of Physics and Astronomy, University of Utah, Salt Lake City, UT 84112, USA}

\altaffiltext{2}{George P. and Cynthia Woods Mitchell Institute for Fundamental Physics \& Astronomy, Texas A. \& M. University, Department of Physics and Astronomy, 4242 TAMU, College Station, TX 77843, USA }

\altaffiltext{3}{Department of Physics and Astronomy, University of Pennsylvania, 209 South 33rd Street, Philadelphia, PA 19104, USA}

\altaffiltext{4}{ 
African Institute for Mathematical Sciences, 6 Melrose Road, Muizenberg, 7945, South Africa}
\altaffiltext{5}{  Department of Mathematics and Applied Mathematics, University of Cape Town, Rondebosch, Cape Town, 7700, South Africa }

\altaffiltext{6}{  South African Astronomical Observatory, Observatory Road, Observatory, Cape Town, 7935, South Africa }

\altaffiltext{7}{Apache Point Observatory, P.O. Box 59, Sunspot, NM 88349, USA}

\altaffiltext{8}{Institute of Astronomy, Madingley Road, Cambridge CB4 0HA}

\altaffiltext{9}{Institute of Cosmology \& Gravitation, Dennis Sciama Building, University of Portsmouth, Portsmouth, PO1 3FX, UK}

\altaffiltext{10}{Center for Particle Astrophysics, Fermi National Accelerator Laboratory, P.O. Box 500, Batavia, IL 60510, USA}

\altaffiltext{11}{Kavli Institute for Cosmological Physics, The University of Chicago, 5640 South Ellis Avenue, Chicago, IL 60637, USA}

\altaffiltext{12}{Institut de F\'{i}sica d'Altes Energies, Universitat Aut\`{o}noma de Barcelona, E-08193 Bellaterra (Barcelona), Spain}

\altaffiltext{13}{Centro Multidisciplinar de Astrofisica, Instituto Superior Tecnico,Av. Rovisco Pais 1, 1049-001 Lisbon, Portugal}

\altaffiltext{14}{Department of Physics, University of Notre Dame, Notre Dame, IN 46556, USA}

\altaffiltext{15}{Department of Astrophysics, Peyton Hall, 4 Ivy Lane, Princeton, NJ 08544, USA}

\altaffiltext{16}{Department of Physics and Astronomy, Rutgers, the State University of New Jersey, 136 Frelinghuysen Road, Piscataway, NJ 08854, USA }

\altaffiltext{17}{Département de Physique Théorique and Center for Astroparticle Physics, Université de Gen\'eve, Quai E. Ansermet 24, CH-1211 Gen\'eve 4, Switzerland}

\altaffiltext{18}{Instituci\'o Catalana de Recerca i Estudis Avan\c{c}ats, E-08010 Barcelona, Spain}

\altaffiltext{19}{
Department of Astronomy and Astrophysics, 525 Davey Laboratory, 
The Pennsylvania State University, University Park, PA 16802, USA.
}

\altaffiltext{20}{
Institute for Gravitation and the Cosmos, 
The Pennsylvania State University, University Park, PA 16802, USA.
}

\altaffiltext{21}{Department of Physics, University of the Western Cape, Cape Town, 7535, South Africa
}

\email{olmstead@physics.utah.edu}

\begin{abstract}
We present the spectroscopy from 5254 galaxies that hosted supernovae (SNe) or other transient events in the Sloan Digital Sky Survey II (SDSS-II).  Obtained during SDSS-I, SDSS-II, and the Baryon Oscillation Spectroscopic Survey (BOSS), this sample represents the largest systematic, unbiased, magnitude limited spectroscopic survey of supernova (SN) host galaxies.  Using the host galaxy redshifts, we test the impact of photometric SN classification based on SDSS imaging data with and without using spectroscopic redshifts of the host galaxies.  Following our suggested scheme, there are a total of 1166 photometrically classified SNe~Ia when using a flat redshift prior and 1126 SNe~Ia when the host spectroscopic redshift is assumed.  For 1024 (87.8\%) candidates classified as likely SNe~Ia without redshift information, we find that the classification is unchanged when adding the host galaxy redshift.  Using photometry from SDSS imaging data and the host galaxy spectra, we also report host galaxy properties for use in future nalysis of SN astrophysics.  Finally, we investigate the differences in the interpretation of the light curve properties with and without knowledge of the redshift.  Without host galaxy redshifts, we find that SALT2 light curve fits are systematically biased towards lower photometric redshift estimates and redder colors in the limit of low signal-to-noise data.  The general improvements in performance of the light curve fitter and the increased diversity of the host galaxy sample highlights the importance of host galaxy spectroscopy for current photometric SN surveys such as the Dark Energy Survey and future surveys such as the Large Synoptic Survey Telescope.

\end{abstract}

\section{Introduction}\label{sec:intro}
\setcounter{footnote}{0}

Measurements of the luminosity distances derived from $\sim 50$  Type Ia supernovae (SNe~Ia) at redshifts $0 < z < 0.7 $ led to the discovery of the acceleration of the Universe \citep{riess98a,perlmutter99a}.  Since that time, a number of surveys have increased the sample of spectroscopically confirmed SNe~Ia to over a thousand and extended the redshift coverage beyond $z=1$.  The Lick Observatory Supernova Search \citep[LOSS;][]{filippenko01a,ganeshalingam10a}, the Carnegie Supernova Project \citep[CSP;][]{hamuy06a}, the Harvard-Smithsonian Center for Astrophysics SN group \citep[CfA;][]{hicken09a}, and the Palomar Transient Factory \citep[PTF;][]{rau09a, law09a} are each designed with an emphasis of discovery and observation of SNe at redshifts  $z \lesssim 0.1$. Similarly, the second generation of the Sloan Digital Sky Survey \citep[SDSS;][]{york00a} featured a program to study SNe in the southern equatorial stripe (Stripe 82) in the intermediate redshift range $0.1 \lesssim z \lesssim 0.4$  \citep{frieman08a, sako08a}.  The Supernova Legacy Survey (SNLS) \citep{astier06a,guy10a,conley11a, sullivan11a}, and the ESSENCE SN Survey \citep{Miknaitis07a,Wood-Vasey07a} target the redshift range $0.3 \lesssim z \lesssim 1.0$ for SNe~Ia.  Finally, the highest redshift SNe currently used for cosmological constraints, those at $z \greatsim 1$, are studied using the Hubble Space Telescope(HST) \citep{riess04a,riess07a,dawson09a,suzuki12a, rodney12a}.  To date, almost all of the SNe~Ia used for cosmological studies are spectroscopically confirmed.

Across these redshifts, the sample size of SNe~Ia has increased to the point where a careful treatment of systematic effects and uncertainties are an increasingly important part of the analysis of cosmological constraints.  The systematic errors come from both instrumental effects and astrophysical effects.   Examples of astrophysical effects include the correlation between SN~Ia properties and host galaxy properties, appearing as a dependence of the rate of SN~Ia events on stellar mass and star formation activity of the host galaxies \citep{sullivan06a}, a study of host star-formation rate and metal abundance versus SN properties \citep{gallagher05a}, and the relationship between host-galaxy mass and SN absolute brightness \citep{gupta11a, conley11a, meyers12a}.  \citet{hayden13a} studied the correlation between metallicity and Hubble residuals.  Efforts to optimally incorporate correlations between host galaxy properties and SN Hubble diagram residuals will benefit from the large sample of SN light curves and host galaxy photometry and spectra presented in this paper.  

Future surveys, including the Large Synoptic Survey Telescope \citep{lsst09a} and the Dark Energy Survey \citep[DES;][]{DES}, will generate deeper optical images  and observe larger volumes of the Universe than the current suite of surveys, resulting in hundreds to thousands of high-quality SN~Ia light curves each year.  A campaign to spectroscopically classify each candidate SN will require an impractically large allocation of existing or new telescope time.  These new surveys will instead require a change in the strategy of SN observation, transitioning to a new phase in which SNe~Ia are classified using only photometric light curves calibrated from a much smaller training set of spectroscopically-confirmed SNe~Ia.  The multi-band light curve data will be used to photometrically type SNe candidates as well as determine redshift and luminosity distance.  There have been a few initial studies of cosmology using SNe~Ia with only photometry \citep{barris04a,bernstein12a}, and recent work has simulated the effect of using photometric redshifts and SN classifications on large samples of data \citep{kessler10a, rodney10a}.  Spectroscopy of host galaxies of 400 candidate SNe from SNLS with the 2dF fiber positioner and spectrograph on the Anglo-Australian Telescope  demonstrates the ability to gather a large number of redshifts of SN host galaxies efficiently \citep{lidman12a}. Obtaining redshifts of a large number of SN candidate host galaxies allows one to improve SN classifications based on the light curve properties.  A large sample of high-quality light curves with accurate photometric typing allows tests of SN astrophysics and cosmology without spectroscopic confirmation.

\subsection{Science from SDSS-II SNe}
The large volume of the SDSS SN Survey has the potential to substantially increase the current sample size of photometrically classified SNe~Ia.  As described in \citet{frieman08a}, there was considerable effort to spectroscopically type SNe during the imaging program.  A  total of 518 spectroscopically-confirmed SNe~Ia were discovered using a number of telescopes including the Subaru telescope \citep{konishi11a}, the New Technology Telescope (NTT) and the Nordic Optical Telescope (NOT) \citep{ostman11a}, the 2.4 m Hiltner telescope at MDM, the Italian Telescopio Nazionale Galileo (TNG), the 10 m Keck I telescope \citep{foley12a}, the 3.5 m Astronomy Research Consortium telescope at Apache Point Observatory, the 9.2 m Hobby-Eberly Telescope \citep[HET;][]{shatrone07a} at McDonald Observatory, and the 4.2 m William Herschel Telescope (WHT) \citep{zheng08a}.  Additionally, 1070 \citep{sako11a} candidates were identified as likely SNe~Ia from their photometric light curves but were never spectroscopically classified.   

\citet{kessler09a} performed an analysis of a spectroscopically-confirmed SNe~Ia sample from 136 SNe~Ia from the first year of the SDSS-II SN Survey and a set of Nearby SN~Ia measurements.  Assuming a flat $\Lambda$CDM cosmology, baryon acoustic oscillations (BAO) \citep{eisenstein05a}, and cosmic microwave background (CMB) measurements from the Wilkinson Microwave Anisotropy Probe \citep{komatsu09a}, they found constraints on the equation of state parameter for dark energy, $w = -0.92 \pm 0.13($stat$)^{+0.10}_{-0.33}$(syst) using MLCS2K2 \citep{jha07a} light curve fits and $w = -0.92 \pm 0.11$(stat)$^{+0.07}_{-0.15}$ (syst) using SALT--II \citep{guy07a} light curve fits.  The first year sample of spectroscopically confirmed SNe~Ia from the SDSS-II SN Survey  was also used in \citet{lampeitl10a}, together with BAO from SDSS and the Two-Degree Field Galaxy Redshift Survey \citep[2dFGRS;][]{colless01a}, redshift-space distortions from 2dFGRS \citep{hawkins03a}, and the integrated Sachs-Wolfe (ISW) effect measured by SDSS \citep{giannantonio08a} to determine the equation of state parameter.  They found $w = -0.81^{+0.16}_{-0.18}$ (stat) $\pm 0.15$ (sys).  Other studies of cosmology using data from the SDSS-II SN Survey include a Bayesian cosmological analysis using photometric candidates from the three year data set \citep{hlozek12a} and constraints on nonstandard cosmological models \citep{sollerman09a}.

In addition to constraints on cosmology, the spectroscopically and photometrically identified SN~Ia have been used to probe SN~Ia astrophysics.  The SDSS SN samples has been used to study SN~Ia rates to $z \lesssim 0.3$ \citep{dilday08a,dilday10a}, rates in galaxy clusters \citep{dilday10b}, and SN~Ia rate as a function of host galaxy property \citep{smith12a}.  The rise and fall times of SN~Ia light curves were studied in \citet{hayden10a}.  \citet{hayden10b} use a sample of 108 spectroscopically-confirmed SNe~Ia to search for shock emission from interaction with a companion star.  Correlations between the metallicity and light curve shape and between the specific star formation rate and the Hubble residuals were found using a sample of spectroscopically-confirmed and photometrically-identified SNe~Ia \citep{dandrea11a}.  \citet{gupta11a} explored the correlation between the Hubble residuals of 206 SNe~Ia and the stellar mass and mass-weighted age of the host galaxies.  \citet{lampeitl10b} demonstrated a correlation between stellar mass and SN properties to high significance using spectroscopically-confirmed and photometric SNe~Ia.  A sample of nearly 200 spectroscopically-confirmed or photometrically-identified SNe~Ia from the SDSS-II SN Survey were used to find relations between distance from the center of the host galaxy and SN properties \citep{galbany12a}. A sample of spectroscopically-confirmed SNe~Ia was used to measure correlations between SNe~Ia properties and host galaxy properties including stellar velocity dispersion, age, metallicity, and element abundance ratios \citep{johansson12a}.

\subsection{Spectroscopy of Host Galaxies of SDSS-II SNe}

As discussed in this paper, the SDSS-II SN Survey data provide a sample to assess the accuracy of photometric typing techniques, by providing a redshift that can be used as a prior to light curve fits.  The host galaxy spectroscopic redshifts remove the ambiguity between cosmological reddening and SN observed color: a high-redshift SN may initially be identified as a low-redshift, highly-extincted SN when no redshift information is used.  The host galaxy redshift furthermore makes the Hubble diagram more reliable by providing precise positions on the redshift axis.

The 2.5-m aperture Sloan Telescope \citep{gunn98a} provides the spectroscopic capabilities required to determine the redshifts of a large, nearly complete SDSS-II SN host galaxy sample. A total of 2256 of these host galaxies were spectroscopically observed as part of SDSS-I/II.  The Baryon Oscillation Spectroscopic Survey \citep[BOSS;][]{dawson12a} in SDSS-III \citep{eisenstein11a} was used in 2009 and 2010 to increase the size of the SN spectroscopic host galaxy sample, obtaining spectra for 2998 galaxies that hosted SDSS-II SN candidates.  With the addition of this new data set, the host galaxy spectra offer a testbed for photometric typing techniques.

This paper is one of three coordinated studies that report results from the combined SDSS-II/BOSS sample of SNe and host galaxies.  The photometric SNe~Ia sample using host galaxy spectroscopic redshifts is used in \citet[][hereafter C13]{campbell12a} to compile a Hubble Diagram solely from photometrically classified SNe~Ia.  C13 tested the efficiency, contamination, and selection effects of photometrically-classified SNe~Ia from simulations.  They found $w = -0.96^{+0.10}_{-0.10}$ (stat) when assuming a constant $w$CDM cosmological model with 753 photometrically identified SNe~Ia in combination with the results of the power spectrum of SDSS luminous red galaxies (LRGs) \citep{reid10a}, the full WMAP7 Cosmic Microwave Background power spectrum \citep{larson11a}, and a measurement of $H_0$ \citep{riess11a}.  The full data release of SDSS-II SNe, host galaxy redshifts, and host galaxy spectroscopic properties is described in this paper and in Sako et al. (in preparation, hereafter S13). We refer to the public database described in S13, and used throughout this paper, as the SDSS-II SN Data Release\footnote{Public Location TBA}.  S13 describes the full three-year spectroscopically-confirmed SNe~Ia sample, the techniques for photometrically typing the SNe, and the resulting classifications from the sample.  This paper presents the new data from BOSS, demonstrates the impact of host galaxy spectroscopic redshifts from SDSS-I/II and BOSS on photometric typing, and presents the photometric and spectroscopic characteristics of the host galaxies.  The description of the data, target selection, observations, and redshift determination are presented in Section~\ref{sec:data}.  In Section~\ref{sec:analysis}, we describe SN typing and the impact of redshifts on those classifications.  Section~\ref{sec:interpretation} describes properties of the host galaxies derived from photometry, derived from the spectra, and trends between SN properties and their host galaxies.  In Section~\ref{sec:snprop}, we examine SN typing failure rates based on light curve properties.  Conclusions are presented in Section~\ref{sec:conclusion}.  Throughout the paper we assume a flat cosmology for determining host galaxy properties with $h=0.7$, $\Omega_M=0.274 $, and $\Omega_\Lambda=0.726$ \citep{white11a}.

\section{Observations and Data}\label{sec:data}

The SDSS-I and the SDSS-II Legacy Survey covered 8,400 deg$^2$ in five optical bands and obtained spectra of more than 1.2 million objects: the data are described in the Seventh Data Release of the Sloan Digital Sky Survey \citep[DR7;][]{abazajian09a}.  The 2.5 m Sloan Telescope \citep{gunn06a} is located at Apache Point Observatory and has a field of view of 7 deg$^2$ ($3^{\circ}$ diameter).  The imaging survey was conducted using a large mosaic CCD camera \citep{gunn98a}, leading to 230 million unique objects detected in $ugriz$ \citep{fukugita96a} and repeated imaging of a 300 deg$^2$ region leading to the SN light curves in the SDSS SN Survey \citep{frieman08a}.  The spectroscopic survey used two double spectrographs, each consisting of 320 optical fibers plugged into aluminum plates with a resolving power $R = \lambda$/FWHM $\sim 2000$ from the near ultraviolet to the near infrared \citep{smee12a}.  As mentioned in the introduction, 3,835 candidates from the SDSS SN Survey were hosted by galaxies that were already observed spectroscopically in SDSS before the commencement of the SN program.  

SDSS-III \citep{eisenstein11a}, which has been operating from 2008 and will continue into 2014, consists of four projects; BOSS is the extragalactic component designed for cosmology.  BOSS will measure the BAO feature using a sample of 1.5 million galaxies (including 150,000 from SDSS-I/II) to $ z < 0.7$ and Lyman-$\alpha$ forest absorption using more than 150,000 quasars at  $ 2.15 < z <3.5$  over 10,000  deg$^2$.  The BOSS targets are more numerous and reach 1--2 magnitudes fainter than the SDSS spectroscopic targets; this improvement is possible due to upgrades to the spectrographs, an increase in the number of fibers from 640 to 1000 \citep{smee12a}, and targeting quasars and luminous galaxies that are easier to identify from imaging data and spectroscopic confirmation.  The spectra from the first two years of BOSS observations, including most of the galaxies hosting SNe discovered in the SDSS-II SN Survey, are found in the Ninth Data Release \citep[DR9;][]{ahn12a}.

The improved sensitivity and expanded wavelength coverage of the BOSS spectrographs relative to SDSS motivated a decision to increase the sample size of SDSS-II host galaxy spectra.  The host galaxy spectroscopy from SDSS-I and BOSS is used in this analysis to determine redshift estimates for the SN candidates, assess the photometric classification using the spectroscopic redshift as a prior, and determine galaxy properties using photometry and spectroscopy.

\subsection{Imaging and Light curves}\label{subsec:Imaging}

The SDSS-II SN Survey occurred over the three Fall seasons of 2005-2007 following a commissioning period in Fall 2004.  The survey observed the five SDSS filters ($ugriz$) to obtain imaging over the 300 deg$^2$ southern equatorial region known as Stripe 82 ($-60^{\circ} < \alpha < 60^{\circ}, -1.25^{\circ} < \delta < 1.25^{\circ}$).  Stripe 82 was observed repeatedly with a cadence of roughly four days from September 1--November 30 in each season except near full moon (see \citet{frieman08a}).  Of the five filters, $gri$ were used in the SN detection algorithms.

During the survey, the data were processed daily to allow rapid response spectroscopy on as many SN~Ia candidates as possible.  Each image was compared to an image template derived from earlier observing seasons and the subtracted frames were searched for variable objects \citep{sako08a}.  Variations of at least three standard deviations above background in at least two contiguous pixels were considered possible variable objects in the first stage of analysis.  Regions of stars and known variability such as active galactic nuclei (AGN) were masked and not included in the search for possible SN candidates.  A visual screening process was performed on these candidates to reject artifacts and objects that were not SNe.  

Objects selected by the visual screening are designated `candidates' and are stored in a database.  The database contains 21,787 transient candidates on Stripe 82.  SN~Ia and core-collapse SN light curve templates were fit to the time-series photometry from each candidate and interesting candidates were again visually inspected.  Spectroscopy was performed on highly probable active SN~Ia candidates, leading to 518 spectroscopically-confirmed SNe~Ia using the same techniques as in the first year of the survey \citep{zheng08a}.  Other types of SNe were also identified during the classification and visual inspection.  A small number of these non-SN~Ia objects were spectroscopically classified: 7 SNe Ib, 11 SNe Ic and 67 SNe II.  The initial sample of 21,787 transients was filtered so that only candidates with light curves constructed from scene model photometry (SMP) are used \citep{holtzman08a}.  SMP uses a stack of images without convolution or spatial resampling to determine the brightness of the time-varying SN.  This sample includes all candidates that were detected on at least two dates, thereby eliminating most artifacts.  The filtered sample contains 10,468 candidates that are likely astrophysical in nature.  It is this sample that is described in detail in S13 and made available in the public database.

Although considerable resources and time were used for spectroscopic investigation of these candidates, many candidates were never observed spectroscopically.  While spectroscopic confirmation is no longer possible for these SNe, spectra of the host galaxies allows for improved cosmology and other studies requiring a precise redshift.  It also increases the sample of higher redshift SNe and can be obtained in a much more efficient manner.

\subsection{Photometric Classification}\label{subsec:TS}

The sample of 10,468 candidates was further decomposed into likely transients and likely SNe based on the classification method ``Photometric SN IDentification'' (PSNID) of \citet{sako08a,sako11a}.  To be photometrically classified as an SN, a candidate must show variability in only one observing season.  When this restriction is applied, the number of valid candidates is reduced to 6,697.  

In PSNID, a  $\chi^2$ value is calculated for each candidate by comparing the observed photometry against a set of SNe~Ia and core-collapse SN light curve templates.  The best matching SN type and parameters are recorded for each object with no redshift information. 

The technique was improved in \citet{sako11a} to include a calculation of the Bayesian probabilities that a candidate is a Type Ia, Ib/c, or II, denoted P$_{\rm Ia}$, P$_{\rm Ibc}$, and P$_{\rm II}$ respectively.  Empirical model errors were folded into the Bayesian probabilities, additional core-collapse templates were added, and a Markov Chain Monte Carlo approach was used to estimate the posterior probability distribution for the five parameters in the Bayesian classification.  A total of 26 candidates with variability in only one year fail to converge on a solution in PSNID and do not have well-defined Bayesian probabilities.  Full details of the photometric typing are found in S13.

  The sample of 6,671 objects fit by PSNID is divided into objects of varying probability, $\chi$$^2$ value, and quality of light curve coverage.  The cuts placed on the sample of 10468 candidates is shown in Table~\ref{tab:cuts}.  The details of the initial target ranking for host galaxy spectroscopy are described below and summarized in Table~\ref{tab:priority}.

\begin{deluxetable}{lr} 
\centering
\tablewidth{0pt}
\tabletypesize{\footnotesize}
\tablecaption{\label{tab:cuts} Sample Selection}
\tablehead{\colhead{Selection } & \colhead{Remaining Candidates} \\
\colhead{} & {For Host Spectroscopy}}
\startdata
 All Candidates &  10468 \\
 Variation only one year & 6697 \\
Have Bayesian SN Probability & 6671 \\
No SDSS spectra & 4415 
\enddata
\end{deluxetable}

{\bf High Probability SNe~Ia:  } 
Using the same criteria as \citet{sako11a} for the photometric SNe~Ia, the highest priority candidates for host spectroscopy have a Bayesian probability P$_{\rm Ia}$ greater than 0.9, a $\chi^2$ per degree of freedom less than 1.8, and a well-sampled light curve.  In this case, we define a well-sampled light curve as one with at least two measurements that constrain peak and fall time; one measurement must be within five days (rest-frame) of peak luminosity of the best fit SN~Ia template and one measurement must be between five and 15 days (rest-frame) after peak luminosity of the best fit SN~Ia template.  A total of 1952 candidates meet these criteria.

{\bf Possible SNe~Ia:  }
We increase the number of possible SNe~Ia candidates by loosening the restrictions from the high-probability photometric SNe~Ia sample.  For the next priority in host galaxy observations, we require that candidates meet two of the three above conditions and have $\chi^2$ per degree of freedom less than 2.0.  There are a total of 652 possible SNe~Ia candidates.

{\bf Low Probability SNe~Ia: }  
The remaining candidates that have Bayesian probability P$_{\rm Ia} >$  P$_{\rm Ibc}$ and  P$_{\rm Ia} > $  P$_{\rm II}$ are designated low probability SNe~Ia.  There are 963 low probability SNe~Ia candidates.

{\bf Core-Collapse:  }
A large sample of possible core-collapse SNe was identified during the SDSS-II SN Survey.  A candidate is classified as likely core-collapse when the best fit Type Ibc or II Bayesian probability is greater than the SNe~Ia Bayesian probability; 3098 candidates are identified as core-collapse candidates and assigned fourth priority for host galaxy spectroscopy.

{\bf Remainder Sample:  }
All 3655 remaining candidates are denoted members of the ``remainder sample''.  This sample includes candidates that show variation in more than one year, and candidates where the classification did not converge in the light curve fits.  These remainder sample host galaxies are given lowest priority in target selection and chosen to examine possible incompleteness in the SDSS SN sample.

The likely host galaxy of each of the 10468 variable candidates was initially identified from DR7 photometry, as described below.  The SDSS spectral database was examined for existing host galaxy spectra: 2256 host galaxy spectra existed in the database and are categorized according to spectroscopic prioritization in Table~\ref{tab:priority}.  The redshifts from SDSS can be used to assess the SN classification of these candidates by using the host galaxy redshift as a prior.  The remaining candidates were prioritized for spectroscopy with BOSS.

\begin{deluxetable*}{lccccc}
\centering
\tablewidth{0pt}
\tabletypesize{\footnotesize}
\tablecaption{\label{tab:priority} Target Priority}
\tablehead{\colhead{Selection} & \colhead{\# Meeting} & \colhead{\# SDSS } & \colhead{\# BOSS} & \colhead{Median $r$--band} & \colhead{Max $r$--band}  \\
\colhead{Method} & \colhead{Criteria} & \colhead{Redshifts} & \colhead{Redshifts} & \colhead{\texttt{fiber2mag}} & \colhead{\texttt{fiber2mag}}  }
\startdata
 High Probability & 1952 & 262 & 991 & 21.05 & 23.94 \\
 SNe~Ia \\
 \\
 Possible SNe~Ia & 652 & 118 & 193 & 20.35  & 23.21   \\
 \\
 Low Probability & 963 & 95 &  585 & 20.74 & 23.65 \\
 SNe~Ia \\
 \\
 Core-Collapse & 3098  & 489 & 823 & 20.61 & 23.55 \\
 \\
 Remainder Sample & 3655  & 1247 & 406 & 19.45 & 23.02 
\enddata
\end{deluxetable*}

\subsection{BOSS Target Selection}\label{subsec:BOSSts}

A majority of fibers ($\sim$ 80\%) in BOSS observations are dedicated to BAO galaxy or quasar targets; some of the remaining fibers were made available to diversify the spectroscopic sample by observing other objects in the BOSS footprint.  Dubbed ancillary programs, these additional fibers were assigned to smaller programs proposed by members of SDSS-III, as explained in the appendix of \citet{dawson12a}.  An ancillary program was approved to obtain spectroscopy of all host galaxies of transients from the SDSS-II SN Survey.

Candidates for host galaxy spectroscopy were chosen from the reduced sample of 10,468 variable objects in the 220 deg$^{2}$ region $\alpha_{\rm J2000}$=[-42.9$^{\circ}$,44.9$^{\circ}$], $\delta_{\rm J2000}$=[-1.26$^{\circ}$,1.26$^{\circ}$] that was covered by BOSS.  A total of 3375 objects were selected using the broad approach of targeting host galaxies of likely SN first, denoted as Algorithm One in C13, and assigning additional fibers to non-SN, denoted as Algorithm Two in C13.   

A restriction on $r$-band \texttt{fibermagnitude} was imposed to ensure a sufficient data quality from the BOSS observations, where \texttt{fibermagnitude} is determined from the integrated flux inside a 3\arcsec\ diameter aperture predicted from SDSS imaging.  BOSS targeting priority was given to those objects with $r$-band \texttt{fibermagnitude} $<21.25$  while an additional restriction was loosely imposed by rejecting targets fainter than $r$-band \texttt{fibermagnitude} of 22. The targets were visually inspected from the SDSS DR7 images to determine the most likely host galaxy of the candidate.  The three nearest galaxies were examined and in most cases the nearest galaxy was considered the most likely host.  Occasionally the second nearest galaxy (4\%) or the third nearest galaxy (1\%) was determined to be the most likely host.  These galaxies were assigned \texttt{ANCILLARY\_TARGET1} flags 36, 37, 38, respectively, where \texttt{ANCILLARY\_TARGET1} denotes the target flag for BOSS ancillary programs on Stripe 82 \citep{dawson12a}.  Selection of the second or third nearest galaxy generally occured if the nearest galaxy was either a mis-classification, bright star or artifact, or a clear background object.  

For most of the targets, the center of the host galaxy was selected as the position of the fiber.  Additionally, 282 candidates occurred in a host galaxy for which there is already existing SDSS spectra.  In only these cases the location of the original variability was assigned as the location of the fiber to study the local environment of the candidate.  These objects are assigned a BOSS \texttt{ANCILLARY\_TARGET1} value equal to 39.  A sample of 52 active SNe~Ia were found in SDSS galaxy spectra \citep{krughoff11a}; two of these host galaxies were selected to be re-observed with BOSS and were given BOSS \texttt{ANCILLARY\_TARGET1} equal to 40.  The remainder sample discussed in Section~\ref{subsec:TS} was selected at random from the parent sample with a loose restriction of $r< 21.5$ to quantify selection and contamination biases.  These targets were also visually inspected to select the most likely host from the nearest three galaxies.

\subsection{BOSS Observations}\label{subsec:BOSS}

\begin{figure}[htbp]
\epsscale{1.1}
\vspace{-1.5in}

\includegraphics[scale=0.45]{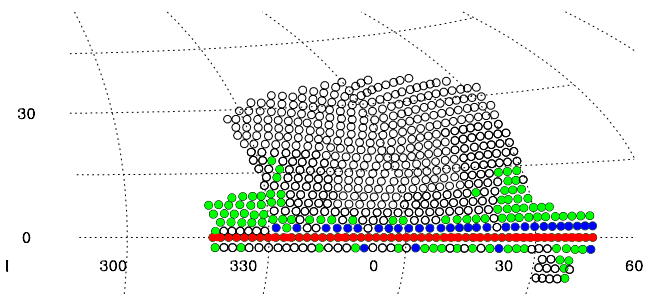}
\caption{\label{fig:footprint}  The projected five year coverage of BOSS in the South Galactic Cap in equatorial coordinates.  The BOSS plates on Stripe 82 used for the SN host galaxy candidates are shown in red. Plates bordering Stripe 82 that contain SN host galaxy candidates have a smaller number of targets per plate and are shown in blue.  Regions included in DR9 are shown in green and regions to be observed in the final three years are shown as open circles.}

\end{figure}

In total 4777 targets were requested for BOSS fiber assignment as part of the SDSS SN ancillary science program.  Due to restrictions and priorities of BOSS fiber assignments (primarily from a finite number of available fibers and collisions with higher priority targets) only 3761 unique candidates were observed.  These targets were distributed over 78 BOSS plates in Stripe 82 and 33 BOSS plates that border Stripe 82.  Figure~\ref{fig:footprint} shows the footprint of the plates that contained SN host galaxy targets relative to the overall BOSS footprint in the South Galactic Cap, the full BOSS footprint is found in Figure~1 of \citet{dawson12a}.

The host galaxy spectroscopy occurred in 2009 and 2010 in a series of 15 minute exposures until a certain depth was reached for each plate. Observations ceased when a real time reduction pipeline yielded a signal-to-noise ratio (S/N), such that (S/N)$^2$/pixel $\ge$ 26 over a synthetic $i$ filter for objects with $i=21$  and (S/N)$^2$/pixel $\ge 16$ at $g$=22 using \texttt{fiber2magnitudes} corrected for Galactic extinction \citep{schlegel98a}.  As with \texttt{fibermagnitude}, \texttt{fiber2magnitude} is determined from the integrated flux inside a 2\arcsec\ diameter aperture predicted from SDSS imaging.  The observations typically consisted of between four and eight 15 minute exposures.  The exposures sometimes extended over multiple nights, depending on various factors including weather conditions, visibility of targets, and time of first exposure.  

BOSS first observed Stripe 82 during the commissioning period in Fall 2009 and included 10 plates containing SN targets.  Full survey observations of Stripe 82 continued into the Spring 2010, producing another 795 targets with a confident redshift.  In Fall 2010 the entire Stripe 82 region was retiled and the remaining targets were either observed for the first time or re-observed if there was not a reliable redshift from the first year observation, see Section~\ref{subsec:redshifting} for details on redshift reliability.  When a target was observed during both commissioning and regular operation, the later observation is considered the primary spectrum for this analysis.

\subsection{Photometric Classification and Host Galaxy Differences Between C13 and the SDSS-II SN Data Release}\label{subsec:campbell}
A cosmological study was performed using the photometrically-classified SNe~Ia from the SDSS-II SN Survey in C13.  These SNe~Ia candidates were classified using spectroscopic redshifts determined by SDSS and BOSS.  While the host galaxies described here and in C13 are the same, the SN classifications without galaxy redshifts changed somewhat between C13 and the SDSS-II SN Data Release. In addition, the host galaxy associations for a number of candidates is different.  We describe the differences here.

The target selection for BOSS spectroscopy of SNe host galaxies consisted of two separate philosophies: observe the host galaxies of as many likely SNe as possible, assigning fibers to likely SNe~Ia with the highest prioritization; and gather spectra for other interesting transients with the allotted fibers. In defining the original SN~Ia samples, C13 uses the SN classification that was available in 2009.  The target classification described in Section~\ref{subsec:TS} uses a newer photometric-only SN classification that will be released in S13.  We choose the updated classification in this paper because that is the classification scheme that best matches future SN campaigns.  This is also the sample that can be recreated using the information in the SDSS-II SN Data Release.  

The main difference between the target classification here and C13 is in the number of objects assigned to the likely SNe~Ia class and the remainder class.  In evaluating the original sample, the objects selected for observation as described in C13 include 2364 objects classified as high probability, possible, or low probability SNe~Ia.  C13 also selected 417 objects initially classified as core-collapse in their observed sample and 980 objects in their ``unbiased'' sample (equivalent to the remainder sample in this paper).  On the other hand, with the new classification of transients without spectroscopic redshifts, we identify samples of 1769 SNe~Ia, 823 core-collapse SN, and 406 remainder sample objects.  Objects typed as SN~Ia in the final Hubble Diagram of the SN cosmology analysis in C13 are flagged with the keyword \texttt{Campbell2013} in the SDSS-II SN Data Release. 

As discussed in Section~\ref{subsec:BOSSts}, the original identification of host galaxies was derived from visual inspections of DR7 images.  Occasionally, galaxies other than the nearest to the transient event were selected as the likely host galaxy.  C13 used the host galaxies from the initial BOSS target selection.  The host galaxy identification was updated for the SDSS-II SN Data Release using deeper imaging from SDSS DR8 \citep{aihara11a}.  All primary objects within 30\arcsec\ of the SN candidate position were selected as possible host galaxies.  The primary object nearest to the SN candidate in terms of light radius was selected as the host galaxy.  The primary benefit of using this automated method with DR8 is that some likely host galaxies were not visible during visual inspection of DR7.  Automated matching sometimes misses candidate host galaxies in which the light radius parameters were not able to be determined (primarily faint galaxies).  The automated matching also fails to identify the correct host galaxy in the case of extended objects which are deblended into several discrete galaxies. 

Host galaxy association changed between C13 and the SDSS-II SN Data Release for nearly one-tenth of all objects.  371 of these objects with host galaxy changes were observed in the BOSS sample.  From the sample of 3,354 objects selected as the nearest galaxy (BOSS \texttt{ANCILLARY\_TARGET1} 36), 360  were no longer identified as the likely host galaxy.  Of the 123 targets initially selected as the second or third nearest host galaxy (BOSS \texttt{ANCILLARY\_TARGET1} 37 and 38), 11 were removed from the final list of likely host galaxies used for the SN classification in this paper and in S13.    

We have continued to improve the sample and data reductions since C13.  Although this has changed the sample,
the conclusions of C13 remain valid.  We have found similar results to C13 as described in Section~\ref{subsec:lightcurveparam} and Section~\ref{subsec:completeness}.

\subsection{Spectroscopic Data Reduction and Redshift Determination}\label{subsec:redshifting}
The BOSS data were reduced in the pipeline referred as \texttt{idlspec2d} described in \citet{dawson12a}\footnote{The analysis of SNe host galaxy spectra obtained by BOSS in determining redshifts was performed during early stages of the BOSS pipeline development, primarily using v5\_4\_14 and v5\_4\_31 of \texttt{idlspec2d}.}.  The two-dimensional images are bias subtracted, flat-field corrected, and masked for pixels with known defects or affected by cosmic rays.  The two-dimensional image is then transformed into a one-dimensional image using a flat-field exposure to both determine the profile of each fiber projected onto the CCD and to normalize fiber-to-fiber throughput variations.  The wavelength solution is determined from an arc exposure.  The arc exposure is performed using a  mercury-cadmium lamp for emission in the blue and a neon (neon-argon upgrade for BOSS) for emission at the red end of the spectrum.  A sky model is created from the spectra of fibers assigned to locations on the sky in which no objects were detected in the original SDSS imaging.  This sky model depends on the fiber position and is subtracted from each one-dimensional spectrum. The collection of one-dimensional spectra for each object on the plate is then combined into the coadded spectrum found in DR9.

\begin{figure*}[htbp]
\begin{center}
\includegraphics[scale=0.95]{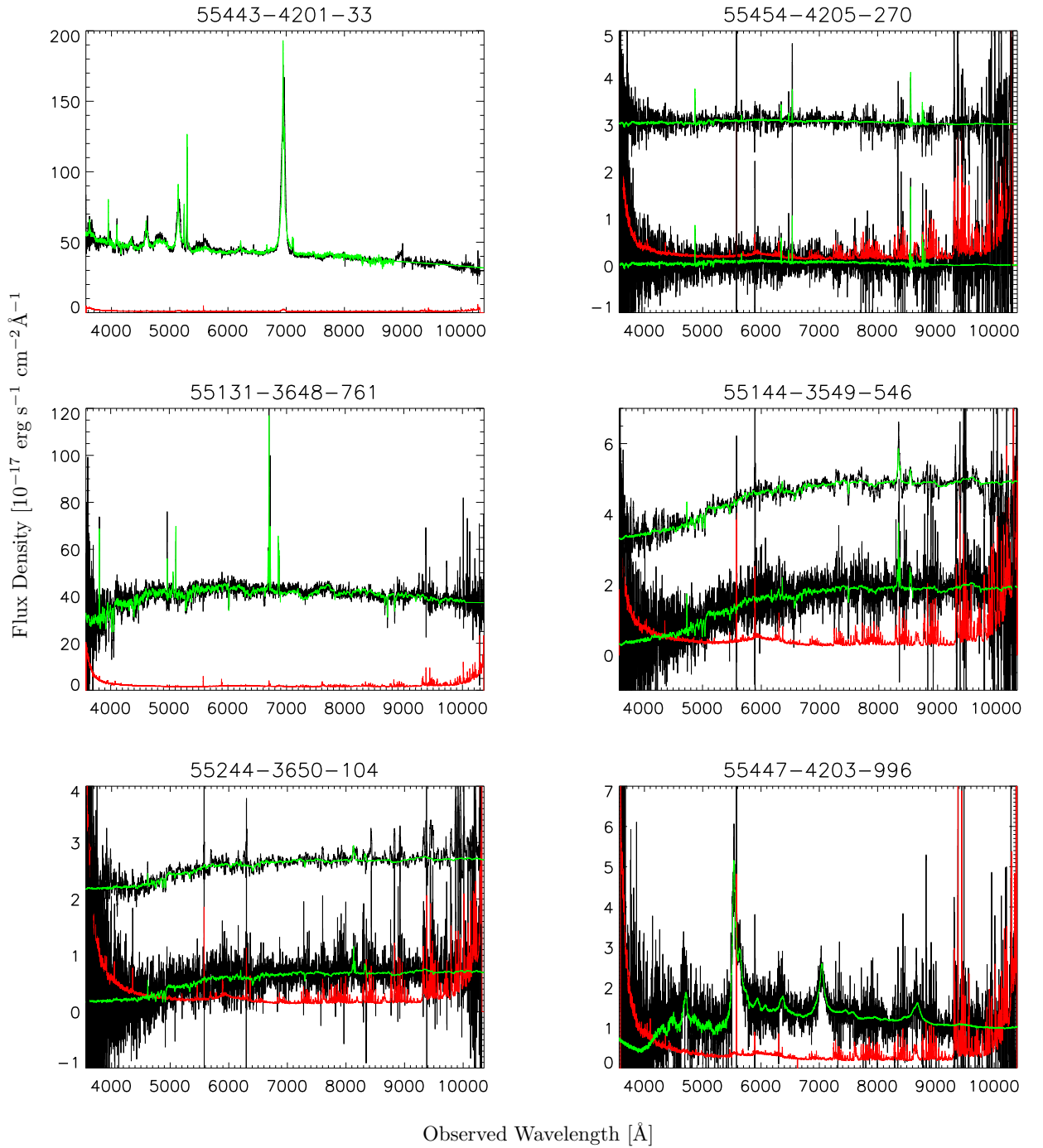}
\vspace{-3 in}
\caption{\label{fig:spec} A selection of BOSS spectra with secure redshift from the observed sample.  The observed flux is shown in in black, the best-fit template in green and the error in red.  For the three plots with two spectra shown, the top spectrum is the observed flux with a five-pixel smoothing and offset by 10 $\times$ 10$^{-17}$ erg s$^{-1}$ cm$^{-2} $\AA$^{-1}$.  The MJD, plate, and fiberid for each object is shown.  TOP LEFT: One of the highest S/N spectra of  a broadline QSO at z=0.0580.  The variability appeared at 0.20\arcsec\ from the center so this candidate was potentially selected by AGN activity.  TOP RIGHT: One of the lowest S/N spectra of a galaxy at z=0.305.  MIDDLE LEFT: One of the brightest objects (r-band cmodelmag 14.062) of a starforming galaxy at z=0.02073.  MIDDLE RIGHT: A faint galaxy (r-band magnitude 24.80) at z=0.238.  BOTTOM LEFT: A galaxy near median S/N, redshift, and magnitude.  BOTTOM RIGHT: The highest secure redshift in our sample, a broadline QSO at z=3.55.  Table~\ref{tab:spec} presents details on the six spectra.   }
\end{center}
\end{figure*}

A sample of spectra from the BOSS SN ancillary program is shown in Figure~\ref{fig:spec}.  These spectra were chosen to demonstrate the range of S/N, magnitude and redshift of the target sample, as detailed in Table~\ref{tab:spec}.
A redshift and galaxy classification is performed by fitting a set of principle component analysis (PCA) templates derived from BOSS data to each coadded spectrum \citep{bolton12a}.  Warnings for suspect spectra are encoded in the parameter \texttt{ZWARNING} found in each data release.  The pipeline had a confident redshift (\texttt{ZWARNING} = 0) for 3447 (91.6\%) of the targets.  The most common redshift failure (\texttt{ZWARNING}=4) occurs when there are at least two different redshifts with templates of similar values of $\chi^2$ per degree of freedom; 91.4\% of the failed redshifts were assigned \texttt{ZWARNING}=4 flags.

\begin{deluxetable*}{lrcccccccr}
\centering
\tablewidth{0pt}
\tabletypesize{\footnotesize}
\tablecaption{\label{tab:spec} Properties of Figure~\ref{fig:spec} Spectra}
\tablehead{\colhead{Objid} & \colhead{Corresponding } & \colhead{Redshift } & \colhead{cmodelmag} & \colhead{r-band} & \colhead{SN Distance }  & \colhead{SN } & \colhead{MJD } & \colhead{Plate } & \colhead{Fiberid }\\ 
\colhead{} & \colhead{ CID} & \colhead{ } & \colhead{($r$)} & \colhead{S/N} & \colhead{ from Center(\arcsec)}  & \colhead{Type} }

\startdata

1237663543148675172 & 20259 & 0.0580 & 17.8 & 50.4 & 0.13 & Ia & 55443 & 4201 & 33 \\
 $\ast$  & 20953 & 0.305 & & 0.27 & 0.22 & Ia & 55454 & 4205 & 270 \\
1237663784750940246 & 13195 & 0.0207 & 17.8 & 24.3 &  6.04 & Ibc & 55131 & 3648 & 761 \\
1237678595929931869 & 17414 & 0.270 & 21.1 & & 0.84 & II & 55144 & 3549 & 546 \\
1237657584950379205 & 20768 & 0.239 & 22.0 & 2.8 & 2.0 & Ia & 55244 & 3650 & 104 \\
1237663479261495791 & 3562  & 3.55 & 21.3 & 5.1 & 0.1 & II  & 55447 & 4203 & 996 

\enddata
\tablenotetext{}{ Properties for the spectra in Figure~\ref{fig:spec}. The SN type here is based from initial Bayesian probability assuming a flat prior used for target selection.   OBJID is the identification number used for matching spectra in DR9.  CID is the SN candidate identification used in the SDSS-II SN Data Release.   * This galaxy does not have an OBJID and can be matched on plate, fiberid, and mjd in DR9.}
\end{deluxetable*}

\begin{figure*}[htbp]
\epsscale{0.85}
\centerline{
\includegraphics[scale=0.45]{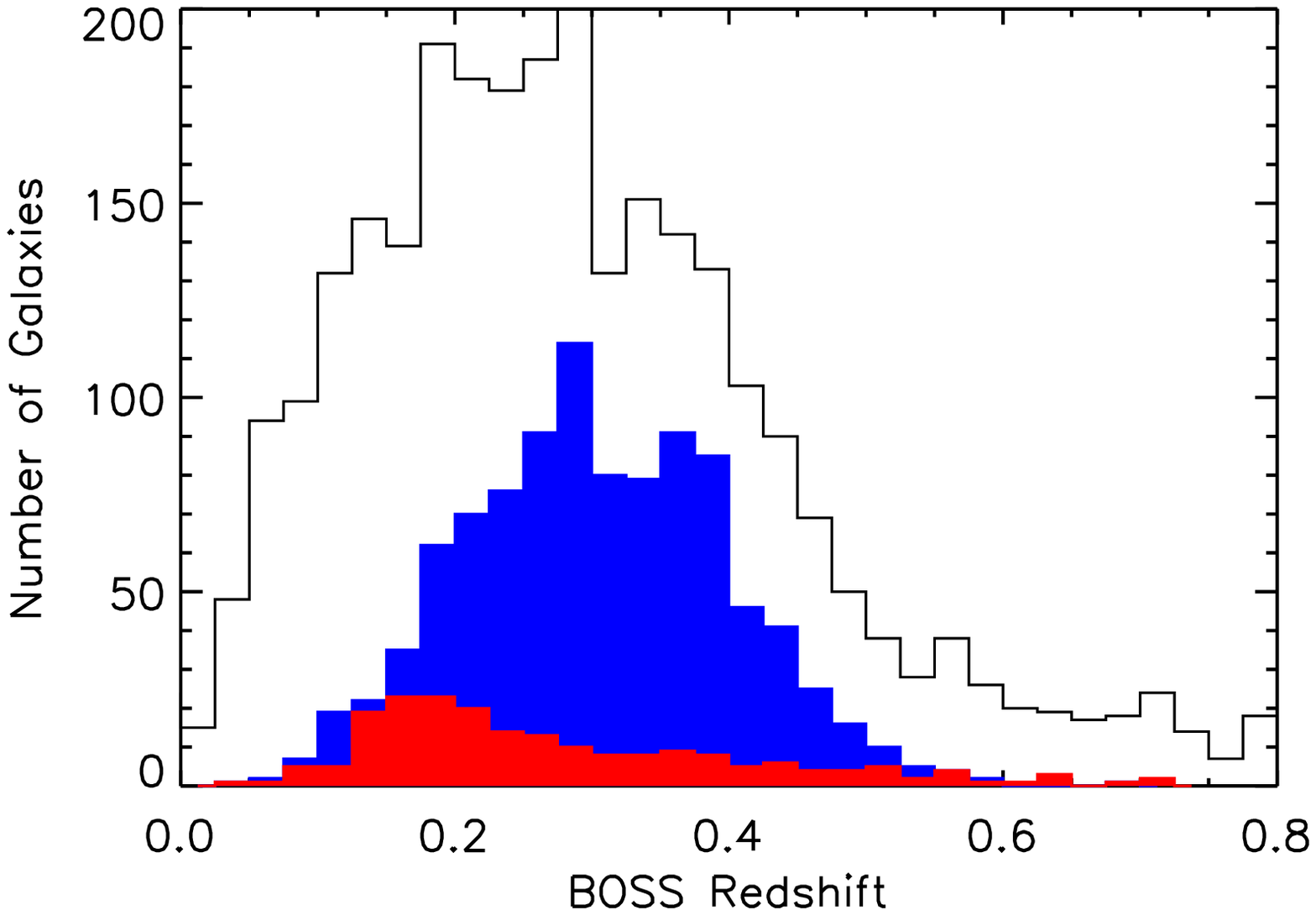}
\includegraphics[scale=0.45]{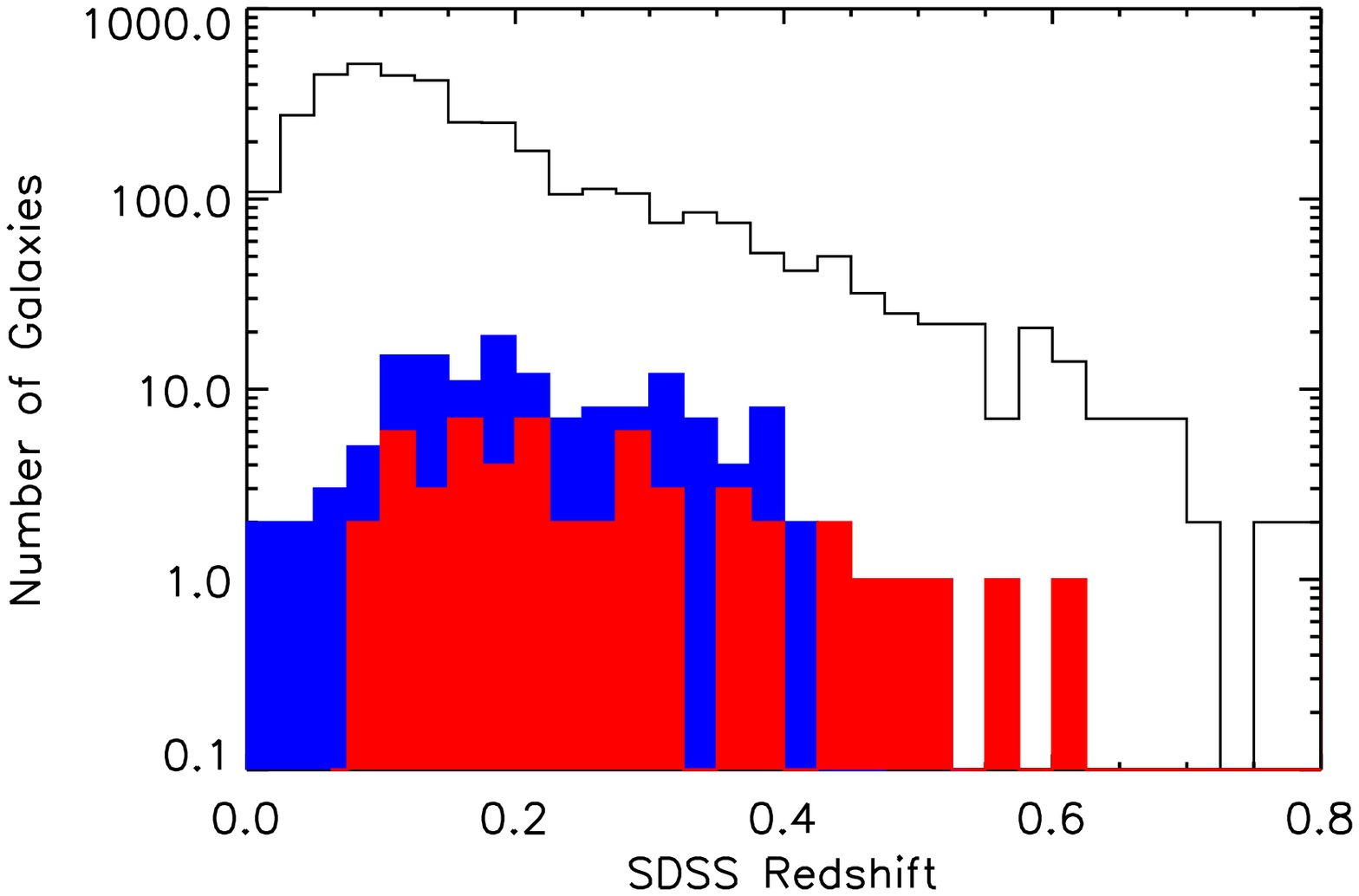}
}
\centerline{
\includegraphics[scale=0.45]{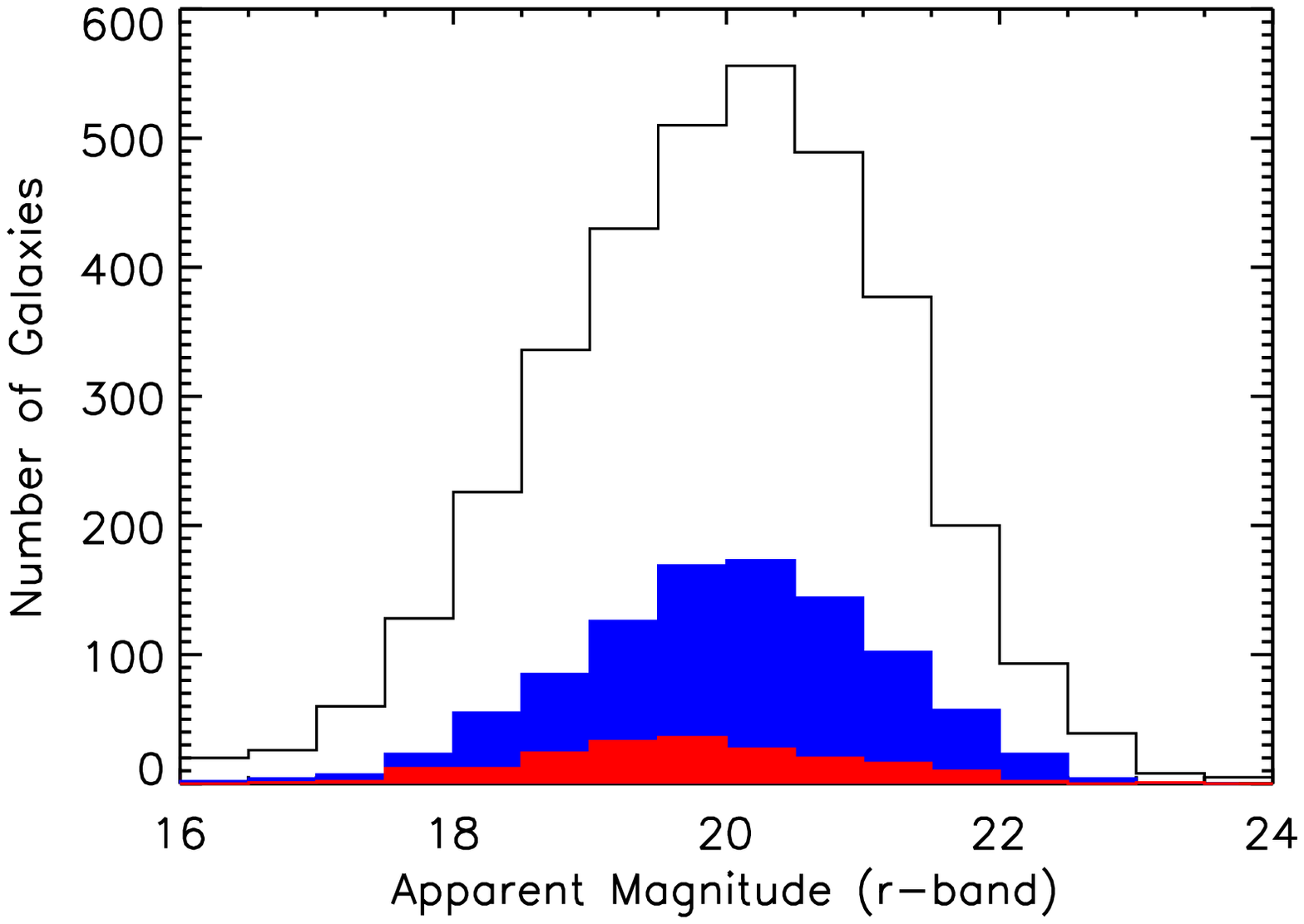}
\includegraphics[scale=0.45]{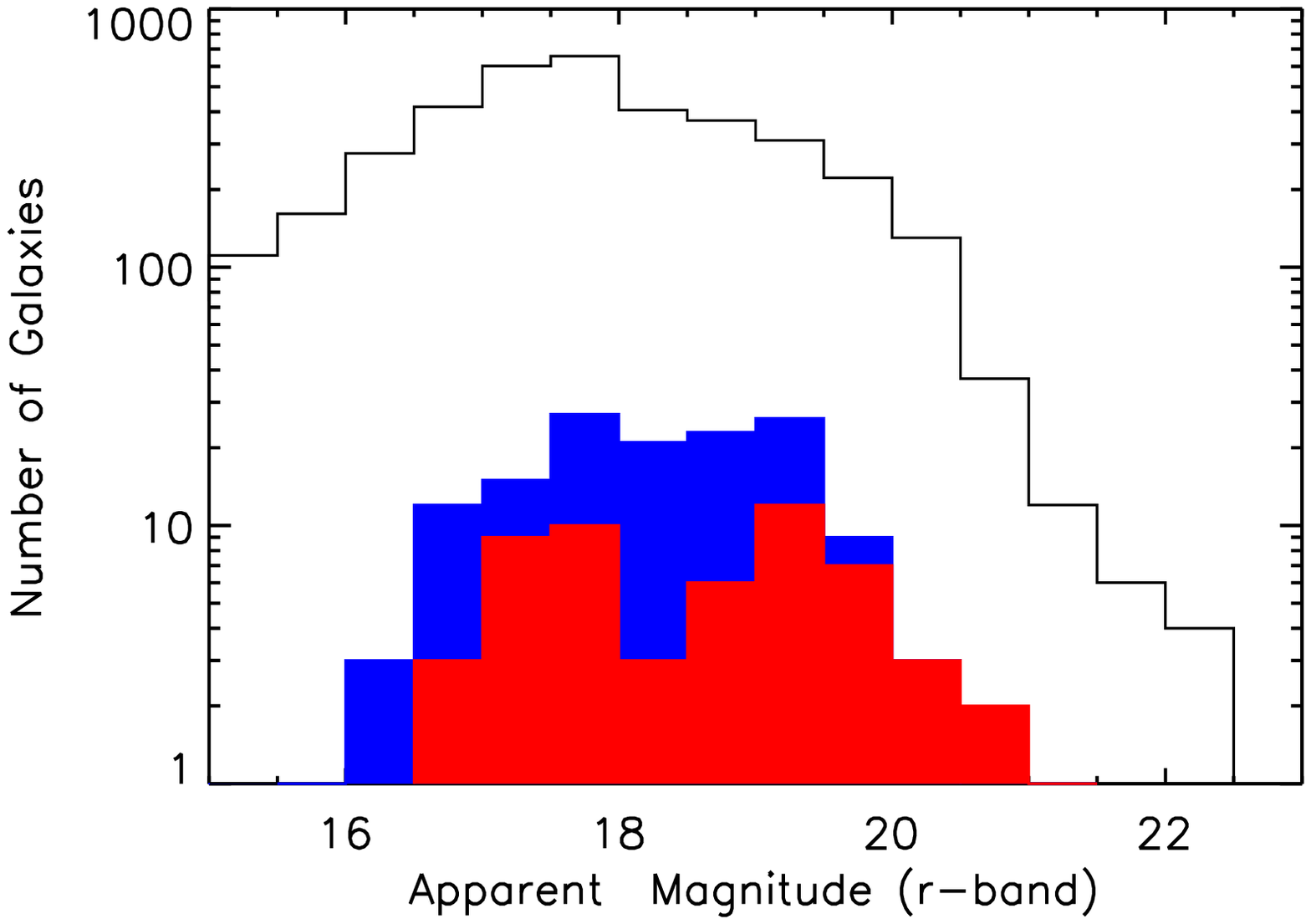}
}
\caption{\label{fig:zhist}TOP LEFT: The distribution of redshifts (for redshifts $<$ 0.8) of the observed BOSS host galaxy sample.  The entire BOSS sample is represented by the solid line, the blue curve shows the distribution of SNe~Ia and the red curve the distribution of core-collapse SNe.  The SNe~Ia and core-collapse SNe are classified using the full photometric classification described in Section \ref{subsec:phottyping}.  TOP RIGHT:  The distribution of redshifts (for redshifts $<$ 0.8) of the observed SDSS SN host sample with the same line scheme as the top panel.  The SDSS sample is put on a logarithmic scale because the number of candidates is dominated by the remainder sample.  BOTTOM LEFT: A histogram of the $r$-band cmodel magnitude for the BOSS sample with the same line scheme as before. BOTTOM RIGHT: A histogram of the $r$-band cmodel magnitude for the SDSS SN host sample with the same line scheme as before.}
\end{figure*}

\begin{figure}[htbp]
\epsscale{0.85}
\includegraphics[scale=0.5]{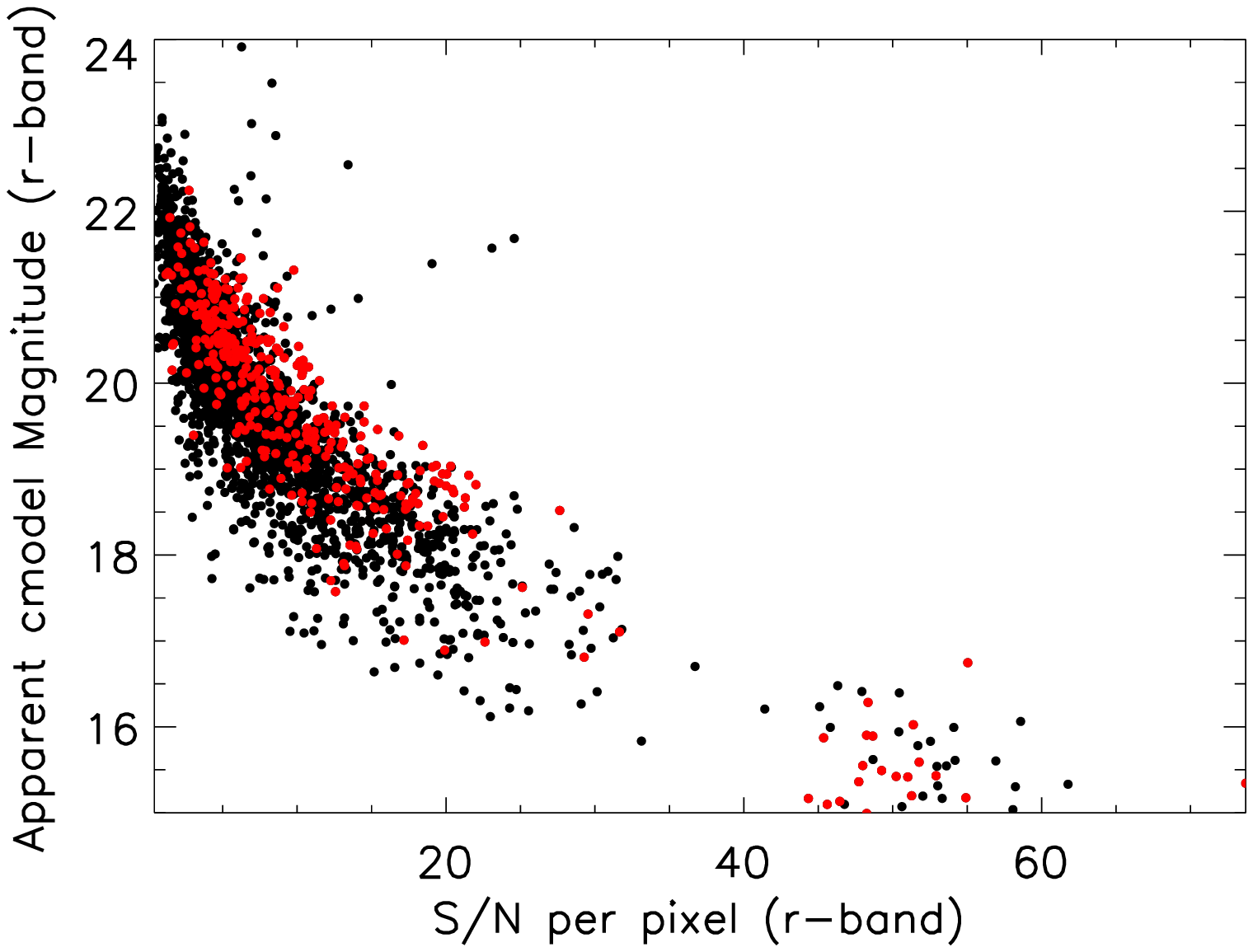}
\includegraphics[scale=0.5]{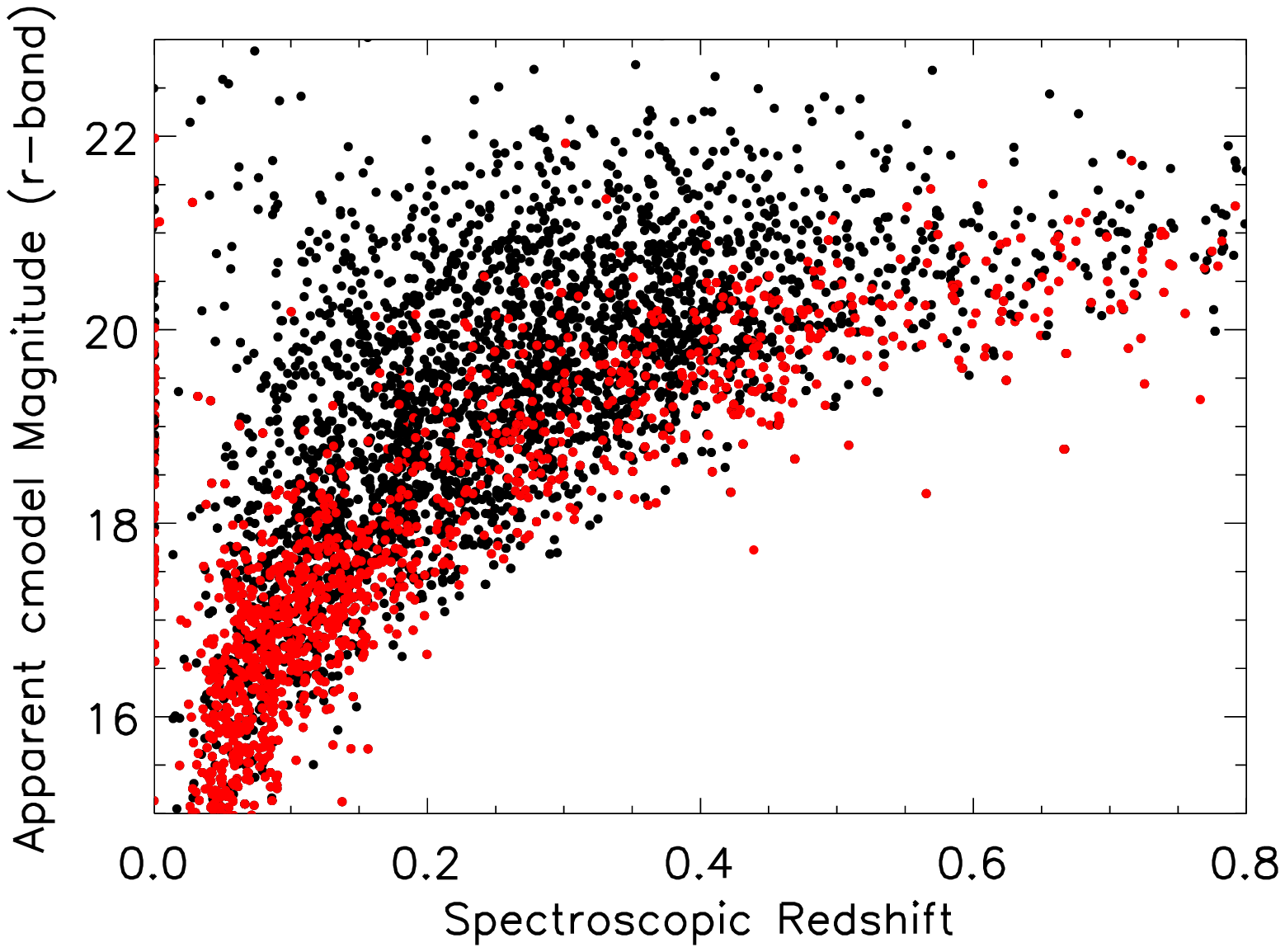}  
\caption{\label{fig:mag-sn} TOP: The $r$-band apparent magnitude (cmodelmag) as a function of synthetic $r$-band S/N per pixel for the BOSS host galaxy spectroscopic sample.  The remainder sample is shown in red.  BOTTOM: The $r$-band apparent magnitude  (cmodelmag) vs redshift for the entire spectroscopic sample with the remainder sample in red.}
\end{figure}

\begin{deluxetable*}{lrccrrr}
\centering
\tablewidth{0pt}
\tabletypesize{\footnotesize}
\tablecaption{\label{tab:target} BOSS Target Properties}
\tablehead{\colhead{m$_\textrm{r}$ Bin} & \colhead{\# of } & \colhead{Median} & \colhead{RMS } & \colhead{ Secure } & \colhead{Median }  & \colhead{RMS } \\
\colhead{} & \colhead{ Objects} & \colhead{Redshift } & \colhead{Redshift} & \colhead{  Redshift } & \colhead{S/N}  & \colhead{ S/N} \\
\colhead{} & \colhead{ with secure $z$} & \colhead{ } & \colhead{} & \colhead{   } & \colhead{per pixel}  & \colhead{per pixel }}
\startdata

      17.0-17.5 &            3 &     0.0441 &     0.02641 &       1.000 &       52.5 &       33.5 \\ 
      17.5-18.0 &           54 &     0.0554 &     0.0177 &       1.000 &       51.0 &       16.4 \\ 
      18.0-18.5 &           10 &     0.0662 &     0.0454 &       1.000 &       40.3 &       9.1 \\ 
      18.5-19.0 &           34 &     0.0966 &     0.0486 &       1.000 &       28.6 &       7.2 \\ 
      19.0-19.5 &           98 &      0.120 &     0.0664 &       1.000 &       21.3 &       8.3 \\ 
      19.5-20.0 &          275 &      0.176 &     0.0889 &      0.996 &       16.8 &       6.2 \\ 
      20.0-20.5 &          399 &      0.220 &      0.100 &      0.998 &       11.9 &       4.3 \\ 
      20.5-21.0 &          540 &      0.265 &      0.120 &      0.987 &       8.3 &       3.1 \\ 
      21.0-21.5 &          609 &      0.312 &      0.147 &      0.990 &       5.8 &       2.3 \\ 
      21.5-22.0 &          522 &      0.351 &      0.161 &      0.987 &       3.9 &       1.5 \\ 
      22.0-22.5 &          359 &      0.340 &      0.160 &      0.932 &       2.6 &       1.1 \\ 
      22.5-23.0 &          143 &      0.347 &      0.154 &      0.872 &       1.7 &      0.6 \\ 
      23.0-23.5 &           56 &      0.392 &      0.187 &      0.727 &       1.0 &      0.7 \\ 
      23.5-24.0 &            8 &      0.536 &      0.163 &      0.364 &      0.86 &      0.16 \\ 
      24.0-24.5 &            1 &  -0.000324 &   &      0.50 &      0.28 &

\enddata
\end{deluxetable*}

The galaxy templates for BOSS have been chosen for optimal redshift performance for the luminous galaxy sample used to constrain BAO in BOSS.  The pipeline templates therefore cover a relatively small parameter space in terms of galaxy properties.  For targets in the primary BOSS sample of color-selected  massive galaxies, the pipeline has a success rate of 98.7\% \citep{bolton12a}.  The SN host galaxy sample, on the other hand, consists of a much more heterogeneous galaxy population and the pipeline efficiency might be suspect.  We verified with a visual inspection the pipeline redshifts for this SN host galaxy sample as described below.  

Another redshifting routine was made available to BOSS by the WiggleZ Survey \citep{drinkwater10a} called ``RUNZ''\footnote{Maintained by Scott Croom.}, allowing us to perform manual redshift estimates.  RUNZ is a modified version of the code used for the Two Degree Field Galaxy Redshift Survey \citep[2dFGRS;][]{colless01a} and the 2dF-SDSS LRG and QSO Survey \citep[2SLAQ;][]{cannon06a} which includes a total of 17 templates; these templates cover the classes of star, galaxy, emission line galaxy, and quasar.  Redshifts are determined through a cross-correlation fit between the spectrum and each template.  RUNZ also determines a best match to emission line features as a second redshift estimate.  The graphical interface is very convenient for manual redshifting because it allows one to assign redshifts to specific spectral features and then identifies the location of additional lines at that redshift.  A template at any given redshift can also be displayed as an overlay with each spectrum for easy comparison.  All 3741 BOSS spectra were manually inspected using this graphical interface.  A redshift was given high manual confidence when there were at least two spectral features identifiable by visual inspection at that redshift.  We then compared the visual redshift with the \texttt{idlspec2d} pipeline redshift.  All objects that had no \texttt{ZWARNING} flags, good manual confidence, and in which the redshifts agreed were considered robust (3363 objects).

There were 398 targets with disagreement between the \texttt{idlspec2d} pipeline and visual redshift, where the \texttt{ZWARNING} flag was not zero, or where there was low confidence in the original visual inspection; these were all visually inspected again.  In this second manual inspection, two co-authors examined the best fits using the \texttt{idlspec2d} pipeline templates and the RUNZ graphical interface.  A redshift was declared good during the second inspection if the two inspectors agreed on the redshift.  

After the second manual inspection, 3,520 of the 3,761 observed objects were determined to have a secure redshift.  Table~\ref{tab:target} shows the efficiency for obtaining a secure redshift as a function of target magnitude and other observational properties.  These objects are found in the online table with \texttt{mwarning} equal zero, where \texttt{mwarning} is a flag that denotes the quality of the final manual inspection.  In total, 80 of the final redshifts used in this analysis do not come from the \texttt{idlspec2d} pipeline with \texttt{ZWARNING} equal zero, but from a combination of RUNZ and additional manual redshifting.  

There are 351 spectra from the BOSS commissioning period in which we have confident redshifts.  The plates containing these 351 spectra are marked as bad in DR9 and are not contained in the \texttt{spAll} summary file.  The reduced data for these plates exists in the individual plate directories, but only contain data for one of the spectrographs.   For this paper, these objects were reduced using an earlier version of \texttt{idlspec2d} including both spectrographs and therefore will be used only to provide host galaxy redshifts and will not be used in further analysis requiring calibrated spectra.  The raw data were released in DR9, but are not recommended for further analysis due to calibration uncertainty associated with commissioning data.  Objects with redshifts from the commissioning data are identified with column \texttt{spectra\_commission} in the public release of SDSS SN light curves described in S13.

As described in Section~\ref{subsec:campbell}, there were 371 objects with confident redshifts observed as part of this BOSS ancillary project that later had a change in host galaxy association.  Each of these galaxy spectra were examined as described above with the original BOSS \texttt{ANCILLARY\_TARGET1} bit and included in DR9 and in the public database.  They are marked in the public database in column \texttt{bad\_host}.  Using only BOSS spectra assigned to the correct final host galaxy reduces the sample size with confident redshift to 3,083 objects.  From the final inspection, BOSS spectra were observed with a confident redshift for 2,764 host galaxies of candidate SNe and 319 host galaxies from the remainder sample.  While the actual targets described here and in C13 are the same, we have redefined the remainder sample due to the change in classification schemes and host galaxy identification criteria as described in Section~\ref{subsec:campbell}.  

The redshift and the apparent magnitude distributions of these objects are shown in Figure~\ref{fig:zhist}.  The target distribution and redshift statistics binned by $r$-band apparent magnitude are presented in Table~\ref{tab:target}.  A scatter plot demonstrating S/N and redshift range is shown in Figure~\ref{fig:mag-sn}.

\subsection{SDSS Spectroscopy of SN Host Galaxies}\label{subsec:sdss-spec}

In addition to the BOSS spectroscopic sample described in Section~\ref{subsec:redshifting}, there are 3,879 candidates with SDSS host galaxy spectra\footnote{The spectra can be found at http://data.sdss3.org/sas/dr9/sdss/spectro/plates/26/2010-05-23/}.  These galaxies were coincidentally observed as part of the original SDSS program.  These SDSS galaxy spectra have higher S/N and lie primarily at a lower redshift than the BOSS sample.  The SDSS pipeline was optimized for analysis of this class of objects and is expected to perform effectively in determining redshift estimates of these galaxies.  

We performed a brief test of the SDSS pipeline performance by choosing a random subset of 100 spectra from the 3,879 host galaxies.  We required only that objects were classified with \texttt{ZWARNING} equal to zero.  This subsample was manually inspected following the same procedure as described in Section~\ref{subsec:redshifting}; all visual redshifts were found to be in agreement with the automated classification.  

There were 65 objects with \texttt{ZWARNING} not equal to zero in the sample of 3,879 objects.  Manual inspection of these objects revealed that the SDSS pipeline redshift estimates were correct for 21 objects even though the \texttt{ZWARNING} flag had been set; these spectra were also included in the sample for a total of 3,835 objects.  Some objects were observed spectroscopically in both SDSS and BOSS.  In these cases of repeat observations, we refer to the BOSS spectrum as the primary redshift if the fiber was positioned at the center of the galaxy due to the larger wavelength coverage of BOSS.  There were 282 objects that were observed with both surveys in which the BOSS \texttt{ANCILLARY\_TARGET1} was set to 39, indicating that the fiber was positioned at the location of variability. In these cases, we use the SDSS spectrum to determine the host galaxy redshift.  For the final sample of 457 candidates with redshifts obtained by both BOSS and SDSS, there is only one candidate with redshift disagreement, defined as $|(z_{\rm BOSS}-z_{\rm SDSS})/(1+z_{\rm BOSS})| > 0.005$, further demonstrating confidence in the SDSS redshifts.

\subsection{Host Galaxy Ambiguity}\label{subsec:host-galaxy-ambiguity}

\begin{figure}[htbp]
\begin{center}
\includegraphics[scale=0.45]{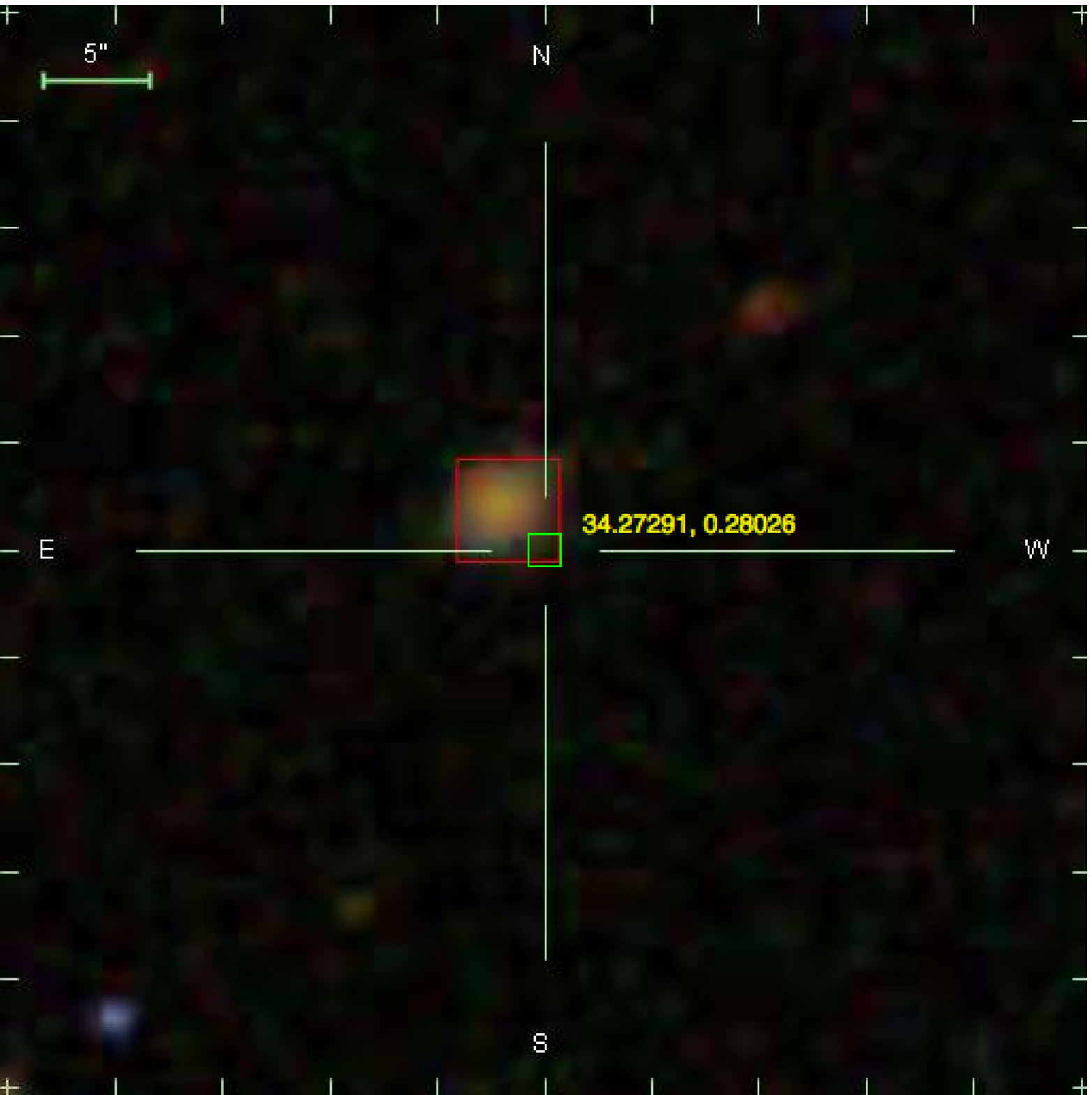}
\includegraphics[scale=0.5]{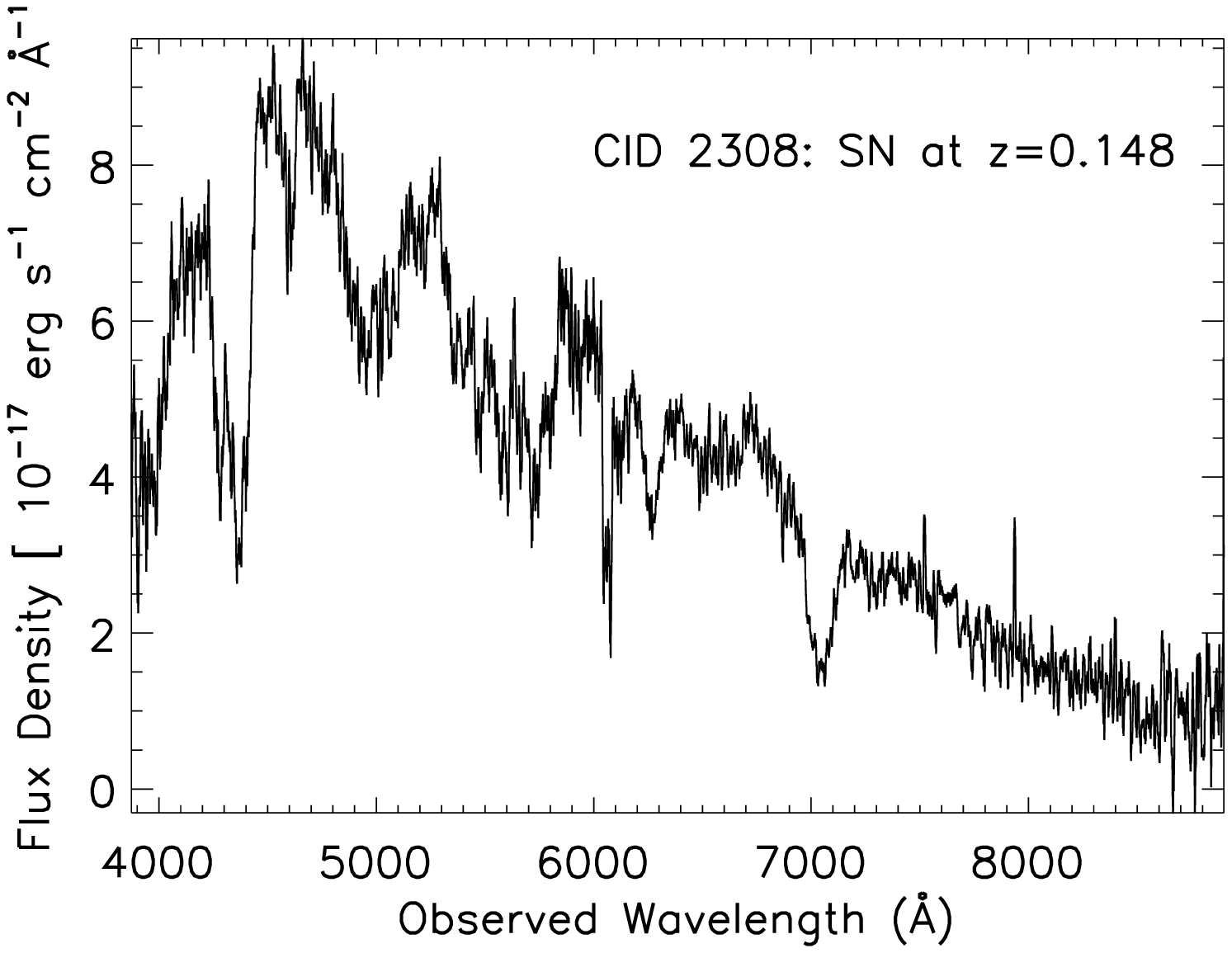}
\includegraphics[scale=0.5]{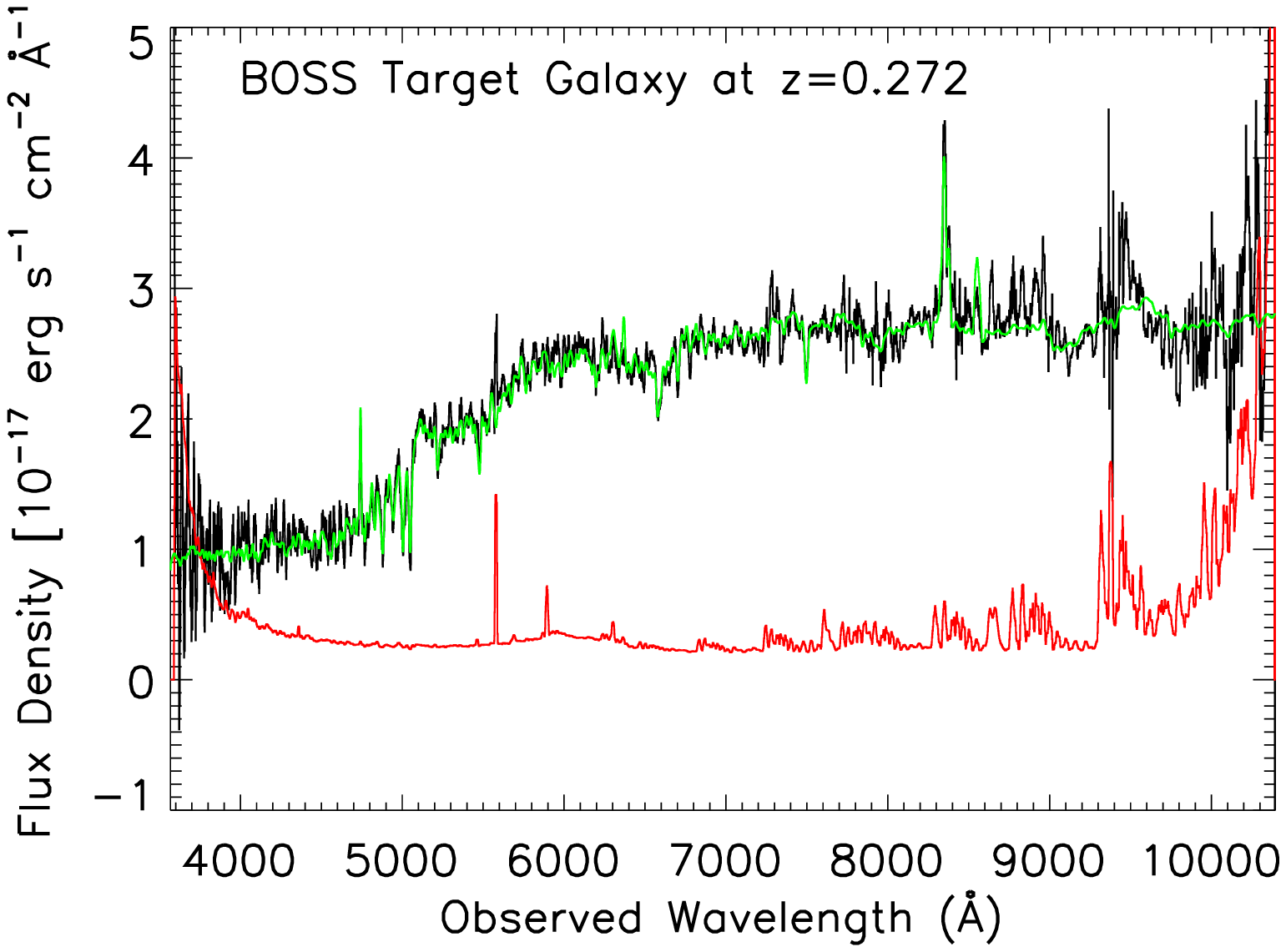}
\caption{\label{fig:2308}  The single case in our data set where the SN redshift and the host redshift differ, see Section \ref{subsec:host-galaxy-ambiguity}.  TOP:  DR9 image  at the location of CID 2308, showing the SN location marked by cross-hairs and boxed in green along with the BOSS targeted galaxy boxed in red.  MIDDLE:  The SN spectrum for 2308 with redshift 0.148.  BOTTOM:  The host galaxy spectrum for 2308 with a template spectrum of redshift 0.272 shown in green and the error spectrum shown in red, all with 10 pixel smoothing.}
\end{center}
\end{figure}

Even with both methods of host galaxy selection described in Section~\ref{subsec:campbell}, there are always cases of host galaxy ambiguity.  These cases can be identified when both the SN and associated host galaxy have discrepant spectroscopic redshifts.  Because of the difficulty in obtaining conclusive redshifts independently for the host galaxy and SN, there is only one instance of such a discrepancy in this sample.  We show the imaging and spectrum of SN 2005ey (CID 2308) in Figure~\ref{fig:2308} as the sole confirmed discrepancy.  The same object shown in the red box was determined as the host galaxy of SN 2005ey from the visual inspection of DR7 and and automated method using DR8.  The SN candidate has a redshift of 0.148 from the SN spectrum and the host galaxy of this candidate has a redshift 0.272.  While this SN appears to visually be hosted by this galaxy, the redshift difference between the SN spectrum and the galaxy spectrum indicate that is a background galaxy.  While we can not quantify the full effect, we point out that the problem is not only present in ground based data, but also exists in HST data.  In a survey of galaxy clusters, an SN with redshift $z =0.98$ was discovered in the foreground of a large spiral galaxy at $z=1.09$ that was mistakenly identified as the host.  Further examination of the deep HST images revealed a very faint galaxy that was not deblended from the background spiral.  While no redshift was ever determined for the faint galaxy, it is believed to be the true host of SCP06C1 \citep{barbary12b}.   

\section{Performance of Photometric SN Classification}\label{sec:analysis}

\begin{figure}[htbp]
\begin{center}
\includegraphics[scale=0.5]{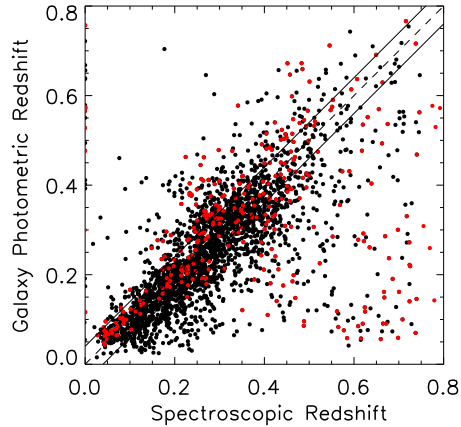}
\caption{\label{fig:zcomp-zcomp}  A comparison between the BOSS spectroscopic redshift and the galaxy photometry redshift, the red points are from the remainder sample.  The solid lines are z $\pm \sigma$/(1+$z$), with $\sigma =0.034$.}  
\end{center}
\end{figure}

The SDSS imaging on Stripe 82 led to precise photometry in $ugriz$ and photometric redshift estimates of the SDSS SN host galaxies released as part of DR8 \citep{aihara11a}.  The redshifts were determined using the DR7 photometric redshift code \texttt{PHOTOZ} which uses the approach of \citet{csabai07a}.  In \texttt{PHOTOZ}, the 100 nearest neighbors of each galaxy are examined in a color-color space populated by the 900,000 spectroscopically confirmed SDSS galaxies.  The redshift is estimated by evaluating the position of the photometric galaxy on the best-fit four color-redshift hyperplane derived from the 100 nearest neighbors.  In Figure~\ref{fig:zcomp-zcomp}, we compare the photometric redshift estimates of SDSS-II SN host galaxies with spectroscopic redshifts from BOSS.  The RMS (defined as $\sigma$/1+$z$) between the spectroscopic and photometric redshift estimates using host galaxy photometry is 0.034 after removing 194 objects (8\%) that have redshift differences $> 3 \sigma$ and are characterized as catastrophic failures.  In what follows, all SDSS and BOSS spectroscopic redshifts are used, regardless of whether the photometric redshift was a catastrophic failure.  

In this section, we update the SN photometric typing as reported in the SDSS-II SN Data Release.  We type each SN using the spectroscopic redshift as a prior, compare the performance of three different classification techniques, and assess performance of photometric classification with varying data quality.

\subsection{Photometric Typing with SDSS and BOSS Redshifts}\label{subsec:phottyping}

In Section~\ref{subsec:TS}, a flat redshift distribution was used as a prior for the photometric typing of SN light curves; here we use the host galaxy spectroscopic redshift as a prior to provide new photometric typing of the candidates.  As before, we use PSNID to perform Bayesian photometric classification.  In comparing the typing of SN light curves using a flat redshift prior to the results using a spectroscopic redshift, we use the following notation.  Using only photometric information and a flat redshift prior, a Bayesian probability of the candidate being a SNe~Ia is reported as P$_{\rm Ia}$, similarly for Type Ib/c (P$_{\rm Ibc}$) and Type II (P$_{\rm II}$).  The same method is performed while using the spectroscopic redshift and recorded as PZ$_{\rm Ia}$, PZ$_{\rm Ibc}$, and PZ$_{\rm II}$, respectively.

\begin{figure}[htbp]
\includegraphics[scale=0.5]{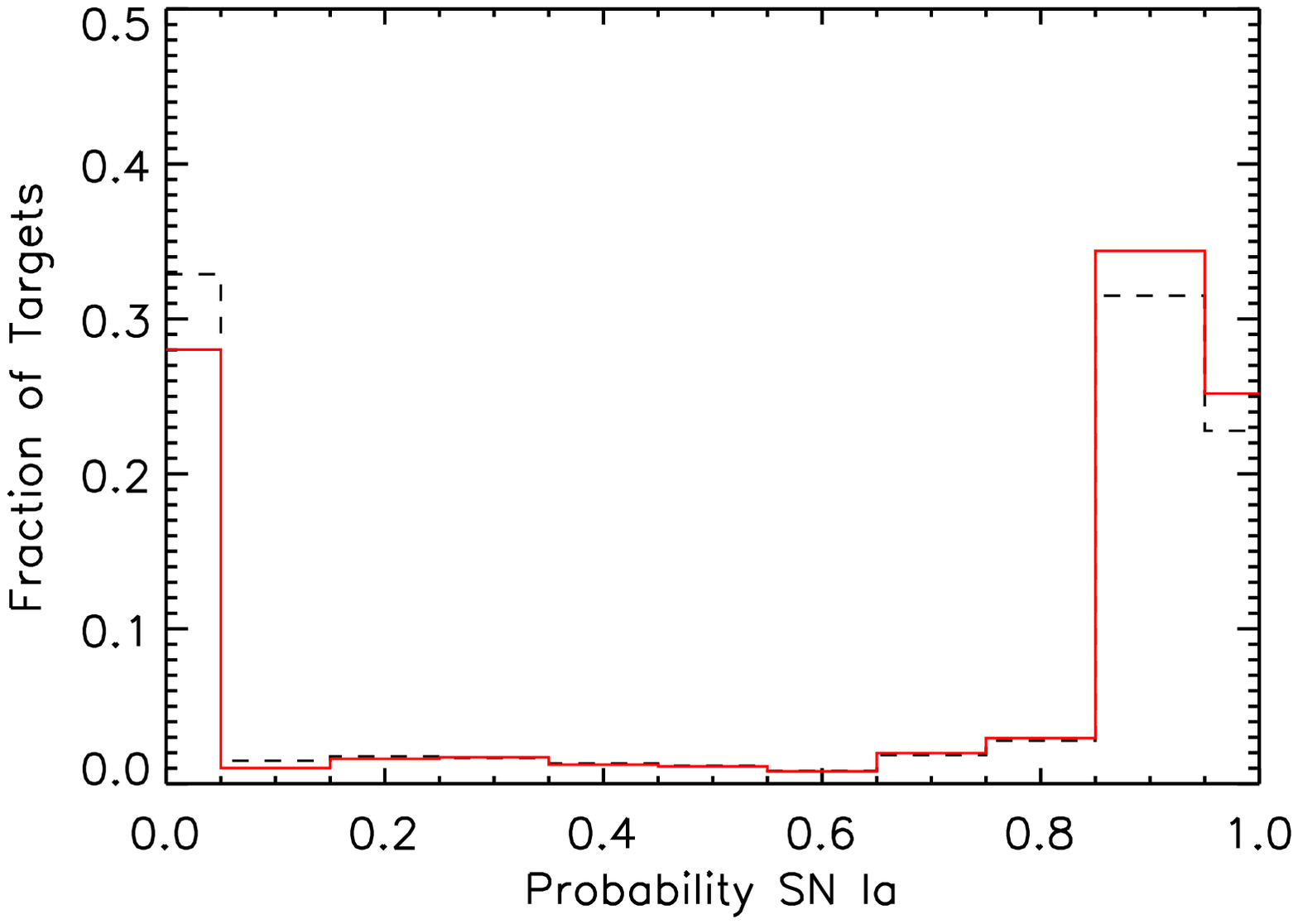} 
\includegraphics[scale=0.5]{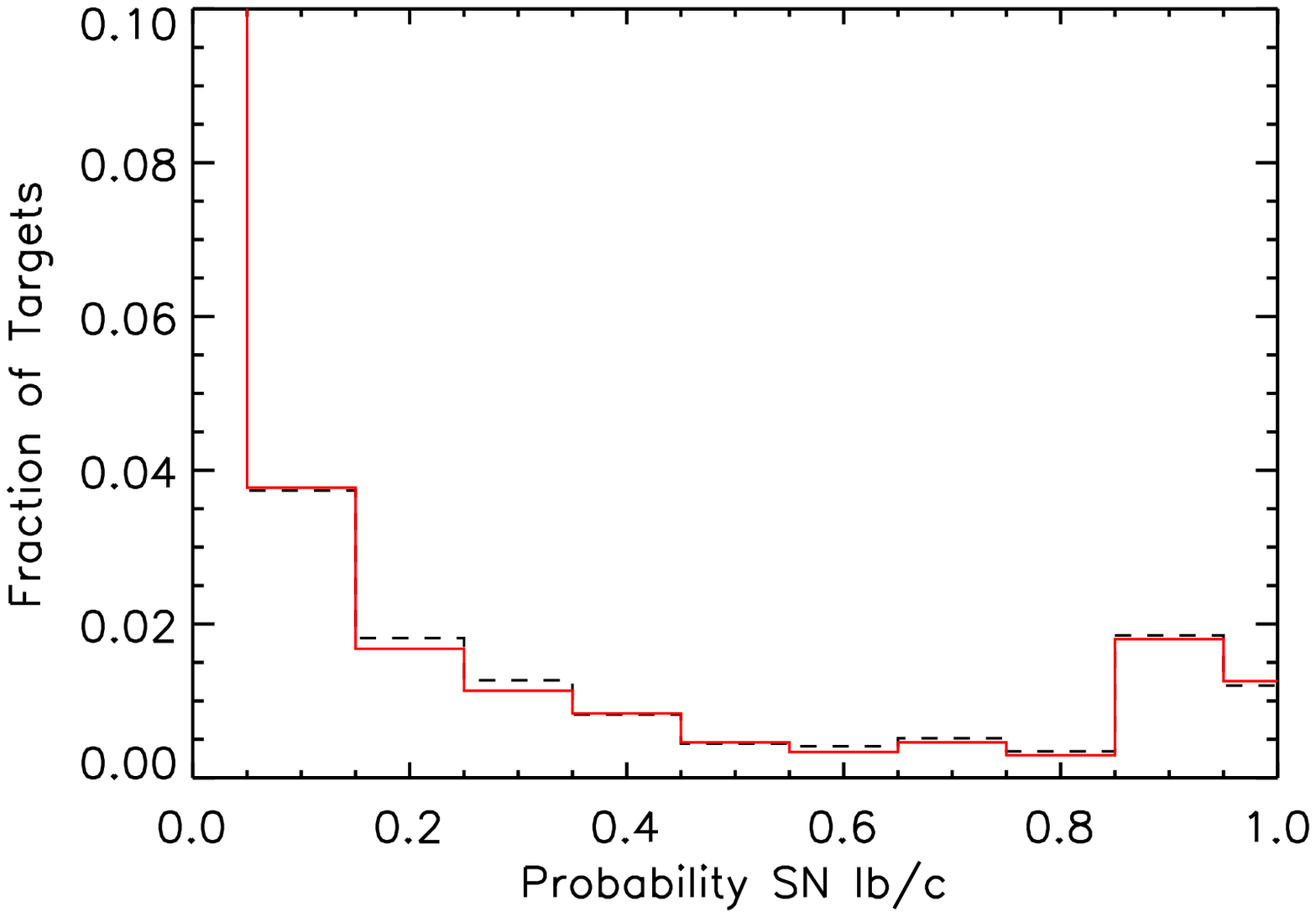} 
\includegraphics[scale=0.5]{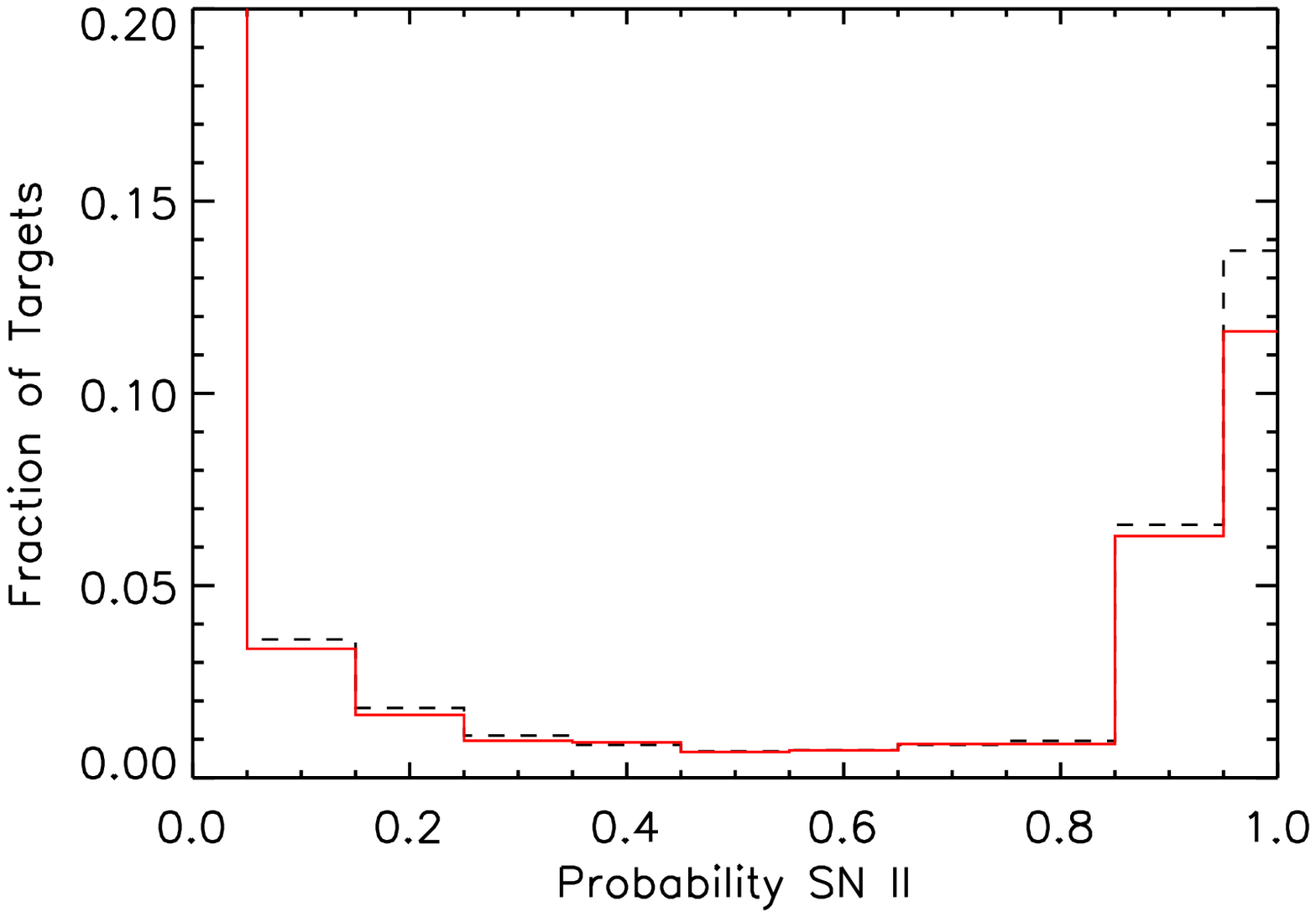} 
\caption{\label{fig:prob-hist} Distribution of Bayesian probability of being a SN~Ia (core-collapse SN) using the host galaxy redshift as a prior.  The black (dashed line) is the entire observed sample including the remainder targets and red (solid) is for the observed sample excluding the remainder sample.  As expected, when the remainder sample is not used, there is a higher probability of a candidate being an SNe~Ia.} 
\end{figure}

The updated Bayesian probability for each transient identified as either an SN~Ia or core-collapse in Table~\ref{tab:priority} is found in the online tables; the typing of a representative subsample is shown in Table~\ref{tab:typing} for the six candidates displayed in Figure~\ref{fig:spec}.  As shown in Figure~\ref{fig:prob-hist}, the probability values for the entire sample computed by Bayesian probability tend to cluster around the most likely SN type.  For this reason, we initially classify a SN based on the highest $P$ value; for example, an object is considered an SN~Ia if the probability of being Ia is higher than both the probability of being Ib/c and core-collapse.  When using a flat redshift prior, there are 2173 candidates that have $P_{\rm Ia} > P_{\rm Ibc}$ and $P_{\rm Ia} > P_{\rm II}$; these candidates are denoted ``B Ia''.  When using a spectroscopic redshift prior, there are 2327 candidates that have $PZ_{\rm Ia} > PZ_{\rm Ibc}$ and $PZ_{\rm Ia} > PZ_{\rm II}$; these candidates are denoted ``BZ Ia''.  This classification does not take into account any information on light curve data quality.  We use this classification for the purpose of separating each candidate with spectroscopic host galaxy redshift into one of three SN classification.  Further requirements on light curve quality must be applied to make this a robust classification.

Toward the goal of robust classification, we test two additional classification techniques.  The first of these is the approach of S11 using PSNID.   In addition to using the Bayesian probability above, there are requirements on the light curve quality and the $\chi^2$ fit probability.   To be classified as an SN~Ia using these restrictions, a candidate must have a light curve containing at least one epoch within five days rest-frame of peak, at least one epoch between 5 and 15 days rest-frame after peak, and a $\chi^2$ fit probability $\ge 0.01$. These 1301 candidates are denoted ``BP Ia''.  There are 1273 objects classified when the spectroscopic redshift is also used; these candidates are denoted ``BPZ Ia''.  The total number of likely SN~Ia is around half those included in the purely Bayesian case; this apparent discrepancy is due to the stringent requirements on light curve coverage and on the quality of fit.

 In the third classification technique, we categorize the objects as likely Type Ia, Ib/c or II SNe using the same criteria as S13 and refer to this selection as the full photometric classification.  A full photometric SNe~Ia is subject to a nearest neighbor classification with additional light curve quality restrictions.  In the nearest neighbor classification, the candidates are compared to a simulated training set in multi-dimensional SN parameter space that includes extinction, light curve shape, and redshift.  Different SN types cluster in different sections of this parameter space.  A candidate is given a probability of being an SN~Ia (P$_{\rm NN\_Ia}$ or P$_{\rm NNZ\_Ia}$ for flat and spectroscopic redshift priors, respectively)  by counting the number of SNe~Ia within a certain distance in this parameter space normalized by the total number of SNe within that distance in parameter space.   

In addition, a full photometric SN~Ia must have variability in only one year, PZ$_{\rm Ia} >$ PZ$_{\rm Ibc}$, PZ$_{\rm Ia} >$ PZ$_{\rm II}$, PZ$_{\rm NN\_Ia} >$ PZ$_{\rm NN\_Ibc}$, PZ$_{\rm NN\_Ia} >$ PZ$_{\rm NN\_II}$, at least 10 nearest neighbors in parameter space, a $\chi^2$ fit probability $\ge 0.01$, a light curve containing at least one epoch within five days rest-frame of peak, and at least one epoch between 5 and 15 days rest-frame after peak.  Similarly, a full photometric core-collapse SNe must have variability in only one year,  PZ$_{\rm Ibc} >$ PZ$_{\rm Ia}$ or  PZ$_{\rm II} >$ PZ$_{\rm Ia}$, PZ$_{\rm NN\_Ibc} >$ PZ$_{\rm NN\_Ia}$ or PZ$_{\rm NN\_II} >$ PZ$_{\rm NN\_Ia}$, and at least 10 nearest neighbors in parameter space.  Unlike the first two classification techniques, we do not report probabilities associated with each classification here.  Instead, we simply refer to the final classification as ``photometric Ia'' or ``photometric CC'' for typing without redshifts.  When adding a redshift, we report the classification as ``photometric Z Ia'' and ``photometric Z CC''.   Employing the photometric classification technique, we find 1166 photometric SNe~Ia and 1126 photometric Z SNe~Ia.  See section \ref{subsec:comp} for a comparison of the techniques.

Of the 518 spectroscopically-confirmed SNe~Ia from SDSS-II, 320 objects have SDSS or BOSS host galaxy spectra; 83.4\%, 54.4\%, and 53.1\% are classified as SNe~Ia using each BZ Ia, BPZ Ia, and the full photometric classification.  While at first these numbers appear low, this result is not surprising because many SNe candidates were targeted for spectroscopic confirmation at the beginning or end of each season and had poor light curve coverage for photometric typing.  When one examines the 239 spectroscopically-confirmed SNe~Ia that meet the requirements of a well-covered light curve, a requirement for having a nearest neighbor classification, 89.9\%, 85.6\%, and 82.3\% SNe are photometrically identified as SNe~Ia using the BZ Ia, BPZ Ia, and full photometric classification, respectively.

Finally, we examine AGN activity in the SDSS and BOSS spectra of the photometric Z SN~Ia hosts.  We find that 22 of these host galaxies are classified in \texttt{idlspec2d} as as likely AGN, likely due to broadline emission features such as MgII or H$\beta$.  Of these, nine were identified in Table~\ref{tab:priority} as high probability SNe~Ia, six as low probability SNe~Ia, and seven from the remainder sample.  Of these, nine high or low probability SN~Ia and three from the remainder sample were within 0.3\arcsec\ of the center of the host. We conclude that the variation is most likely due to AGN activity rather than SN.  There are five SNe~Ia hosted in quasars that are separated by more than 1.0\arcsec\ from the center of the host.  These numbers are small compared to the overall sample.

\begin{deluxetable}{lrrrrrr}
\centering
\tablewidth{0pt}
\tabletypesize{\footnotesize}
\tablecaption{\label{tab:typing} SN Classification}
\tablehead{\colhead{CID} & \colhead{$PZ_{\rm Ia}$} & \colhead{$PZ_{\rm Ibc}$} & \colhead{$PZ_{\rm II}$}& \colhead{$P_{\rm Ia}$} & \colhead{$P_{\rm Ibc}$} & \colhead{$P_{\rm II}$}  }
\startdata
       20259 &      0.999 &       0.000 &   0.0003 &      1.00 &       0.00 &      0.00         \\ 
      20953 &      0.998 &   0.0007 &    0.0017 &      0.956 &     0.0385 &    0.0067      \\ 
       13195 &       0.000 &       1.00 &       0.000 &       0.000 &       1.00 &       0.000 \\ 
       17414 &      0.954 &     0.046 &       0.000 &      0.536 &      0.465 &       0.000 \\ 
       20768 &      0.993 &    0.007 &       0.000 &      0.967 &     0.033 &   0.0001 \\ 
        3562 &       1.00 &       0.000 &       0.000 &       0.000 &       0.000 &       1.00 

\enddata
\tablenotetext{}{SN probabilities for the six spectra in Figure~\ref{fig:spec}.}
\end{deluxetable}

\subsection{Comparison of Classification Techniques}\label{subsec:comp}

As shown in Table~\ref{tab:typingstatsphot}, each SN candidate has been assigned a type using three different classification schemes.  The table demonstrates the significance of redshift on photometric classification.  The inclusion of redshift decreases the number of core-collapse SN from 1047 B~II objects to only 844 BZ~II objects.  While the photometric Ia and photometric Z Ia samples are nearly the same size, 1166 and 1126 objects respectively, the photometric CC and photometric Z CC samples are significantly different, 1068 and 2262 objects, respectively.  The addition of spectroscopic redshift helps control the contamination.

\begin{deluxetable}{lr}
\centering
\tablewidth{0pt}
\tabletypesize{\footnotesize}
\tablecaption{\label{tab:typingstatsphot}  Classification Statistics}
\tablehead{\colhead{Class} & \colhead{Number} }
\startdata
B Ia & 2173  \\
B CC  & 265   \\
B II  & 1047  \\
BP Ia  & 1301  \\
BP CC  & 1312  \\
Photometric Ia  &  1166 \\
Photometric CC  &  1068 \\
BZ Ia  &  2327 \\
BZ Ibc  &  295 \\
BZ II  &  844 \\
BPZ Ia  &  1273 \\
BPZ CC  & 1139  \\
Photometric Z Ia  &  1126 \\
Photometric Z CC  &  262 

\enddata

\end{deluxetable}

While it is not possible to do a full analysis of purity or completeness on each of these classification schemes due to the lack of complete spectroscopic classification, we evaluate the performance of each by comparing the results with and without spectroscopic redshifts.  One may consider the most robust technique to be the one that produces consistent typing with and without using redshift priors.  We search for signs of contamination in each classification scheme by examining the number of objects classified as an SN~Ia without redshift and classified as a core-collapse with redshift.  We search for signs of completeness by examining the number of objects classified as an SN~Ia with redshift that were not identified as an SN~Ia without redshift.  The comparison performed here can be considered a recipe to filter samples into higher purity for any classification scheme that may be used in the future. 

 In the Bayesian sample, 1984 objects were consistently classified as B Ia and BZ Ia, while 2327 objects were classified as BZ Ia.  We therefore define the completeness as 85.3\%.  Similarly the completeness for the PSNID classification and full classification are 91.9\% and 91.0\%, respectively.  The consistency of classification for each technique relative to the total number of SNe~Ia classified with redshift as a prior are shown in Table~\ref{tab:cont}.  In the bayesian approach, we count 189 objects that were classified as B Ia that were later classified as BZ Ibc or BZ II.  In total, 2327 transients were classified B Ia, corresponding to a contamination of 8.7\%.  As shown in Table~\ref{tab:cont}, the same analysis reveals 10.0\% and 12.2\% contamination in the PSNID and full classification results, respectively.  

The contamination in the Bayesian approach appears to be the lowest of the three techniques.  As described in Section~\ref{subsec:phottyping}, this result should be regarded with caution because the total number of B Ia and BZ Ia appears artificially increased due to the more stringent requirements on the light curve quality and on the fit-probability placed on the other two classification schemes.  The PSNID approach performs comparably to the full photometric classification.  However, the full photometric classification approach has the higher percentage of completeness, the lower level of contamination, and the smaller number of SN~Ia.  For these reasons, the full photometric technique appears to be more restrictive and will be used for the remainder of the paper.

\begin{deluxetable}{lccc}
\centering
\tablewidth{0pt}
\tabletypesize{\footnotesize}
\tablecaption{\label{tab:cont} Typing Approach Contamination and Completeness}
\tablehead{\colhead{} & \colhead{Bayesian} & \colhead{PSNID } &  \colhead{Full } \\
\colhead{} & \colhead{} & \colhead{ } &  \colhead{ Classification}}
\startdata

Contamination  & 189   & 131   & 142  \\
               & 8.7\%  & 10.0\%   & 12.2\%  \\
\\
Completeness   &  2327  & 1273   &   1126  \\
               &  85.3\%  &  91.9\%   & 90.9\%  

\enddata

\end{deluxetable}

\begin{figure}[htbp]
\epsscale{0.85}
\includegraphics[scale=0.5]{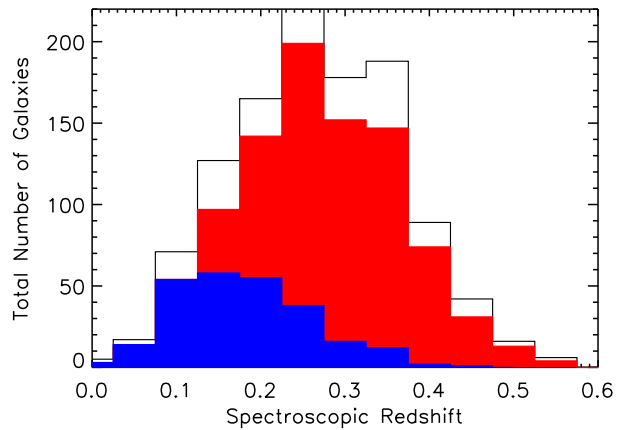}
\caption{\label{fig:sn-completeness} The redshift distribution of galaxies hosting SNe~Ia in redshift bins of 0.05.  The black curve is for candidates typed as SNe~Ia with spectroscopic redshift, the red curve is for photometric SNe~Ia and the blue curve is for the spectroscopically confirmed sample of SNe~Ia.  }
\end{figure}

Figure~\ref{fig:sn-completeness} illustrates the sizes of the different samples.  The photometric Ia sample used here includes only those candidates that are photometric Z Ia.  At low redshift ($0<z<0.1$), the photometric Z Ia sample is 61.6\% larger than the photometric Ia sample.  At higher redshifts ($0.2<z<0.4$), the sample of photometric Z Ia sample is 14.4\% larger.  The photometric Z Ia sample is larger than the spectroscopically confirmed sample and extends to significantly higher redshift.

\subsection{Effects of Data Quality on Contamination and Completeness}\label{subsec:dataqual}

We examine contamination and completeness of the full photometric classification technique as a function of data quality.  We evaluate the performance on subsets of data selected on the minimum S/N, total number of data points on the light curve, and light curve coverage relative to maximum SNe~Ia luminosity.  In the first series of tests, we require that at least one exposure meet minimum S/N measurements in each of $gri$.  We refer to this restriction as the ``S/N'' cut.  In the second series of tests, we require a minimum number of epochs to meet the condition S/N $> 3$ in each of $gri$.  We refer to these subsamples as ``Number of epochs''.  In the final series of tests, we examine the number of SNe~Ia in which at least one S/N $>3$ epoch is separated by a finite amount of time from maximum light.  We refer to these as ``$\Delta$t''.

The contamination and completeness fraction for each data quality cut are shown in Table~\ref{tab:dataqual}.  The contamination decreases from 0.121 to 0.108 when imposing the most restrictive cuts on S/N.  As shown in the final row of Table~\ref{tab:dataqual}, the contamination decreases from 0.121 to 0.095 when requiring that a SN light curve have both early time coverage and late time coverage.  Requiring additional epochs with S/N $>3$ appears to have the largest impact, reducing contamination from 0.121 to 0.090.  Similarly, requiring 10 epochs leads to the largest improvement in completeness, although at a cost of sacrificing 85\% of the total sample.  One may argue that a good compromise requires early and late time coverage, leading to improved statistics on contamination and completeness while only losing half of the SNe~Ia.

\begin{figure*}[htbp]
\begin{center}
\includegraphics[scale=0.45]{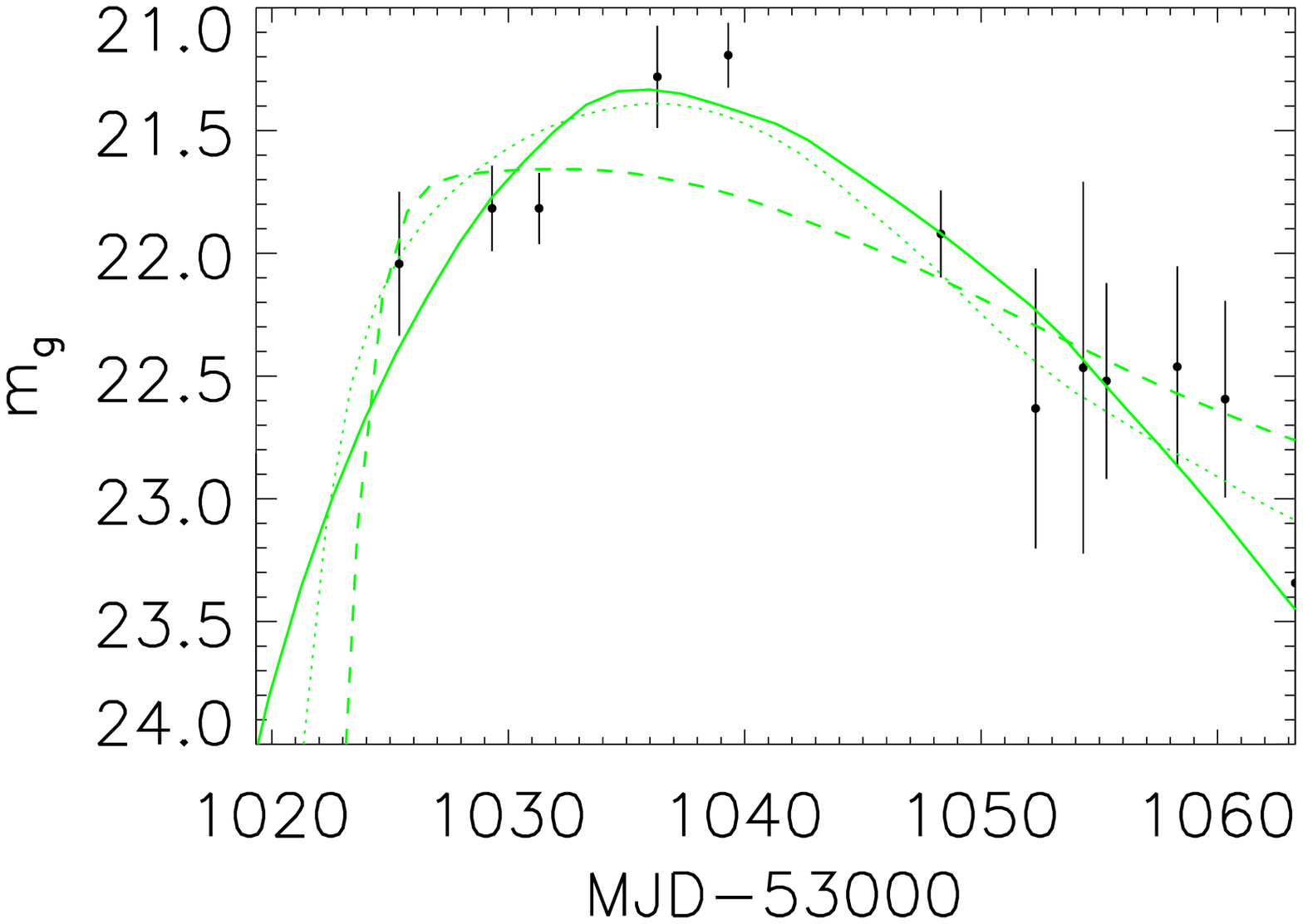}
\includegraphics[scale=0.45]{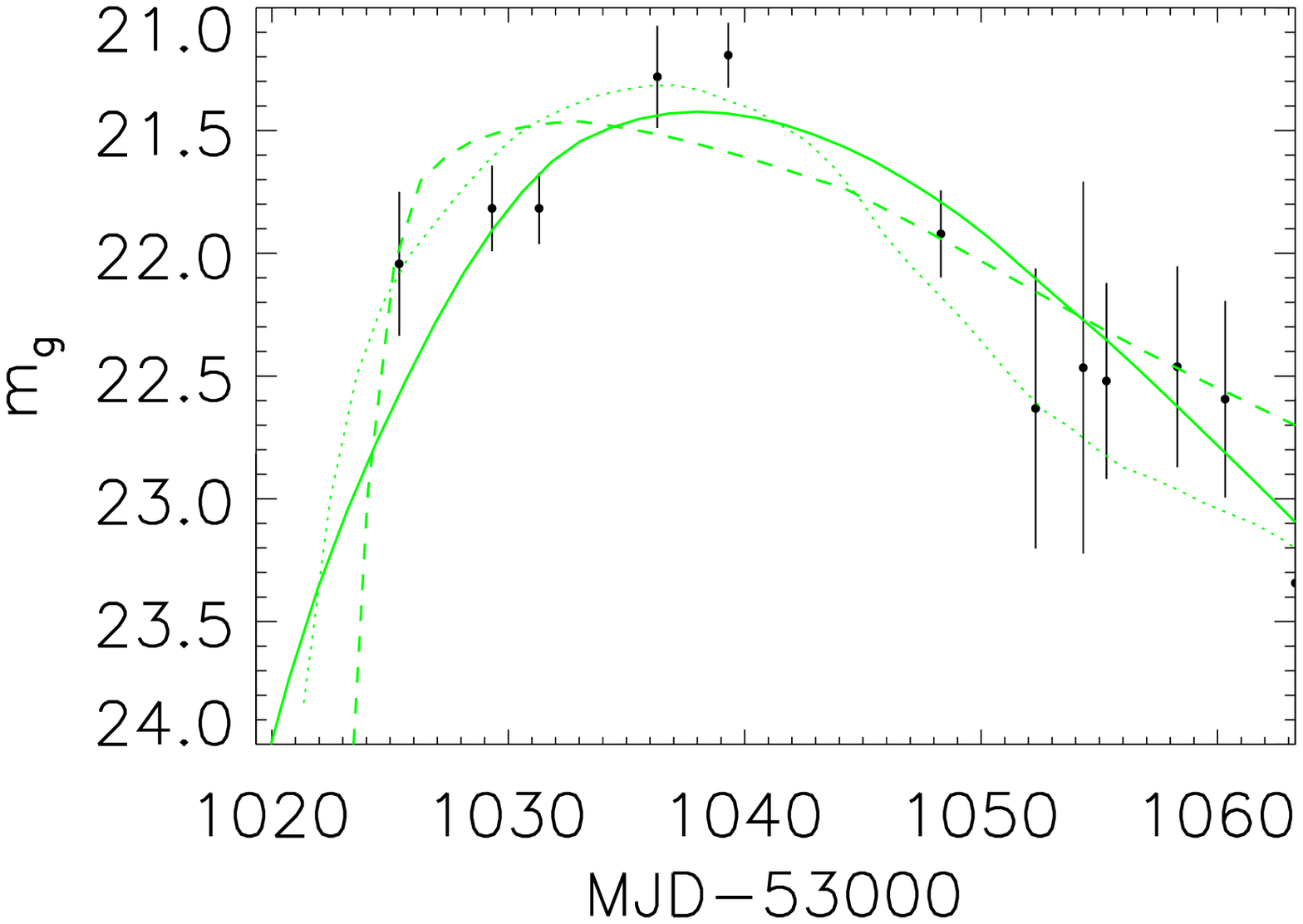}
\includegraphics[scale=0.45]{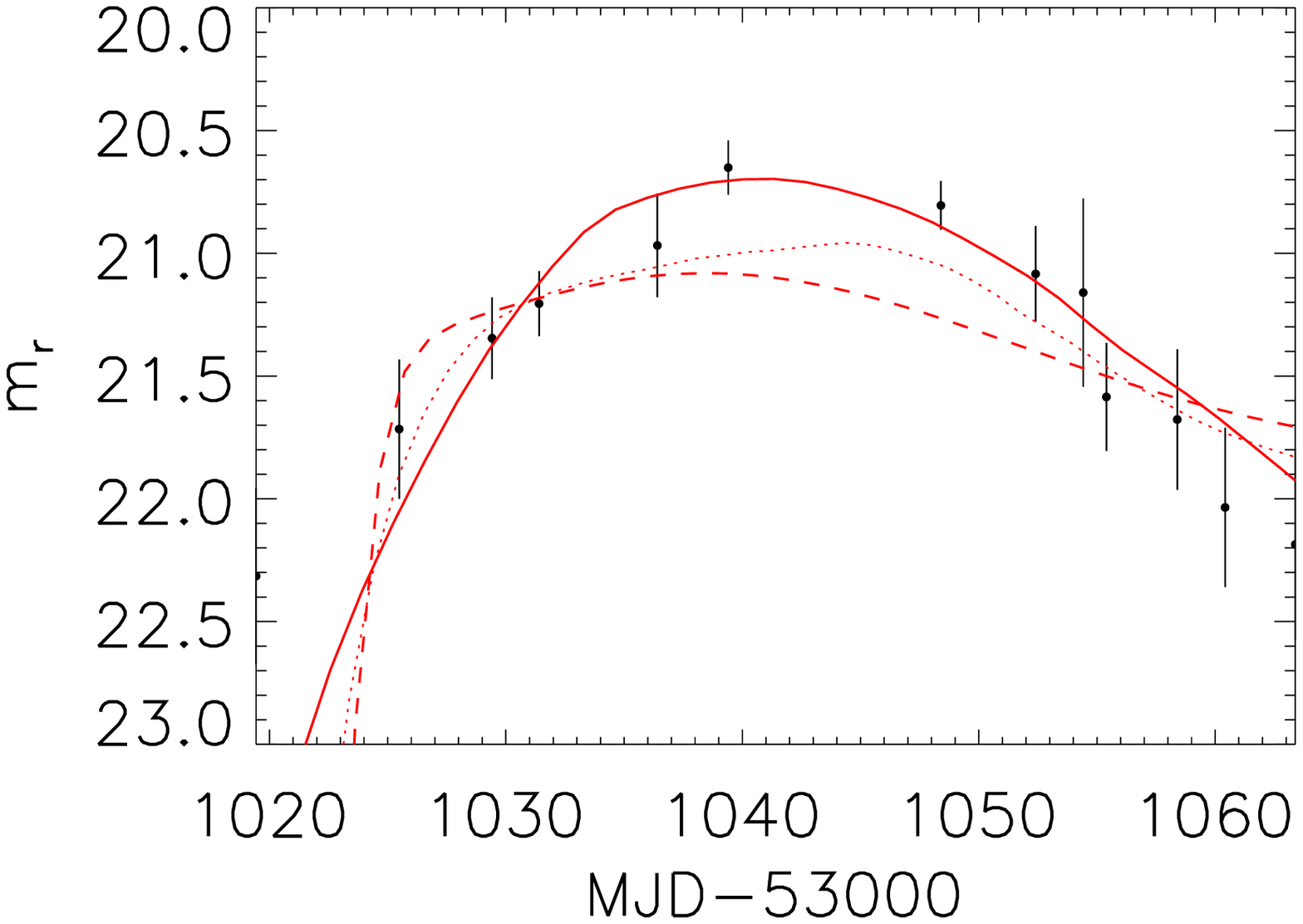}
\includegraphics[scale=0.45]{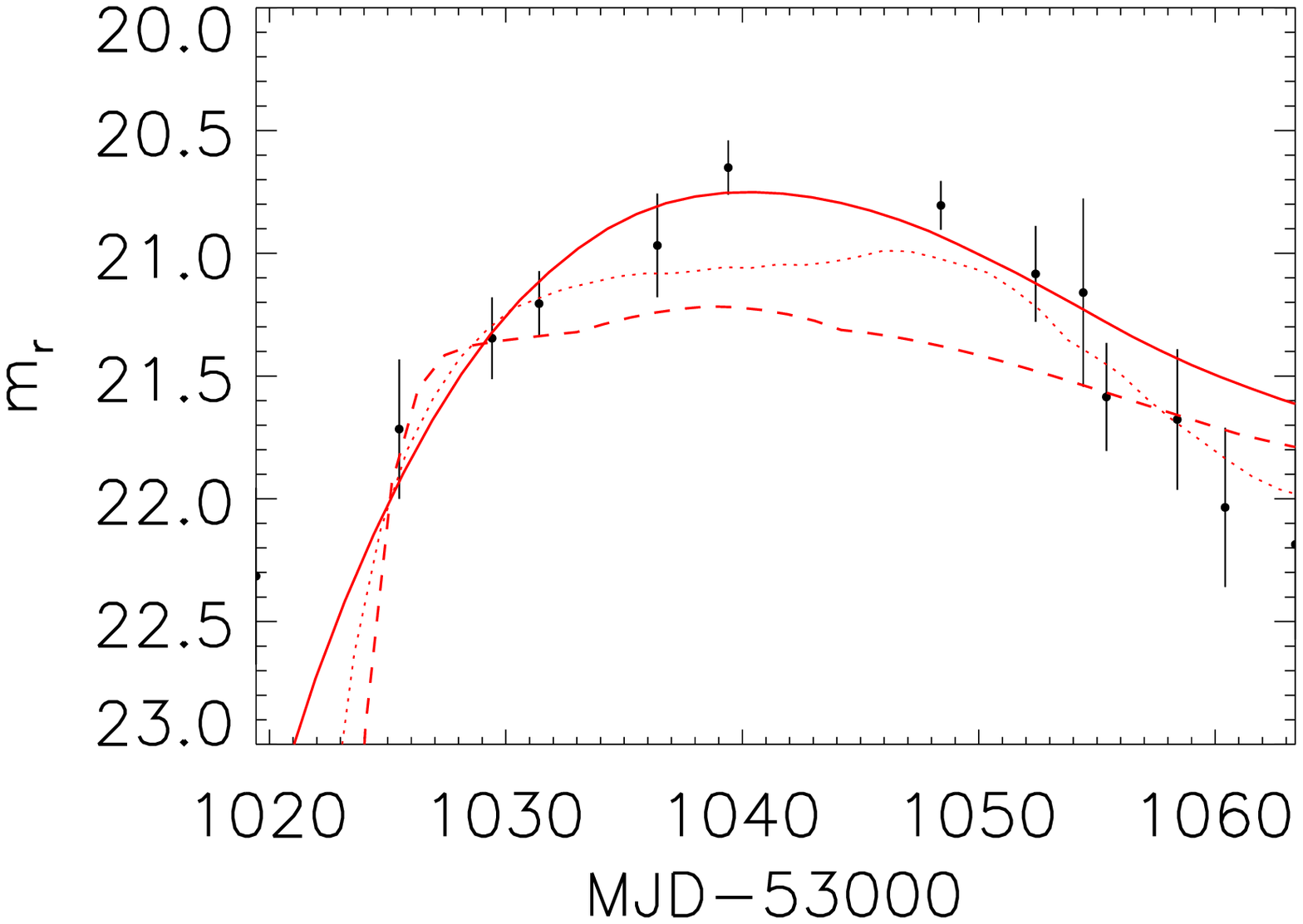}
\includegraphics[scale=0.45]{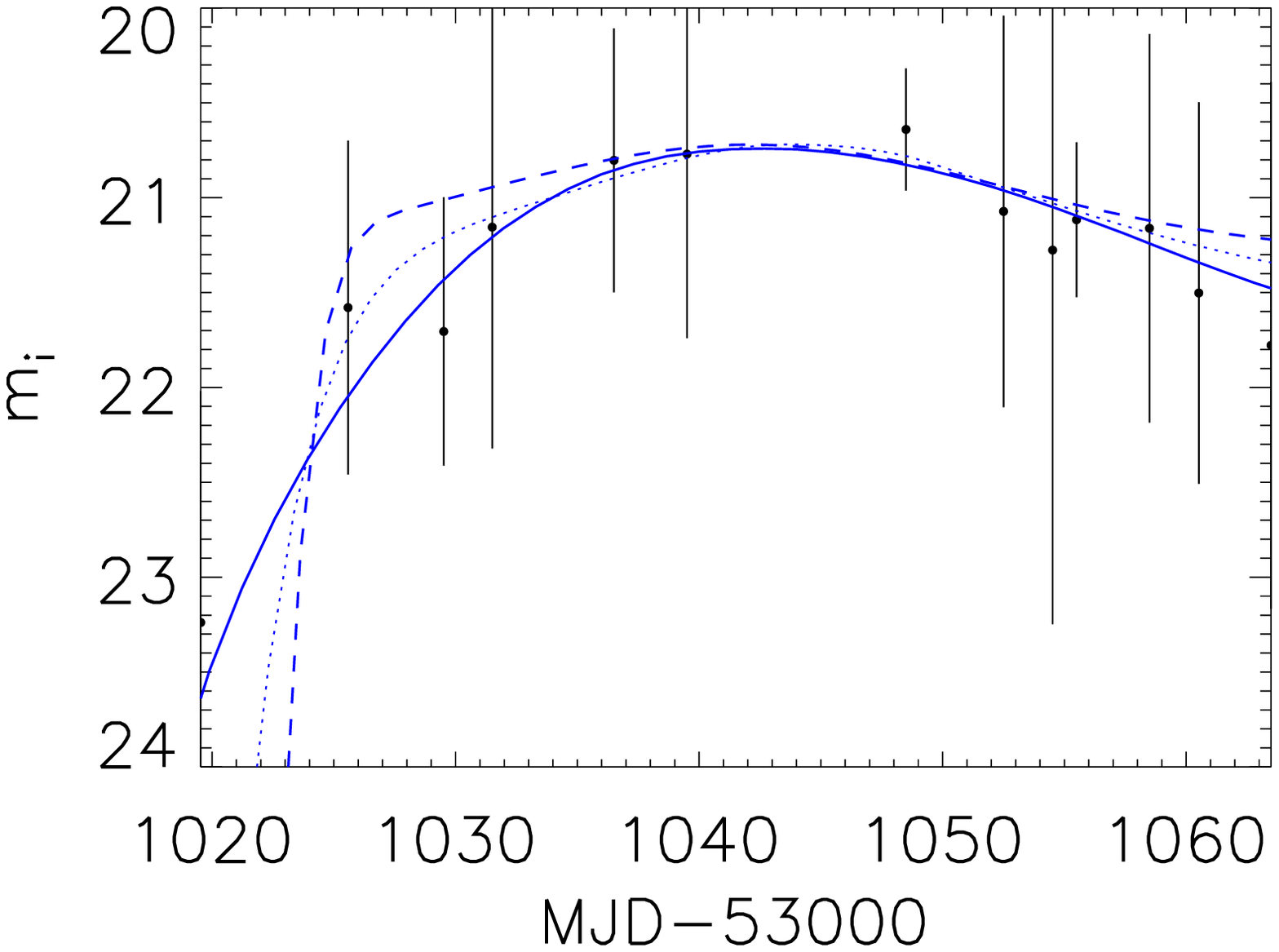}
\includegraphics[scale=0.45]{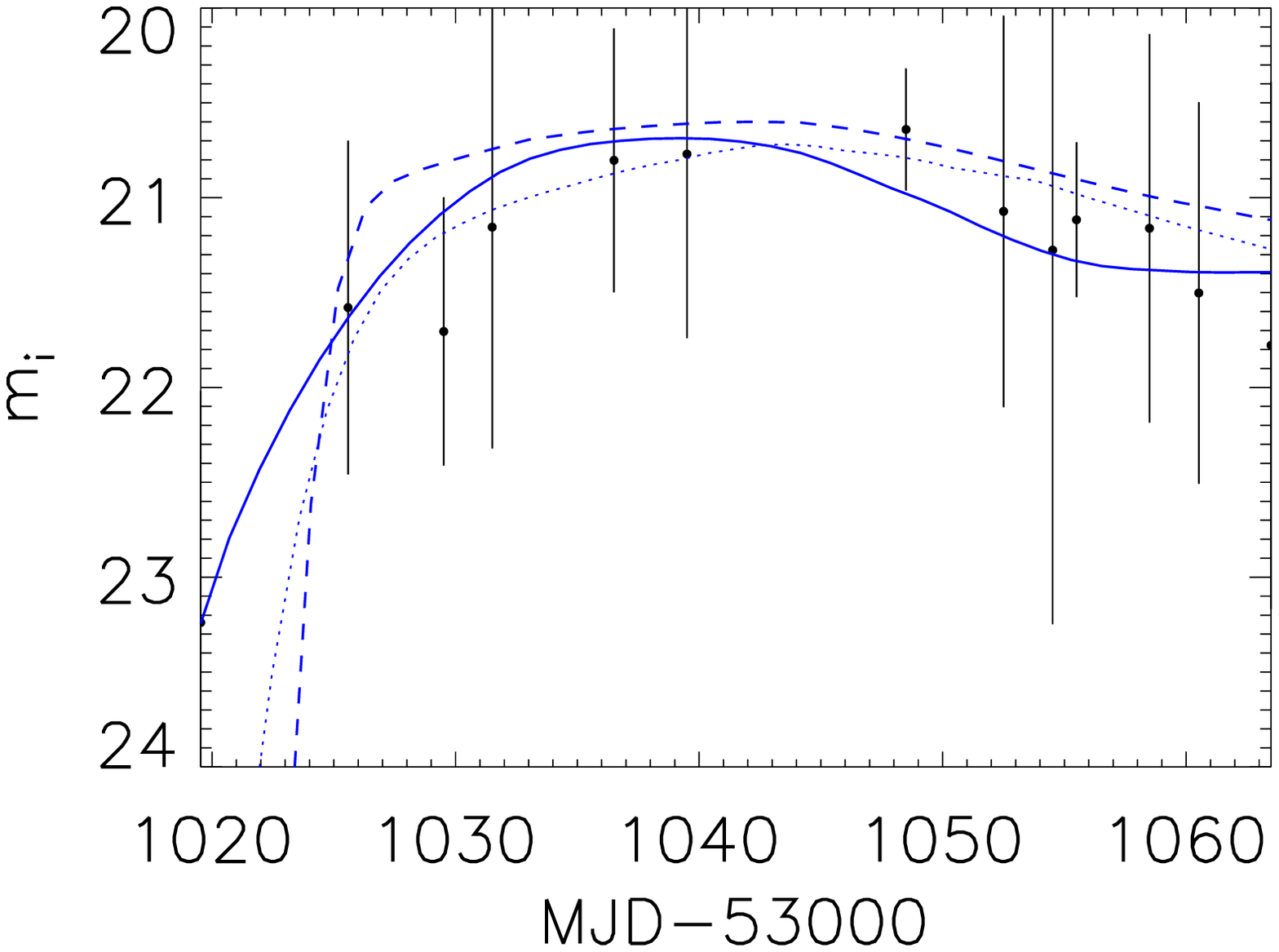}
\includegraphics[scale=0.45]{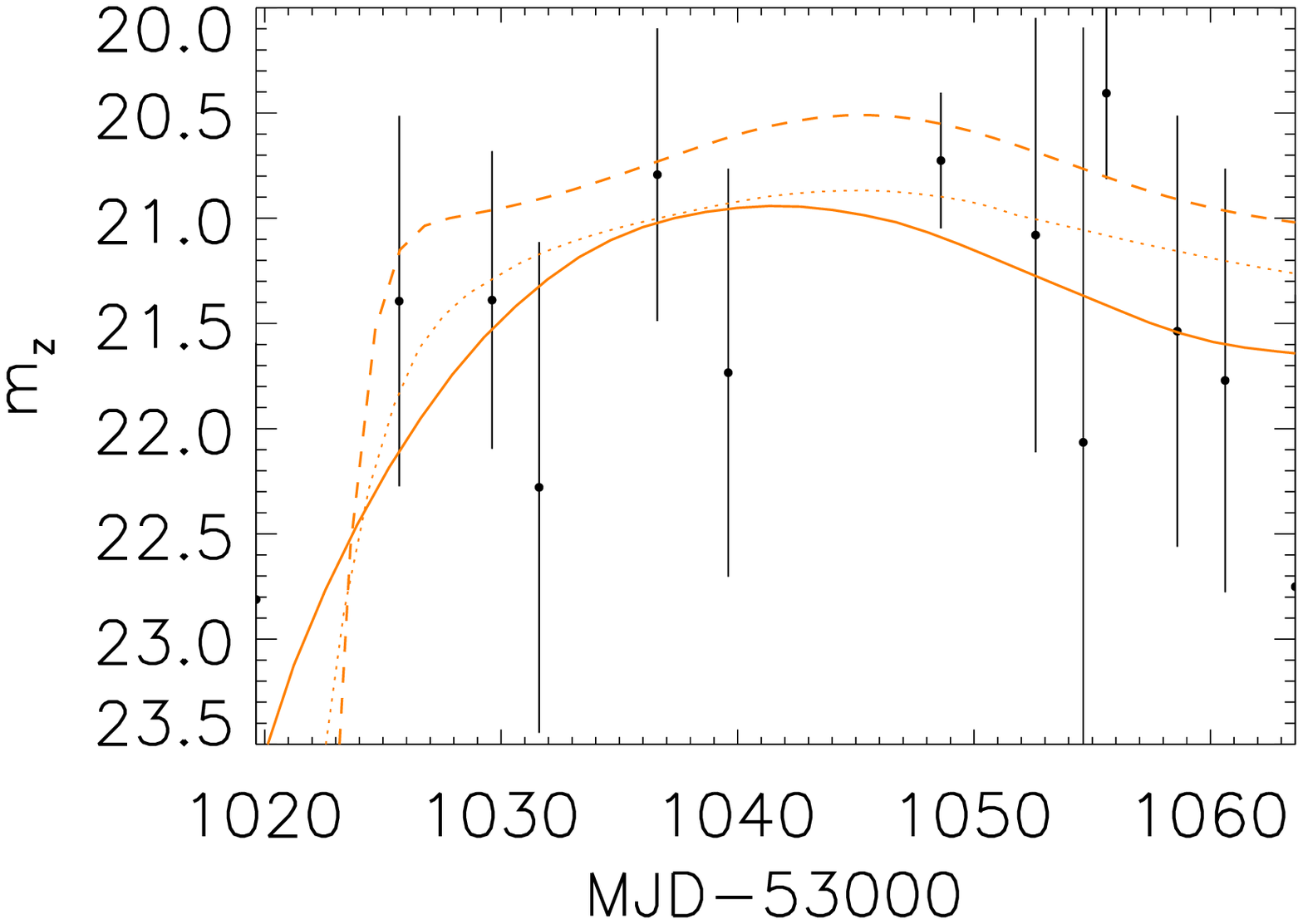}
\includegraphics[scale=0.45]{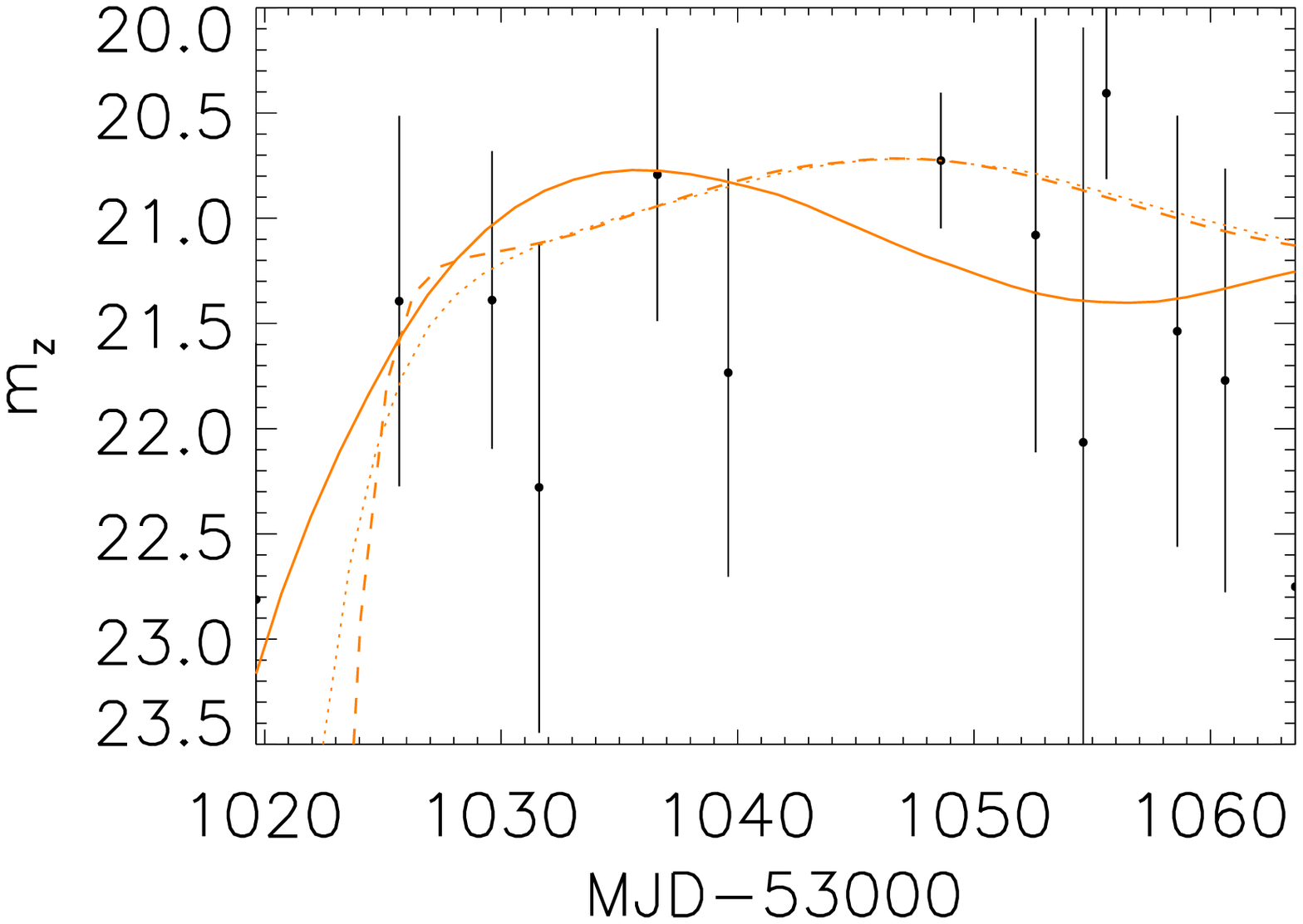}
\caption{\label{fig:lc} Example of the light curves for one SN candidate at $z_{\rm spec}=0.116$ that is in the contamination sample when there are at least 10 epochs.  The solid line shows the best fit SNe~Ia template, the dotted line shows the best fit Ibc template, and the dashed line shows the best fit SNe~II template. The left four plots show best fit templates when using a flat redshift prior and the right four plots show the best fit templates when the spectroscopic redshift is used as part of the classification.  } 
\end{center}
\end{figure*}

Somewhat surprisingly, even SNe that meet the most restrictive conditions outlined in Table~\ref{tab:dataqual} can be misclassified.  An example is CID 15486 as shown in Figure~\ref{fig:lc}.  This SN has 11 epochs at S/N $>3$, coverage as early as 14 days before peak luminosity and as late as 18 days after peak luminosity.  The SN is classified photometric Ia and photometric Z CC, likely due to redshift uncertainty.  The photometric estimate of the best-fit SN~Ia redshift is $z=0.350$ while the host galaxy spectrum revealed a redshift of $z=0.116$.

\begin{deluxetable*}{llllc}
\centering
\tablewidth{0pt}
\tabletypesize{\footnotesize}
\tablecaption{\label{tab:dataqual} Data Quality Effects of Full Photometric Classification}
\tablehead{\colhead{} & \colhead{Contamination}   & \colhead{Total Number } & \colhead{Completeness}  & \colhead{Total Number } \\  
\colhead{} & \colhead{}  & \colhead{in photometric } & \colhead{}  & \colhead{in redshift } \\
\colhead{} & \colhead{}   & \colhead{ classified sample} & \colhead{}  & \colhead{ classified sample} }
\startdata
No additional & 0.122  & 1166  & 0.909 & 1126 \\
quality cuts  \\
\\
S/N $>$ 3  &  0.111  & 1148  & 0.909  &  1122  \\
S/N $>$ 5  & 0.102  & 1040  & 0.931  & 1003   \\
S/N $>$ 10 & 0.102   & 477  & 0.977  & 438   \\
\\
3 epochs    & 0.099 & 985  & 0.939  & 945  \\
5 epochs    & 0.093   & 691   & 0.926  & 652    \\
7 epochs    &  0.093  & 421  & 0.977   & 391  \\
10 epochs  &  0.041  & 172  & 0.994  &  166 \\
\\
$\Delta$t $< -5 $ d   & 0.073   & 674  & 0.933     & 670    \\
 $\Delta$t $>$ 15 d   &  0.061  & 654  & 0.923     & 665    \\
$\Delta$t $< -5 $ d,   & 0.112   & 134  &  0.895    & 133   \\
 and not $ \Delta$t $>$ 15 d \\
\\
 $ \Delta$t 15  $>$ d,  &  0.052  & 114  & 0.843     & 128   \\
and not $\Delta$t $< -5 $ \\
\\
$\Delta$t $< -5 $ d   & 0.063    & 540  &  0.942    &  537  \\
 and $ \Delta$t $>$  15 d

\enddata

\end{deluxetable*}

\section{Galaxy Properties}\label{sec:interpretation}

In this section, we present the statistics of the host galaxies divided into two categories, SNe~Ia and core-collapse, based on final photometric typing with redshift using the full photometric nearest neighbor technique, the photometric~Z~Ia sample.  The redshift and apparent cmodel magnitude distribution of host galaxies of likely Type Ia and likely core-collapse SNe are shown in Figure~\ref{fig:zhist}.  The median redshift of the galaxies of the combined SN~Ia sample of SDSS and BOSS is 0.29.  The median redshift of the core-collapse sample is 0.21.  Similarly, the apparent \texttt{cmodel magnitude} distributions
have a median of 20.6 for the host galaxies of the SN~Ia sample and 20.2 for the core-collapse sample.  With photometry, redshifts, and spectra, we report the stellar mass, star formation, absolute magnitudes, velocity dispersion, and H$\beta$ equivalent linewidths for all host galaxies.  With this information, one can perform analysis of the correlation between SN light curve properties and host galaxy properties.  We conclude the section by performing simplified analysis of these correlations as a demonstration of the power of the data.

\subsection{Photometric Galaxy Properties with SDSS and BOSS Redshifts}\label{subsec:photgalprop}
We examine the stellar mass and specific star formation rates (sSFR) of all photometric Z Ia and all photometric Z CC host galaxies using the spectroscopic redshift and the SDSS imaging photometry.  To determine the galaxy properties using only photometry and spectroscopic redshift, we have chosen to use the ``Granada FSPS'' product (Montero-Dorta et al. in preparation).  These fits are based on the publicly available Flexible Stellar Population Synthesis (FSPS) code~\citep{conroy09a} and were computed for several models of stellar formation history, dust attenuation, and initial mass function (IMF).  The Granada FSPS fits to all galaxies in SDSS and BOSS were released as part of DR10 in Summer 2013 \citep{ahn13a}.  These data may be found in the galaxy product\footnote{http://data.sdss3.org/sas/dr10/boss/spectro/redux/galaxy (BOSS data) and \\ http://data.sdss3.org/sas/dr10/sdss/spectro/redux/galaxy (SDSS data) } at the SDSS-III Science Archive Server (SAS).  The best fits to the BOSS data are recorded in the file named ``v1\_0/granada\_fsps\_krou\_wideform\_nodust-v5\_5\_12.fits.gz''  and the best fits to the SDSS data are recorded in the file named ``granada\_fsps\_krou\_wideform\_nodust-dr10-v1\_0.fits.gz''.  One can match the results of Granada FSPS to the SDSS-II SN Data Release by matching on the keywords plate, fiberid, and mjd, or by matching on SPECOBJID.  

The Granada FSPS stellar masses reported in this paper for the host galaxies were obtained assuming no dust and the Kroupa (2001) IMF. The computation was performed using an extensive grid of models, with varying formation times, e-folding times ($\tau$) and metallicities. In addition, we report absolute magnitudes (K+E corrected to $z=0.55$) and estimated values for other stellar population properties, such as sSFR.  Table~\ref{tab:galaxy-summary} provides a summary of stellar mass, sSFR, and $r$-band absolute magnitude for the sample described in this paper.

\begin{figure}[htbp]
\epsscale{0.85}
\includegraphics[scale=0.48]{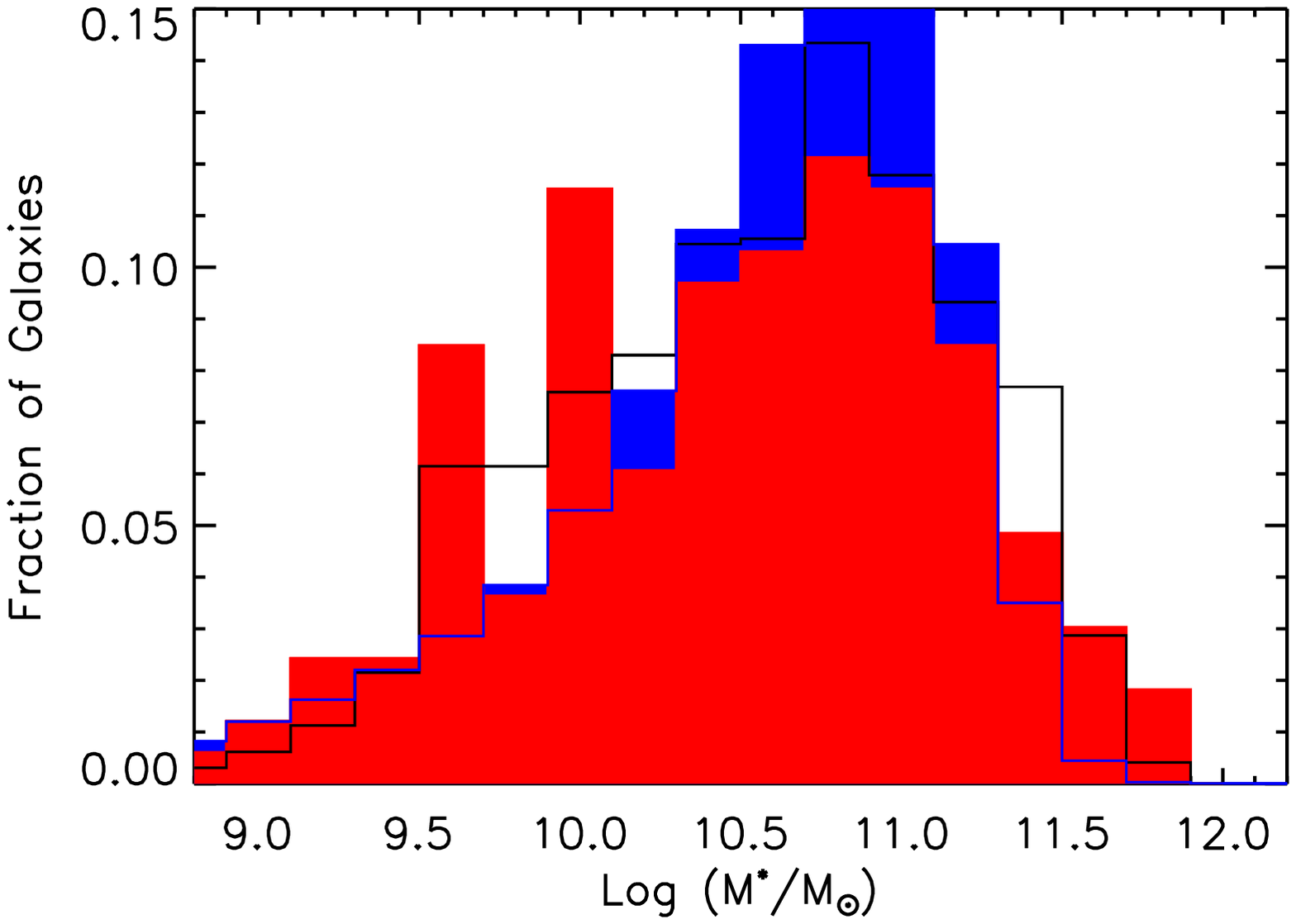}
\includegraphics[scale=0.48]{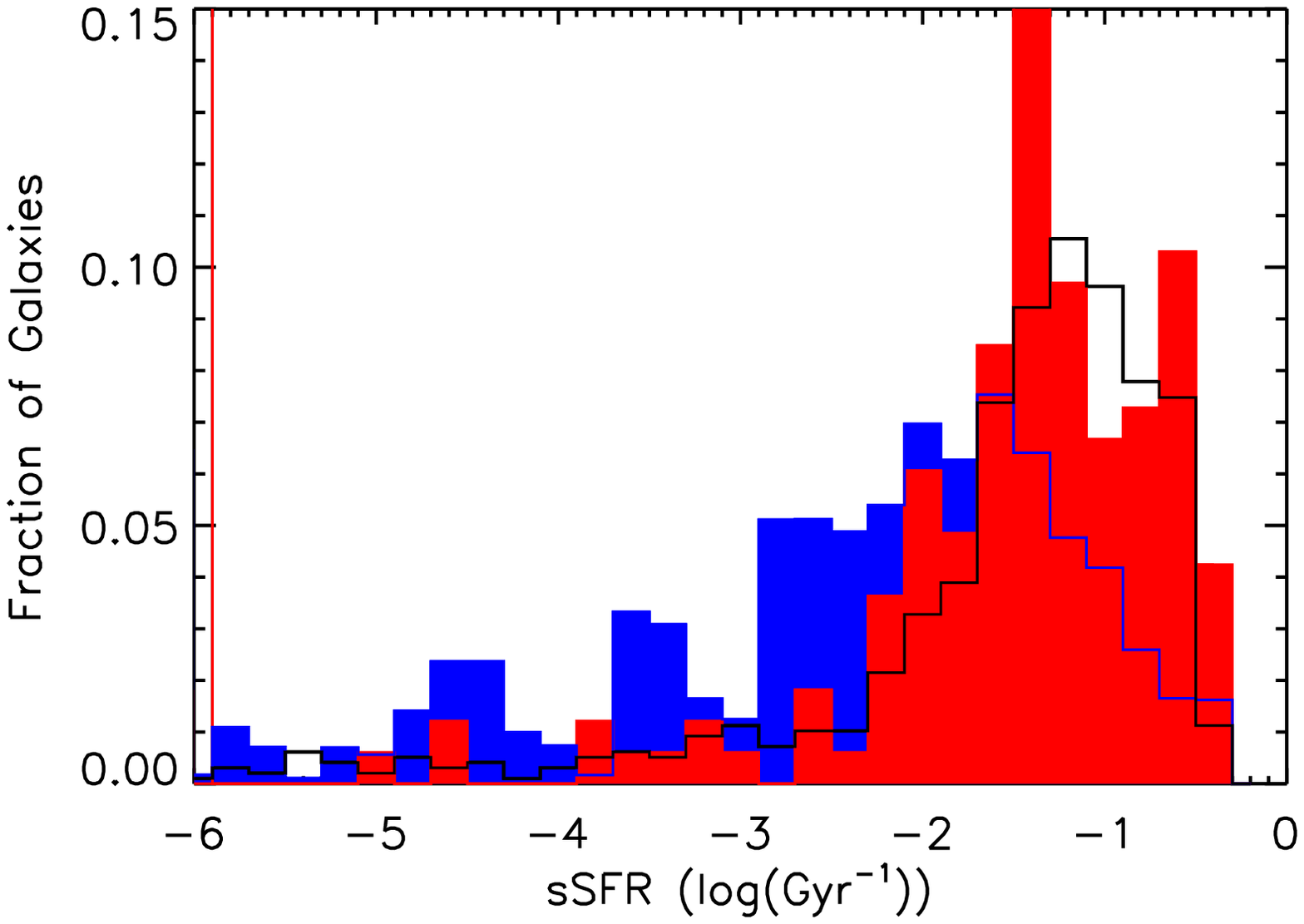}
\caption{\label{fig:mass-comp}TOP: The stellar mass distribution of galaxies hosting SNe~Ia in black and core-collapse in red using spectroscopic redshift priors.  The entire SDSS DR8 galaxy sample is shown in blue. BOTTOM:  The distribution of the sSFR for the galaxies that hosted SNe~Ia is in black and the core-collapse SNe is in red. The entire SDSS DR8 galaxy sample is shown in blue.  The grouping of galaxies shown with sSFR$= -6$ are simply galaxies in which the fit was not allowed to go to smaller values of sSFR.  } 
\end{figure}

 In Figure~\ref{fig:mass-comp}, we present the distribution of the stellar mass and sSFR for the photometric Z Ia and photometric Z CC host galaxy samples.  The SNe~Ia have host galaxies with a log(mass/$M_\Sun$) mean (RMS dispersion) equal to $10.7 \pm0.6$ and the core-collapse sample has mean $10.4 \pm0.6$.  For this analysis, we consider all galaxies with sSFR $< -3 (-6) $ log(Gyr$^{-1}$) to be likely non-star forming galaxies.  By this definition, 49.6\% (21.2\%) of SNe~Ia are found in a non-star forming galaxy and 32.3\% (13.7\%) of core-collapse are found in a non-star forming galaxy.

\begin{figure}[htbp]
\includegraphics[scale=0.5]{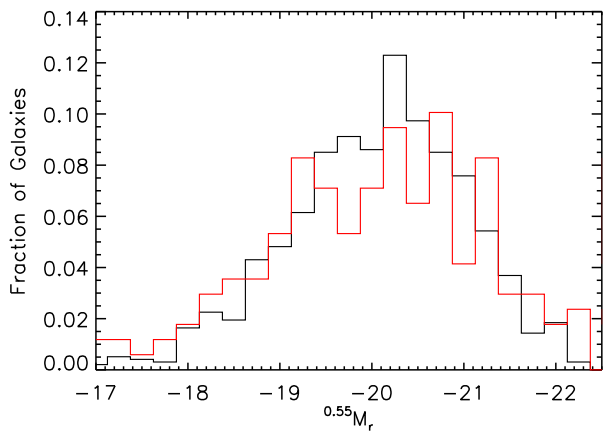}
\caption{\label{fig:abs-mag}  The $r$-band absolute magnitude (derived from cmodelmag) distribution for galaxies hosting SN~Ia (solid) and core-collapse SN (red, dashed).  The information in DR10 is sufficient for one to generate similar figures for $ugiz$.  }
\end{figure}

\begin{figure}[htbp]
\includegraphics[scale=0.55]{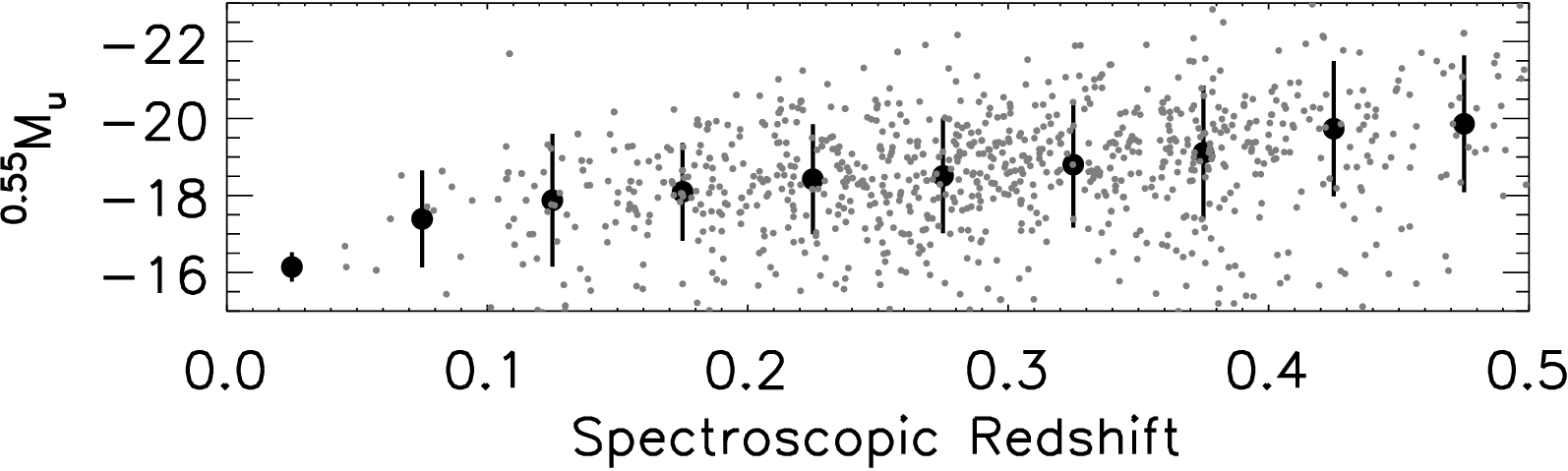}
\includegraphics[scale=0.55]{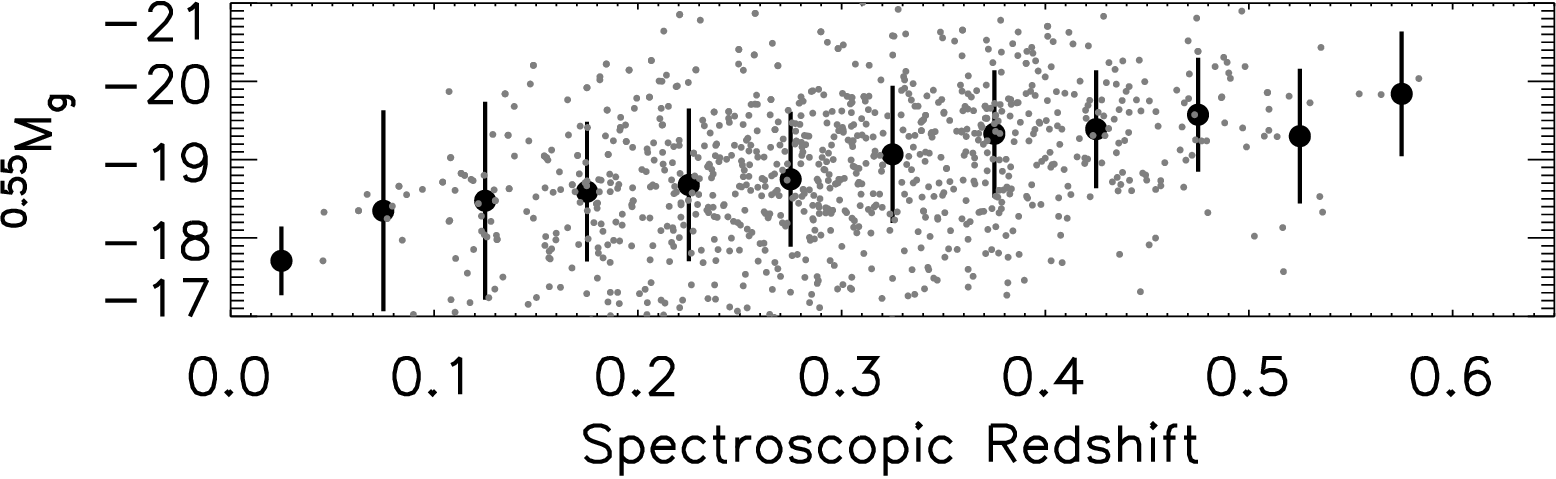}
\includegraphics[scale=0.55]{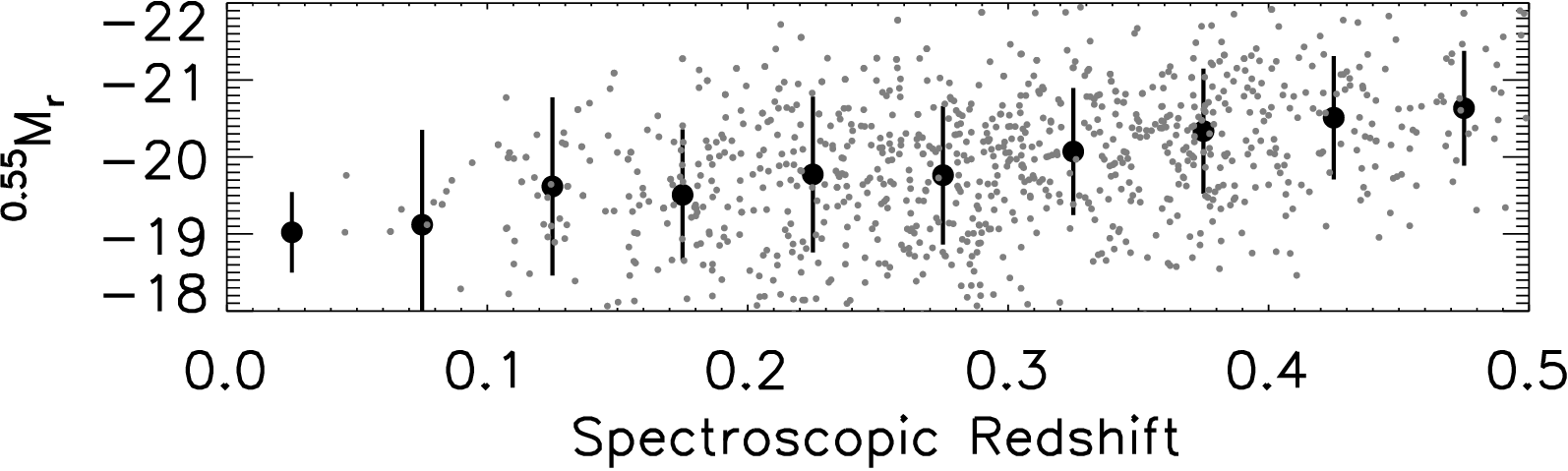}
\includegraphics[scale=0.55]{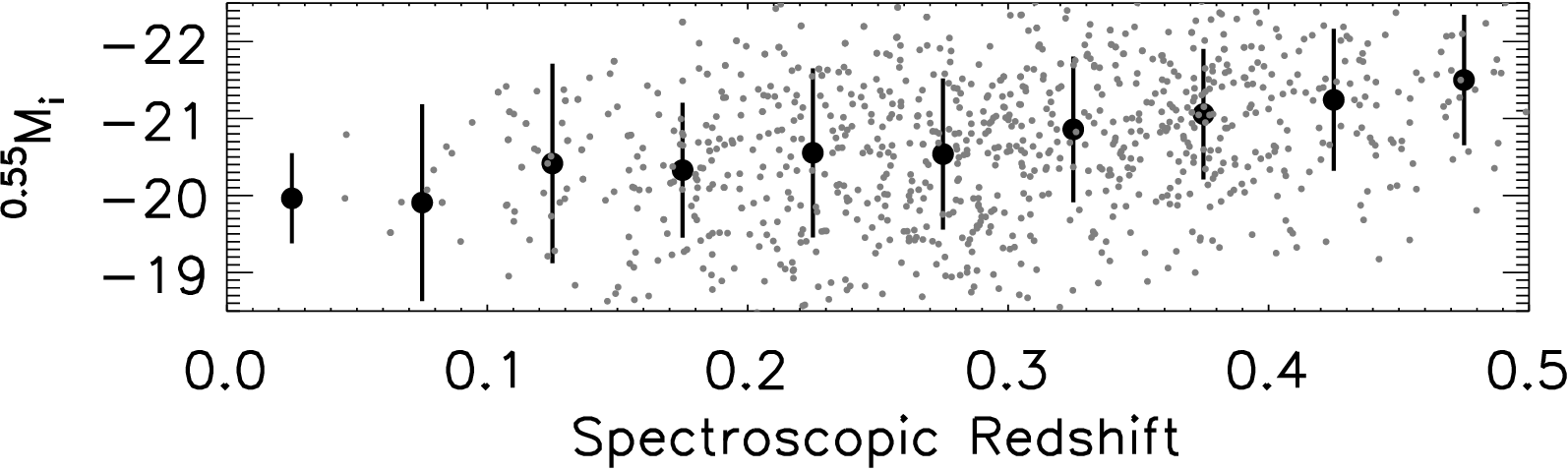}
\includegraphics[scale=0.55]{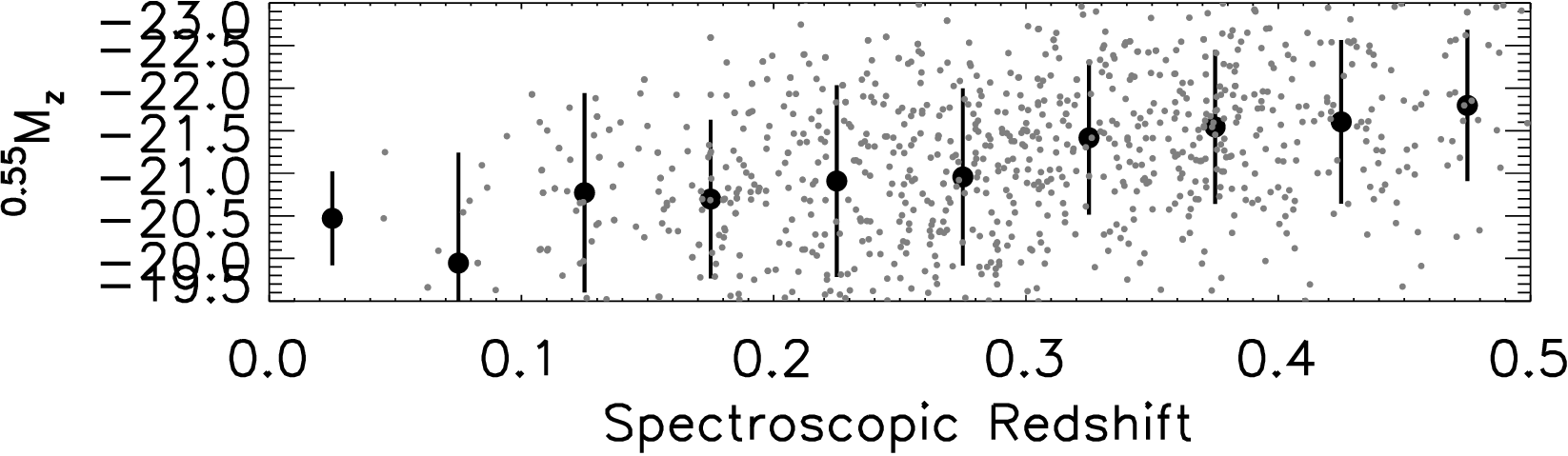}
\caption{\label{fig:abs-magz}The absolute magnitudes of the SN~Ia host galaxy sample by filter.  The scatter plot consisting of small dots describes each galaxy in the sample.  The median value binned by redshift (intervals of $\Delta z =$0.05) is shown by large filled circles with error bars representing the RMS within the bin. }
\end{figure}

Figure~\ref{fig:abs-mag} shows the distribution of the $r$-band absolute magnitude between galaxies hosting SNe~Ia and core-collapse SNe.  The distribution of galaxies hosting SNe~Ia has median $r$-band absolute magnitude of $-20.1$ and RMS of 0.9.  The distribution of galaxies hosting core-collapse has median $r$-band absolute magnitude of $-19.9$ and RMS of 1.1.  Figure~\ref{fig:abs-magz} shows the absolute magnitude of the host galaxies as a function of redshift.

\subsection{ Galaxy Properties from SDSS and BOSS Spectra}\label{subsec:specgalprop}
The SDSS and BOSS spectroscopy offers us the opportunity to derive galaxy properties directly from the host galaxy spectra.  This information is provided for the first time in DR9 for BOSS data using two different algorithms.  The first algorithm is the Portsmouth Stellar Kinematics and Emission Line Fluxes~\citep[][hereafter T12]{thomas12a}.  The second algorithm is the Wisconsin PCA method \citep{chen12a}.  In DR10, both the Portsmouth Stellar Kinematics and Emission Line Fluxes and the Wisconsin PCA computations were applied to SDSS spectra to provide a consistent catalog of spectroscopic galaxy properties across SDSS-I, -II, and -III.  

T12 derive emission line properties using stellar kinematics based on the Gas and Absorption Line Fitting code~\citep[GANDALF v1.5;][]{sarzi06a}.  T12 simultaneously fit a stellar population model~\citep[][hereafter M11]{maraston11a} and Gaussian emission line templates to the galaxy spectrum to separate stellar continuum and absorption lines from the ionized gas emission.  The line-of-sight velocity distribution is fit using the Penalized PiXel Fitting code~\citep[pPXF;][]{cappellari04a} code.

The Wisconsin PCA method decomposes each galaxy spectrum into best-fit PCA components derived from a library of model spectra based on the stellar population models of \citet{bruzual03a}.  A probability distribution function is built from the composite PCA fit to estimate stellar masses and velocity dispersions for each galaxy.  In DR10, the Wisconsin PCA code was extended to include the M11 population models.  

While both the Wisconsin PCA and T12 algorithms provide a characterization of the galaxy spectra for each SN host galaxy in our sample, we report only the T12 estimates here.  We choose the T12 approach because it provides a more empirical measure of galaxy properties through characterization of emission lines and velocity dispersion.  The Wisconsin PCA models were derived from specific models of stellar evolution and tested primarily on the more homogeneous BAO galaxy samples in BOSS \citep{chen12a,dawson12a} and may not be appropriate for the diverse population of galaxies presented here.  The DR10 T12 fits are found the BOSS galaxy product in the file named ``v1\_0/portsmouth\_emlinekin\_full-v5\_5\_12.fits.gz''.  The fits to the SDSS data are found in the file ``portsmouth\_emlinekin\_full-dr10-v1\_0.fits.gz''.  As described in Section~\ref{subsec:photgalprop}, the sample from the SN database can be matched to these files through use of the plate, fiberid, and mjd, or by matching on SPECOBJID. 

\begin{figure}[htbp]
\epsscale{0.85}
\includegraphics[scale=0.48]{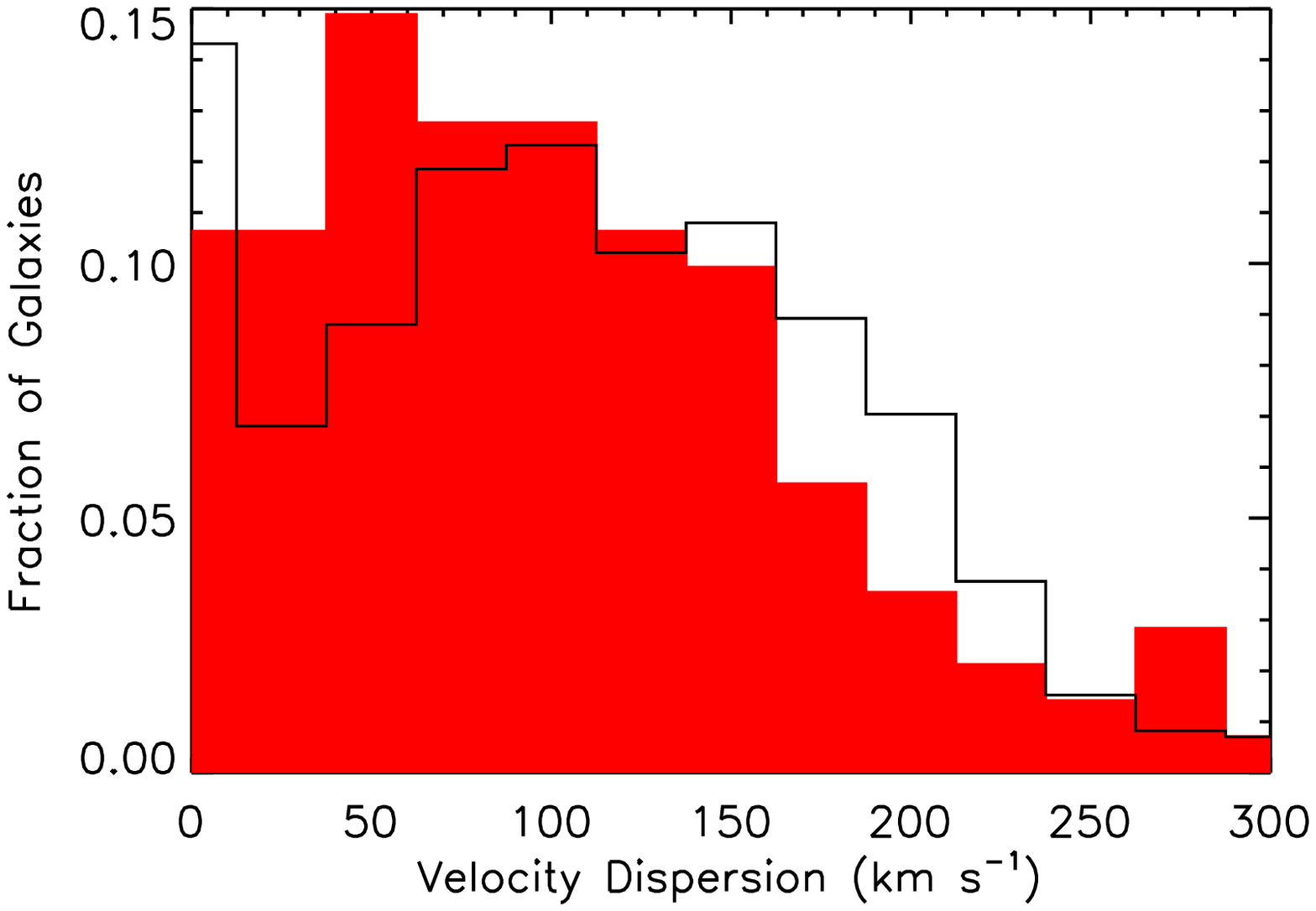}
\includegraphics[scale=0.48]{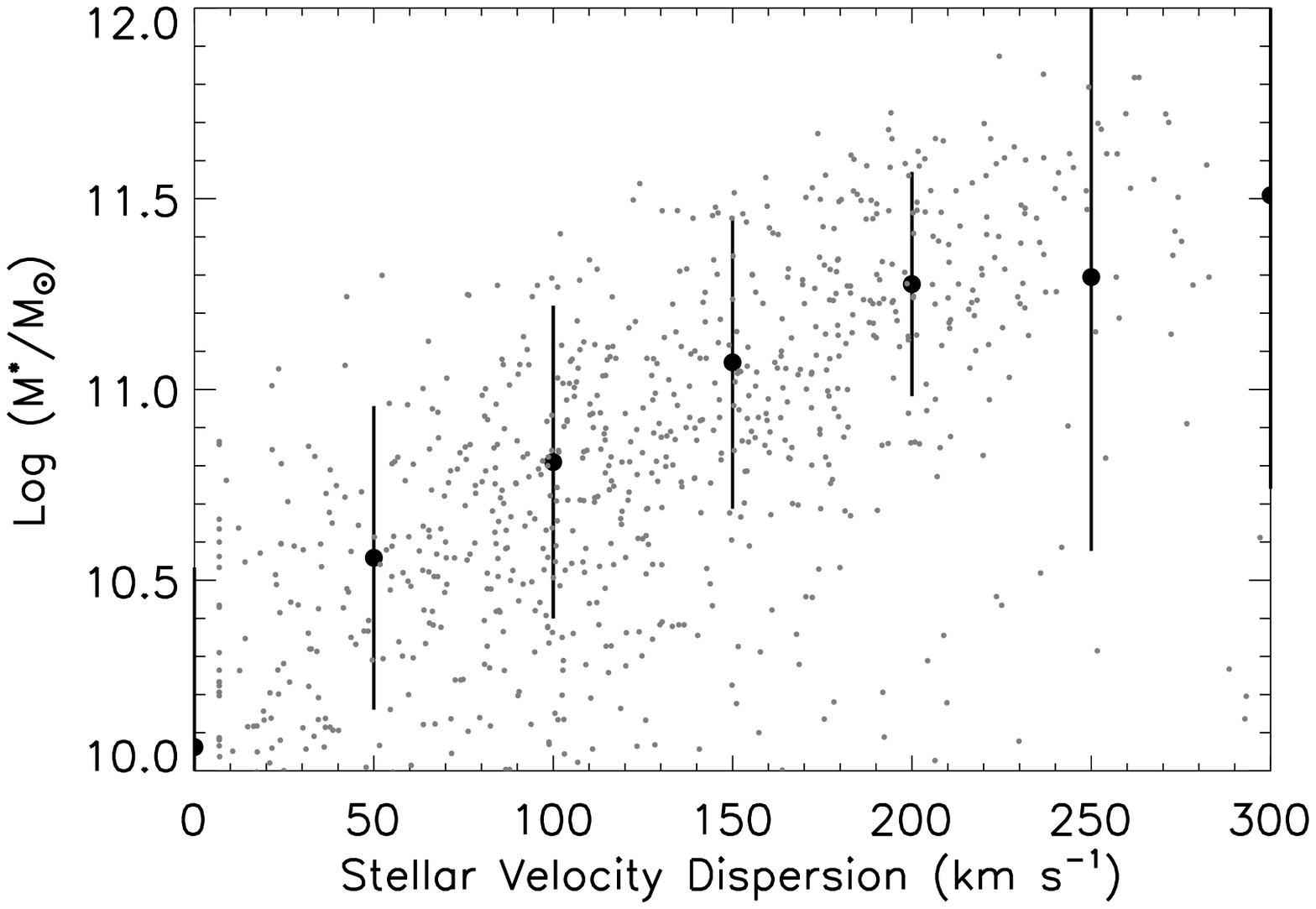}
\caption{\label{fig:spec-mass-vdisp}TOP: The distribution of stellar velocity dispersion for galaxies hosting SNe~Ia shown in solid black and those hosting core-collapse in red. BOTTOM:  The photometric stellar mass estimates as a function of velocity dispersion estimates for the sample of 842 likely SN~Ia host galaxies.  The median value binned by velocity dispersion (intervals of $\Delta \sigma =$50 km s$^{-1}$) is shown by large filled circles with error bars representing the RMS within the bin.}
\end{figure}

\begin{figure}[htbp]
\epsscale{0.85}
\includegraphics[scale=0.48]{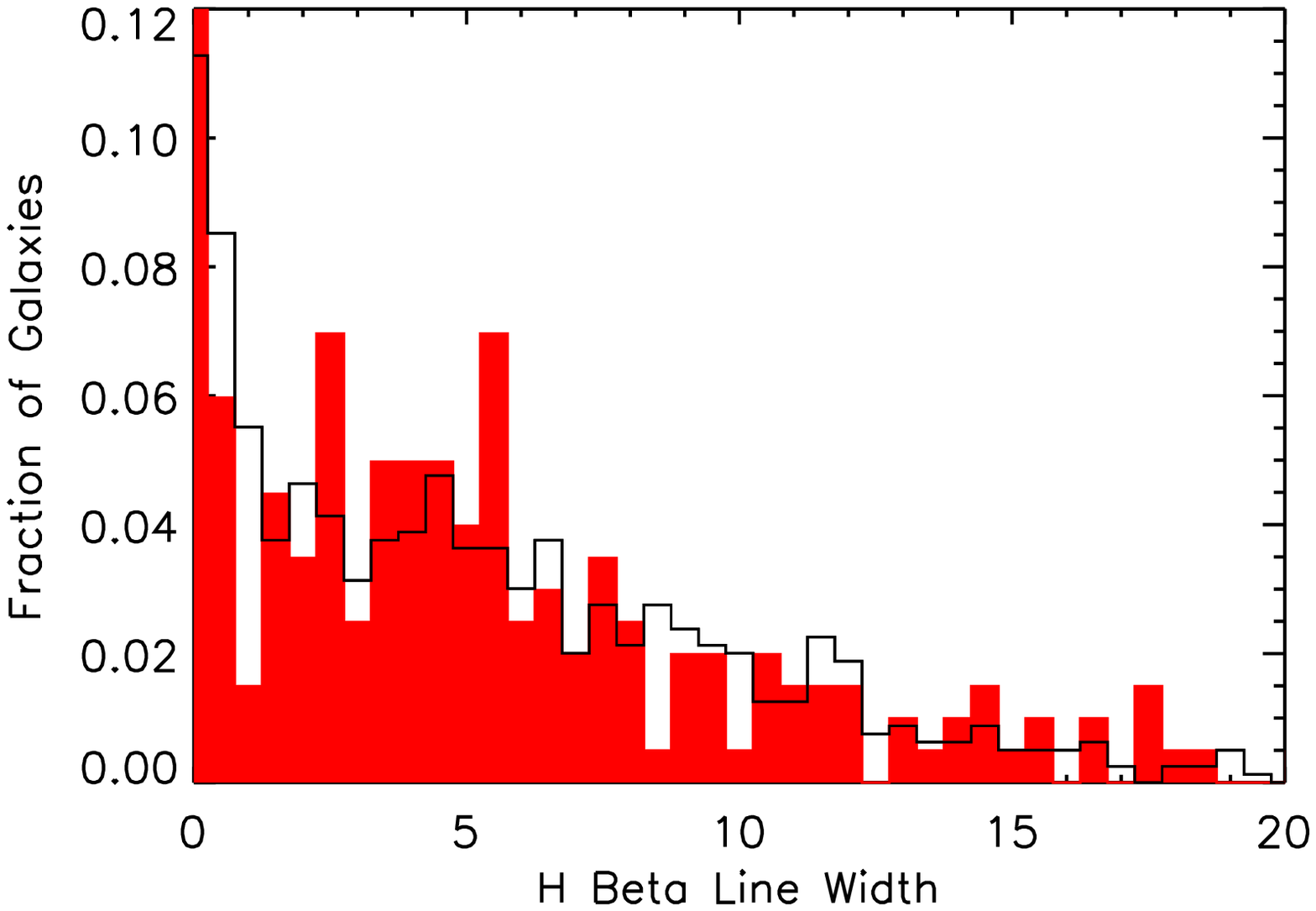}
\includegraphics[scale=0.48]{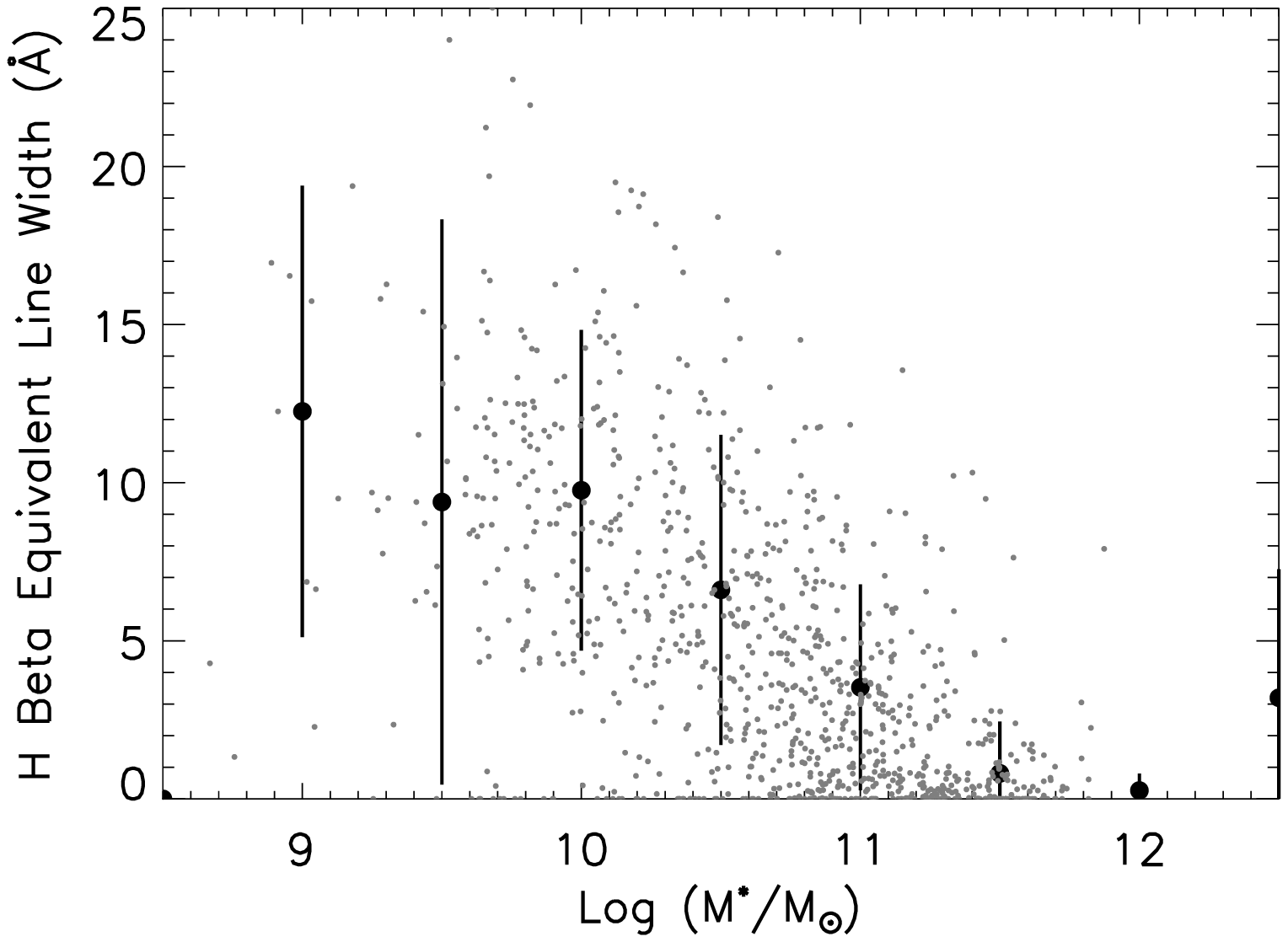}
\caption{\label{fig:spec-hbeta-mass}TOP:  The distribution of H$\beta$ equivalent width estimates for galaxies hosting SNe~Ia shown in solid black and those hosting core-collapse in red.  BOTTOM:  The spectroscopic H$\beta$ equivalent width estimates as a function of photometric stellar mass estimates for the likely SN~Ia host galaxies.  The median value binned by stellar mass (intervals of $\Delta log(m) =$0.5) is shown by large filled circles with error bars representing the RMS within the bin.}
\end{figure}

As before, we present a comparison of statistics from the T12 fits for the photometric Z Ia and photometric Z CC host galaxy samples.  Estimates of velocity dispersion and H$\beta$ equivalent linewidth can be used as proxies for galaxy mass and star formation history, respectively.  We present the distribution of stellar velocity dispersion and rest frame H$\beta$ equivalent linewidths for the entire sample of SN host galaxies in Figure~\ref{fig:spec-mass-vdisp} and Figure~\ref{fig:spec-hbeta-mass}.  The mean of H$\beta$ is 5.1 \AA\ with an RMS of 4.2 \AA.  The velocity dispersion has mean 122 km s$^{-1}$ with an RMS of 74 km s$^{-1}$.  As before, we present the results of the SNe~Ia and core-collapse populations separately using the final photometric classification.  A summary of the SNe~Ia host galaxy spectroscopic properties is included in Table~\ref{tab:galaxy-summary}.  

\begin{deluxetable*}{lrcccccrrc}
\centering
\tablewidth{0pt}

\tabletypesize{\footnotesize}
\tabletypesize{\scriptsize}
\tablecaption{\label{tab:galaxy-summary} Summary of SN~Ia Host Galaxy Properties}
\tablehead{\colhead{Redshift } & \colhead{Number} & \colhead{Median }  & \colhead{RMS }  & \colhead{Median } & \colhead{Median }  & \colhead{RMS} & \colhead{Median } & \colhead{RMS} & \colhead{Median H$\beta$}\\
\colhead{Range} & \colhead{} & \colhead{ Mass}   & \colhead{Mass}    & \colhead{ sSFR}   & \colhead{M$_\textrm{r}$} & \colhead{M$_\textrm{r}$} & \colhead{ Velocity}  & \colhead{ Velocity}  & \colhead{Equivalent } \\
\colhead{}      & \colhead{} & \colhead{} & \colhead{} & \colhead {} & \colhead{} & \colhead{} & \colhead{Dispersion}  & \colhead{Dispersion}  & \colhead{Linewidth} \\
\colhead{} & \colhead{} & \colhead{(log(mass))} & \colhead{(log(mass))} & \colhead{log(Gyr$^{-1}$))} & \colhead{} & \colhead{} & \colhead{(km s$^{-1}$)} & \colhead{(km s$^{-1}$)} & \colhead{(\AA)}}
\startdata

      0.00--    0.05 &            2 &       10.7 &      0.2 &      -8.6 &      -19.4&      0.5 &       115.268 &       9.1 &       1.2  \\ 
    0.050--     0.10 &           16 &       10.4 &      0.8 &      -1.7 &      -19.4 &       1.4 &       72.6 &       51.0 &       4.8  \\ 
     0.10--     0.15 &           64 &       10.7 &      0.7 &      -1.4 &      -19.6 &       1.2 &       93.2 &       51.3 &       5.0 \\ 
     0.15--     0.20 &          109 &       10.6 &      0.6 &      -1.2 &      -19.6 &      0.9 &       99.2&       63.0 &       4.6  \\ 
     0.20--     0.25 &          145 &       10.6 &      0.7 &      -1.1 &      -19.8 &       1.0 &       111.2 &       74.3 &       4.0  \\ 
     0.25--     0.30 &          190 &       10.6 &      0.6 &      -1.0 &      -19.8 &      0.9 &       112.2 &       74.3 &       4.3  \\ 
     0.30--     0.35 &          160 &       10.9 &      0.6 &      -1.4 &      -20.1 &      0.8 &       140.5 &       73.2 &       1.9  \\ 
     0.35--     0.40 &          158 &       10.9 &      0.5 &      -1.4 &      -20.3 &      0.8 &       128.4 &       72.4 &       1.9  \\ 
     0.40--     0.45 &           82 &       10.9 &      0.6 &     -1.0  &      -20.5 &      0.8 &       134.3 &       88.2 &       4.3  \\ 
     0.45--     0.50 &           39 &       11.0 &      0.5 &      -1.2 &      -20.6 &      0.7 &       152.3 &       86.2 &       1.7  \\ 
     0.50--     0.55 &           14 &       11.0 &      0.6 &      -1.8 &      -20.5 &      0.7 &       206.5 &       100.2 &       1.28  \\ 
     0.55--     0.60 &            4 &       10.8 &      0.6 &      -2.2 &      -20.7 &      0.3 &       198.7 &       333.2 &       2.0

\enddata
\end{deluxetable*}

\subsection{Correlating SN Properties With Galaxy Properties }\label{subsec:galprop}

This large SN host galaxy sample can be used for a number of investigations of possible relationships between SN~Ia light curve properties and host galaxy properties.  Many previous studies have searched for a dependence of SN~Ia rate and light curve properties on host galaxy mass and metallicity.  Here we show general examples of how these investigations can be enhanced with this new sample of well-characterized SN~Ia light curves and host galaxies.

The rate of SN~Ia events per cosmological volume element has been shown to depend on both mass and star formation history \citep{scannapieco05a}.  Studies of SN~Ia rates have been performed at all redshifts with samples of various sizes and completeness \citep[e.g.][]{dilday10a,barbary12a}.  Recent analysis have been performed using the subsample of spectroscopically confirmed SDSS-II SNe combined with mass determined by photometry \citep[342 total SNe~Ia;][]{smith12a} and spectroscopy \citep[53 total SNe~Ia;][]{gao13a}.  In the top panel of Figure~\ref{fig:mass-comp}, we compare the distribution of 935 photometric Z Ia host galaxy masses to the nearly magnitude-limited sample of SDSS galaxy masses.  Similarly, in the bottom panel of Figure~\ref{fig:mass-comp}, we compare the distribution of sSFR in those same 935 SN~Ia host galaxies the sSFR in SDSS galaxies.  While a detailed analysis requires a full reconstruction of the SDSS-II SN detection efficiency, one clearly sees trends towards production of core-collapse and SNe~Ia in galaxies with lower mass and higher sSFR. The sheer increase in sample size relative to previous studies highlights the potential of using these results for improved rates analysis.

We next use the galaxy properties found in Sections \ref{subsec:photgalprop} and \ref{subsec:specgalprop} to examine the potential for future analysis on the relationship between host galaxy metallicity and SN~Ia light curve properties.  Previous analysis has been performed on smaller SN~Ia samples using H$\beta$ and iron absorption strengths \citep{gallagher08a,johansson12a} and estimations of metal abundance from host mass and star formation rates derived from SEDs \citep{hayden13a}.  The T12 algorithm produces estimates of emission line strengths for hydrogen, helium, nitrogen, oxygen, neon, sulfur, and argon, but does not fit absorption lines such as iron.  However, the various line fits do allow computation of proxies for gas phase metallicity.  As an example, we here report an estimate of metallicity using one of the proxies derived from oxygen-nitrogen provided in \citet{foster12a}.  We selected the method of \citet{pettini04a} to determine the O3N2 ratio defined as
\begin{equation} 
 O3N2\equiv\log{\left( \frac{[OIII]\lambda5006/H\beta}{[NII]\lambda6583/H\alpha}\right)}. 
\end{equation}
This proxy is empirically calibrated to an O/H estimate of metallicity according the the following relation 
\begin{equation}\label{eq:met}
\log(O/H)+12 = 8.73-0.32 \times O3N2.
\end{equation}
We use the relation 
\begin{equation} 
12+log(O/H)_{\rm T04}=0.103+1.021 \times (12+log(O/H)_{\rm PP04}).
\end{equation}
found in \citet{foster12a} to convert between the metallicity estimates found in Equation~\ref{eq:met} and those found using oxygen abundances with the SDSS data \citep{tremonti04a}.

\begin{figure}[htbp]
\epsscale{0.85}
\includegraphics[scale=0.48]{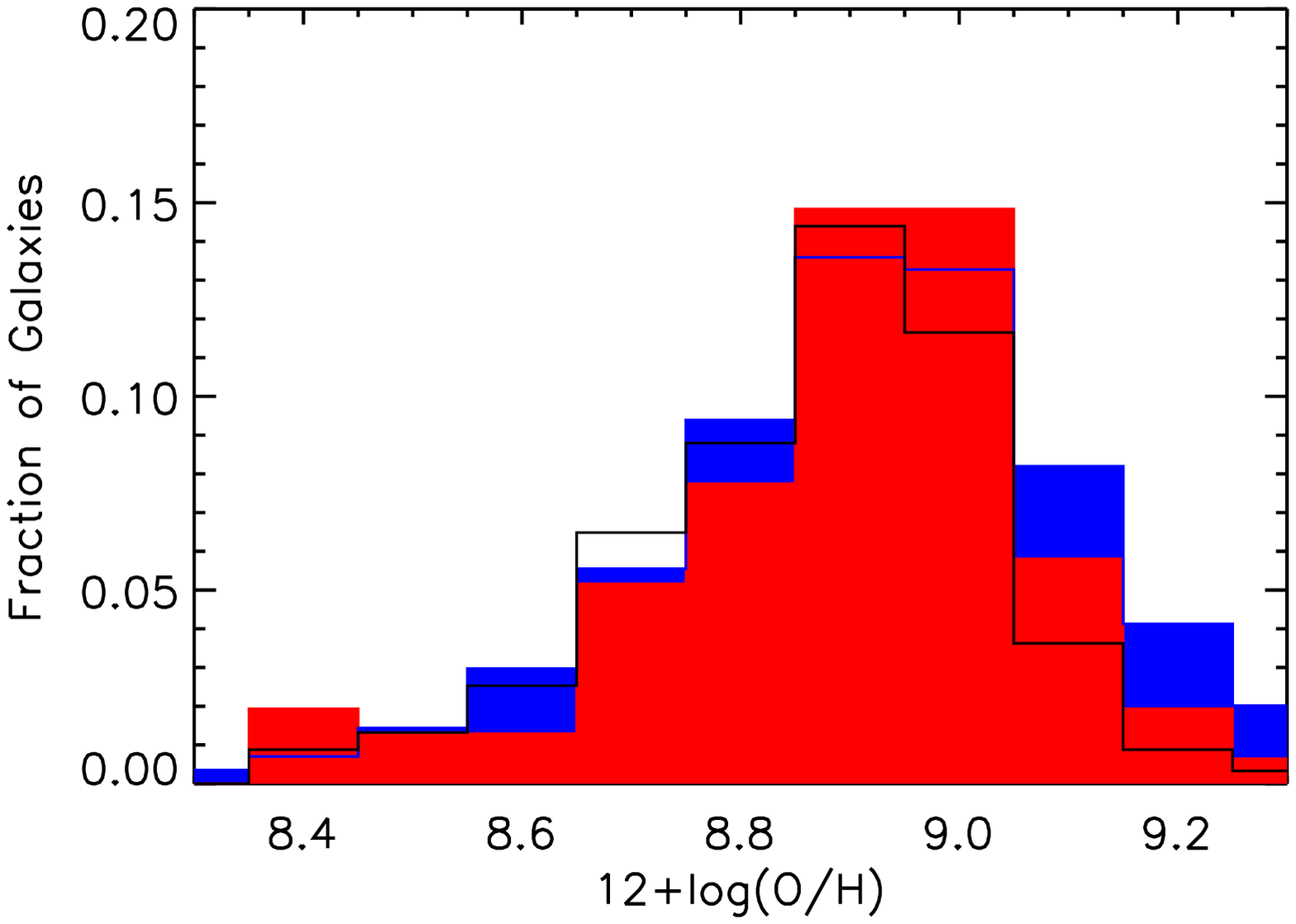}
\includegraphics[scale=0.48]{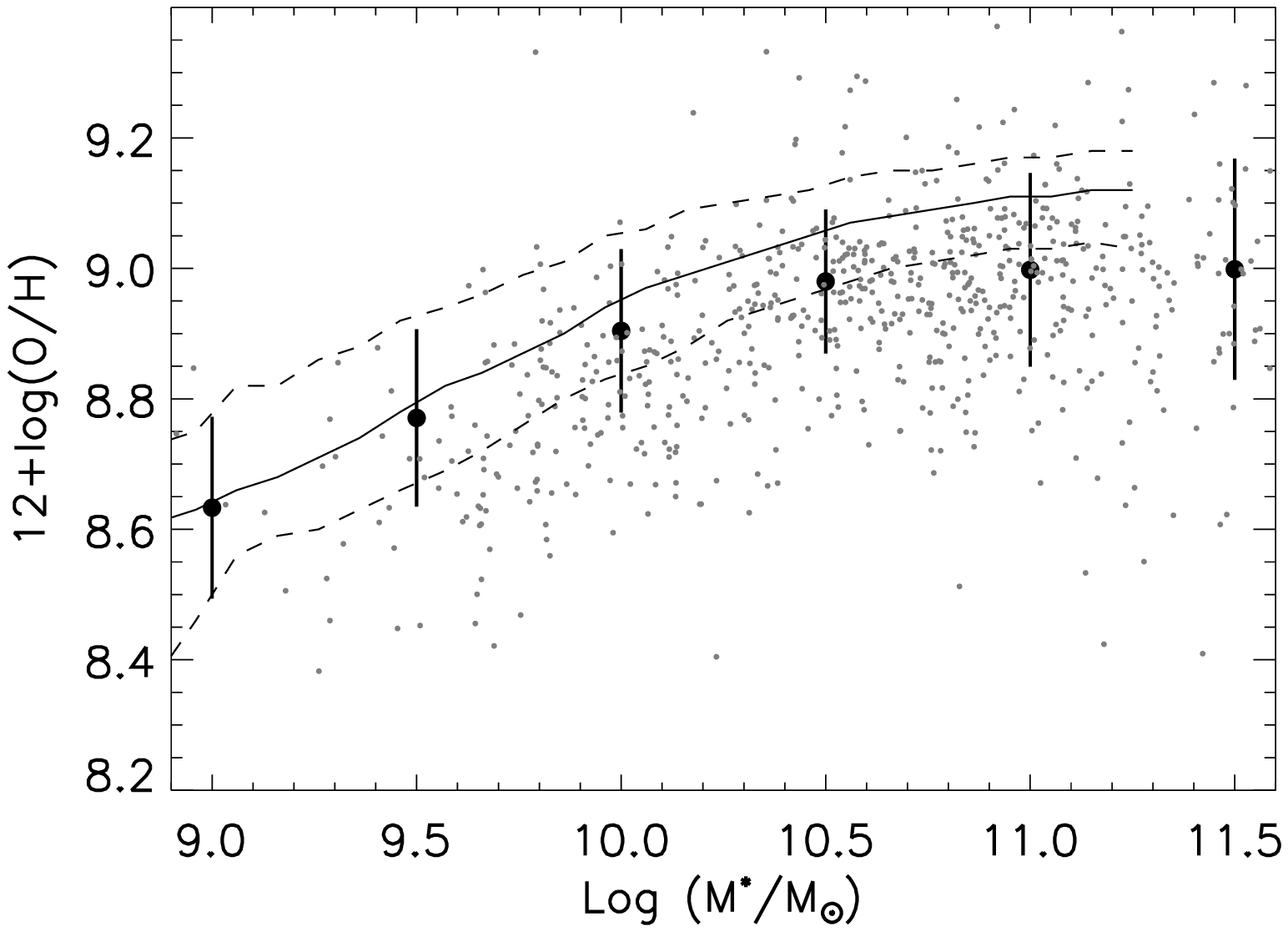}
\caption{\label{fig:mass-metallicity}
TOP:  The metallicity distribution of galaxies hosting SNe~Ia in solid black and core-collapse in solid blue using spectroscopic redshift priors.  In addition, the entire magnitude limited SDSS DR8 galaxy sample is shown in blue.
BOTTOM:  The metallicity as a function of stellar mass estimates for the likely SN Ia host galaxy sample. The median value binned by stellar mass (intervals of $\Delta log(m) =$0.5) is shown by large filled circles with error bars representing the RMS within the bin.  The 68\% confidence contour of the Tremonti relation for local SDSS field galaxies is shown in the dashed lines.   }
\end{figure}

Using this relation, we present the metallicity as a function of stellar mass of the SNe~Ia host galaxy sample in Figure~\ref{fig:mass-metallicity}.  In the top panel, we present the overall distribution of metallicity in the SN~Ia sample; the mean metallicity is  8.63 and the RMS dispersion is 0.15.  In the bottom panel,  we see a degeneracy between host galaxy metallicity and stellar mass; more massive galaxies also appear to have higher metallicity.  This degeneracy, the Tremonti relation for host galaxies \citep{tremonti04a}, will need to be taken into account in detailed analysis of the effects of host galaxy metallicity on SN~Ia light curves.

\begin{figure*}[htbp]
\epsscale{0.85}
\includegraphics[scale=0.48]{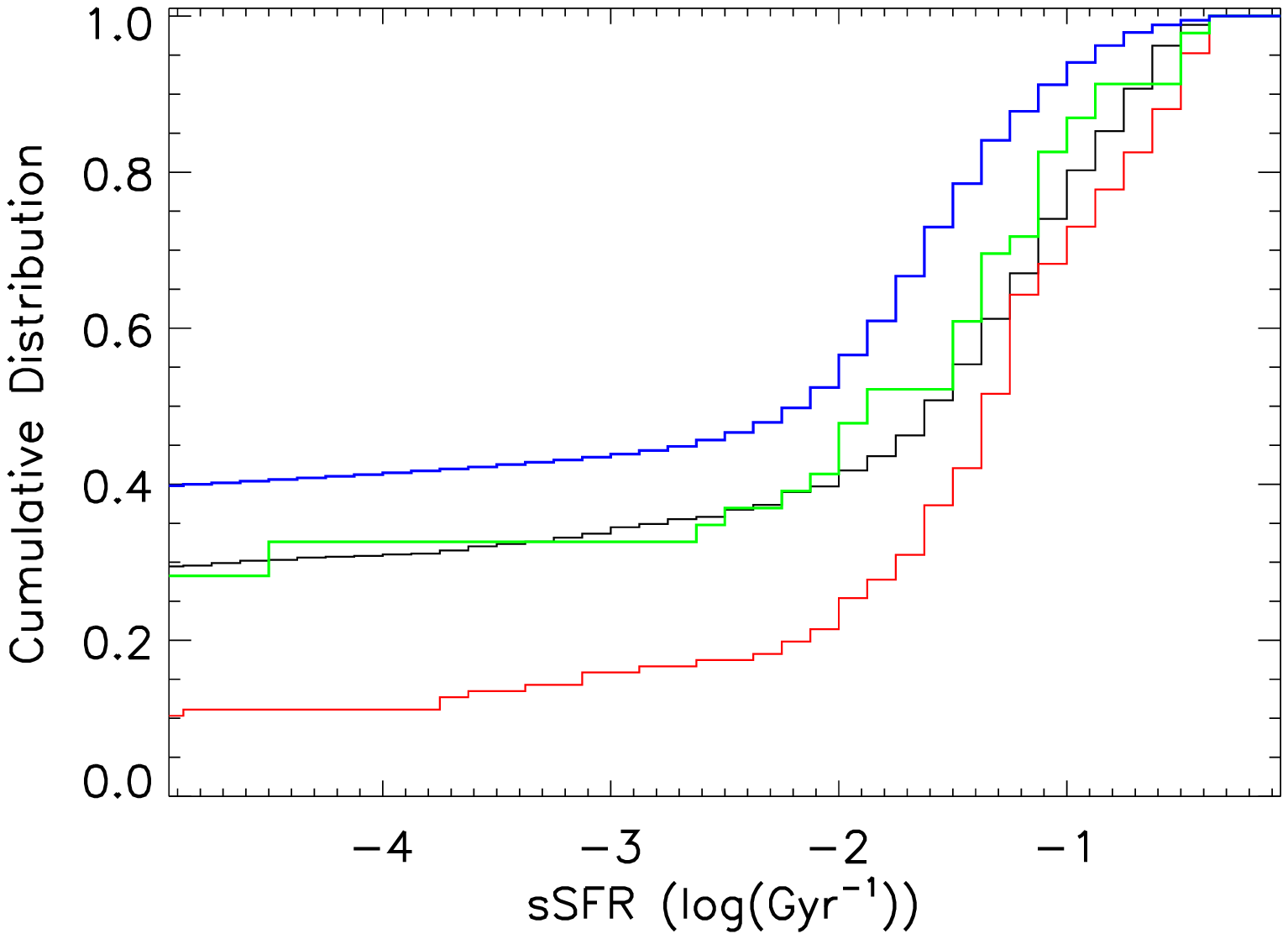}
\includegraphics[scale=0.48]{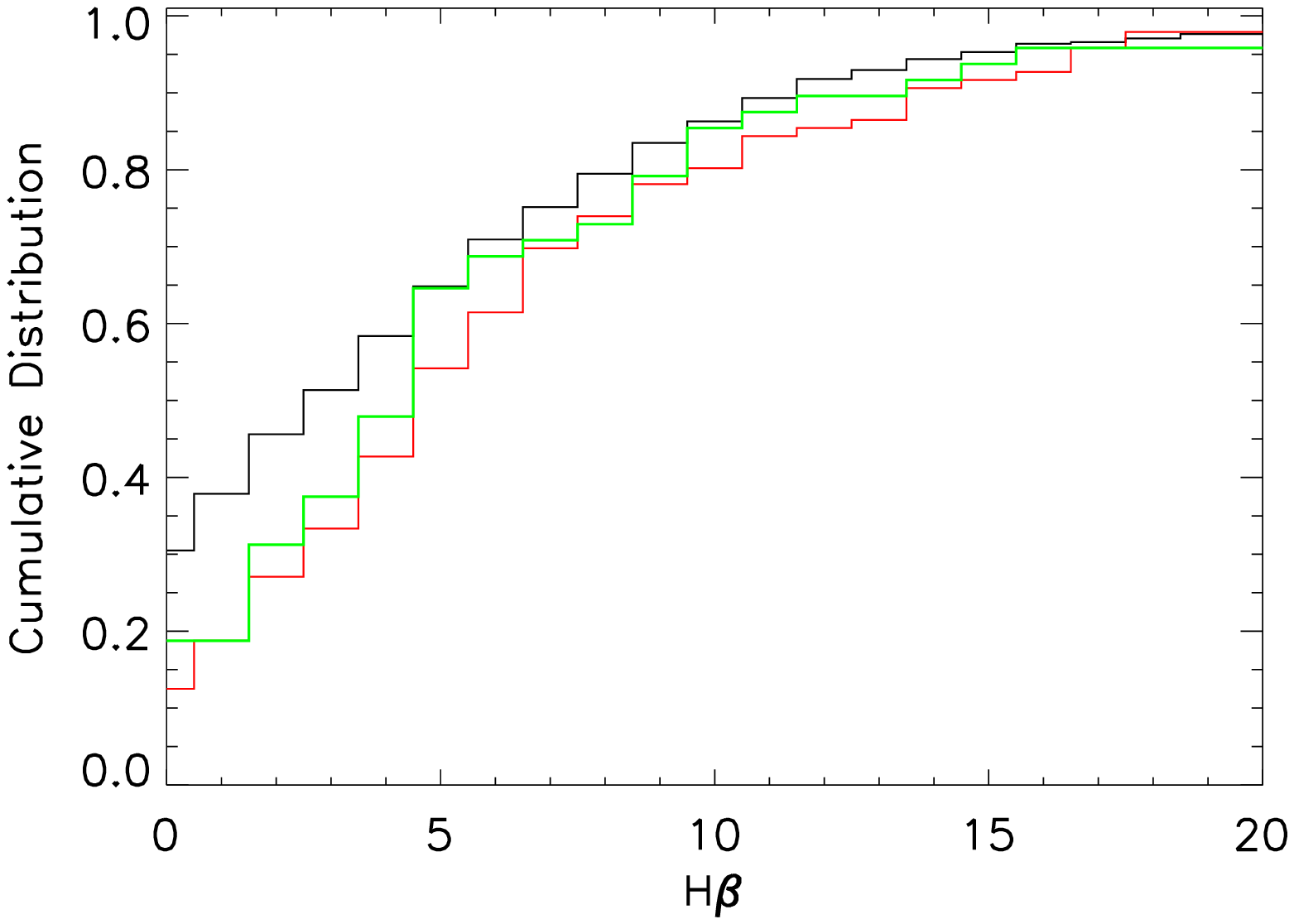}
\caption{\label{fig:cumulative-dist}
LEFT:  The sSFR cumulative distribution of galaxies hosting SNe~Ia in solid black, SNe~Ibc in green, and SNe~II in red (using spectroscopic redshift priors).  In addition, the entire magnitude limited SDSS DR8 galaxy sample is shown in blue.  RIGHT:  The H$\beta$ cumulative distribution. 
 }
\end{figure*}

As a final demonstration, we have divided the SNe sample into SNe~Ia, SNe~Ibc, and SNe~II and evaluated the SN yield in terms of host galaxy sSFR and H$\beta$ equivalent line width.  The left panel of Figure \ref{fig:cumulative-dist} shows the cumulative distribution of sSFR derived from photometry.  As expected, there is a clear separation between the SN~Ia and SN~II populations.  This separation is likely due to the fact that SNe~II preferentially occur in regions of active star-formation.  Surprisingly, the SNe~Ia and SNe~Ibc show similar distributions, contrary to the expectation that SN~Ibc are also associated with star-formation.  We have looked at these trends using different IMFs, assumptions on dust, and star-formation history and the confusion between SN~Ia and SN~Ibc remained.  This apparent contradiction may be due to the limitations of determining star-formation using only five-band photometry, the small sample size of SNe~Ibc (48 SNe~Ibc used here), selection of the wrong host galaxy for photometric properties, or difficulty in distinguishing between SNe~Ia and SNe~Ibc in the photometric classification.  However, the right panel shows a possible resolution to this contradiction; SNe~Ia appear twice as often in galaxies with no measurable H$\beta$ equivalent line width.  As H$\beta$ is an indicator of star-formation, it is likely that the discrepancy in the left panel is due to limitations in the photometric estimates of star-formation.

\section{SN~Ia Light Curve Properties}\label{sec:snprop}

Using the full photometric classification method from Section~\ref{subsec:comp}, we examine properties of the sample based on the actual light curves.  We use the SALT2 \citep{guy10a} approach to fit all of the candidate light curves, with a flat redshift prior and a spectroscopic redshift.  We determine the light curve shape parameter $x_1$, the color parameter $c$, and, in the case of a flat redshift prior, the best fit redshift.  As in Section~\ref{subsec:dataqual}, we examine trends in contamination and completeness with the position of each transient in $x_1$, $c$, and redshift space.  We also report surprising behavior of the SALT2 fitter to underestimate redshifts in cases of poor data quality, leading to biases in interpretation of SN properties without spectroscopic redshifts of the host galaxies.  

\subsection{Effects of Light Curve Parameters on Contamination and Completeness}\label{subsec:lightcurveparam}

Here we estimate the rates of contamination and completeness in a purely photometric SN survey as a function of light curve parameters.  As a baseline, we use the sample of 1268 transients that were classified as either photometric Ia or photometric Z Ia.  For each of these transients, we then fit SALT2 light curves to the time domain photometry, assuming no redshift information.  We use these fits to derive estimates of the $x_1$ parameter, the color $c$, and the photometric redshift derived from the light curve.  In 304 cases, the light curve fits fail selection cuts.  These failures appear to be caused by low S/N data points and can not be used in this analysis.  The resulting sample of 964 possible SN~Ia is now examined for contamination and completeness after performing various cuts on the distributions of $x_1$, $c$, and $z$ parameters.  As before, we define contamination as the fraction of events classified as photometric Ia that were not classified as a photometric Z Ia.  Similarly, completeness is defined as the fraction of events classified both as a photometric Ia and photometric Z Ia relative to the sample of photometric Z Ia.

We evaluate the performance on various subsets of this sample selected on photometric $z$, $x_1$, $c$, and an ellipse in $x_1-c$ space.  The contamination and completeness fractions for each subset are shown in Table~\ref{tab:lcqual}.  Imposing a restriction on the best fit photometric redshift does little to change the contamination or completeness except when requiring $z<0.2$.  Surprisingly, the contamination increases at these low redshifts, possibly due to higher likelihood of observing core-collapse SNe.  On the other hand, cuts on the $x_1$ and $c$ parameters do reduce contamination as one would expect.  Requiring $|x_1|<2$ reduces contamination from 0.108 to 0.080 while keeping over 83\% of the sample.  A greater improvement on contamination arises when requiring $|c| <0.2$; these tighter color restrictions decrease contamination from 0.109 to 0.074 and improve completeness from 0.924 to 0.941.  C13 used simulations to determine the parameters of an ellipse in $x_1-c$ space to maximize a figure of merit based on efficiency and purity.  Using the new photometric classification, we reproduce their cuts; we find values of contamination and completeness of 0.063 and 0.936, respectively.  We determined two additional elliptical cuts for our sample: one to maximize completeness and the other to minimize contamination.  The ellipse to minimize contamination has semi-major axis $x_1=2.0$ and semi-minor axis $c=0.225$.  To maximize completeness, we find an ellipse with semi-major axis $x_1=2.3$ and semi-minor axis $c=0.2$.  These two cuts are very similar in SALT2 parameter space and have very similar contamination, completeness, and sample size.  

We compare the performance of these light curve parameter cuts with the performance of data quality cuts from Section~\ref{subsec:dataqual}.  The best performance found in Section~\ref{subsec:dataqual} occurred when requiring both early time and late time light curve coverage, represented as the final entry in Table~\ref{tab:dataqual}.  The contamination, completeness, and total sample size were found to be 0.095, 0.965, and 511, respectively. The C13 cuts in Table~\ref{tab:lcqual} reduce contamination by a factor of 1.5, reduce completeness by a factor of only 1.015, and produce a sample size that is only 2.6\% smaller.  

C13 estimated contamination and the efficiency of selecting SNe~Ia by tuning cuts on a large sample of simulated light curves.  The primary limitation of this technique is that it requires a full understanding of SN rates to accurately populate the simulations.  In this paper, we estimate contamination and efficiency based on the consistency of SN photometric classifications with and without host galaxy redshift information.  The main limitation of this approach is that the true classification of each transient is not known.  These caveats aside, we compare the contamination and efficiency of the C13 results to our results.  For ease of comparison, we define efficiency in our sample as the number of SNe~Ia in a subset relative to the total number of SNe~Ia in the sample (first entry in Table~\ref{tab:lcqual}).  C13 find a contamination of 8.3\% and an efficiency of 71.6\% using their cuts modeled on simulated data when applying the SALT2 cuts.  We find a contamination of 6.3\% and an efficiency of 57.0\% (498/874) by applying those same cuts in $x_1-c$ parameter space.

\begin{deluxetable*}{lcccc}
\centering
\tablewidth{0pt}
\tabletypesize{\footnotesize}
\tablecaption{\label{tab:lcqual} Light Curve Restrictions}
\tablehead{\colhead{} & \colhead{Contamination}  & \colhead{Number in }& \colhead{Completeness}  & \colhead{Number in } \\  
\colhead{}  & \colhead{} & \colhead{photometric  } & \colhead{}  & \colhead{redshift  } \\
\colhead{}  & \colhead{} & \colhead{ classified sample} & \colhead{}  & \colhead{ classified sample } }
\startdata
No Cuts   & 0.086 & 895  & 0.922 & 888 \\
\\
$z_{\rm phot} < 0.5 $ & 0.079  & 872  & 0.924  & 869  \\
$z_{\rm phot} < 0.4 $ & 0.081  & 815  & 0.929  & 806  \\
$z_{\rm phot} < 0.3 $ & 0.080  & 699  & 0.939  & 685  \\
$z_{\rm phot} < 0.2 $ & 0.090  & 476  & 0.933  & 464  \\
\\
$-4 < x_1<4$ & 0.077 & 858  & 0.938  & 845  \\
$-3 < x_1<3$ & 0.073 & 804  & 0.944  & 818  \\
$-2 < x_1<2$ & 0.064 & 708  & 0.947  & 700  \\
\\
$-0.4 < c < 0.4$&  0.030 & 725 & 0.936 & 751 \\
$-0.3 < c < 0.3$&  0.022 & 635 & 0.931 & 667 \\
$-0.2 < c < 0.2$&  0.008 & 492 & 0.940 & 519 \\
\\

C13 $x_1-c$ ellipse &  0.063  & 497   & 0.936  & 498 \\
($x_1=3.0, c=0.25 $) \\
%\\
 $x_1-c$ ellipse to  & 0.050  & 422  & 0.950  & 423 \\
minimize contamination \\
($x_1=2.0, c=0.225 $) \\
%\\
 $x_1-c$ ellipse to  & 0.054  & 423  & 0.952   & 420 \\
maximize completeness \\
($x_1=2.3, c=0.2 $)

\enddata

\end{deluxetable*}

\subsection{Distribution of SN Light Curve Properties With Redshifts From the Host Galaxies}\label{subsec:completeness} 

As a demonstration of the value of the photometric Z Ia sample for cosmology, we now perform an analysis on the sample properties as a function of redshift.  For each of these 1126 transients, we again fit SALT2 light curves to the time domain photometry, this time using the spectroscopic redshift information.  We use these fits to derive estimates of the $x_1$ and $c$ parameters.  The light curve fits fail to meet selection cuts for 52 of these transients, again due to low S/N.  The failure rate, however, is much smaller than the rate found in Section~\ref{subsec:lightcurveparam} (304/1268).  The much smaller failure rate is likely due to fewer free parameters in the light curve fits, demonstrating yet another advantage of host galaxy redshifts, and less stringent selection criteria.  The remaining sample of 1074 photometric Z Ia SN are used to observe the distribution of the light curve properties with spectroscopic redshift.

\begin{figure}[htbp]
\includegraphics[scale=0.5]{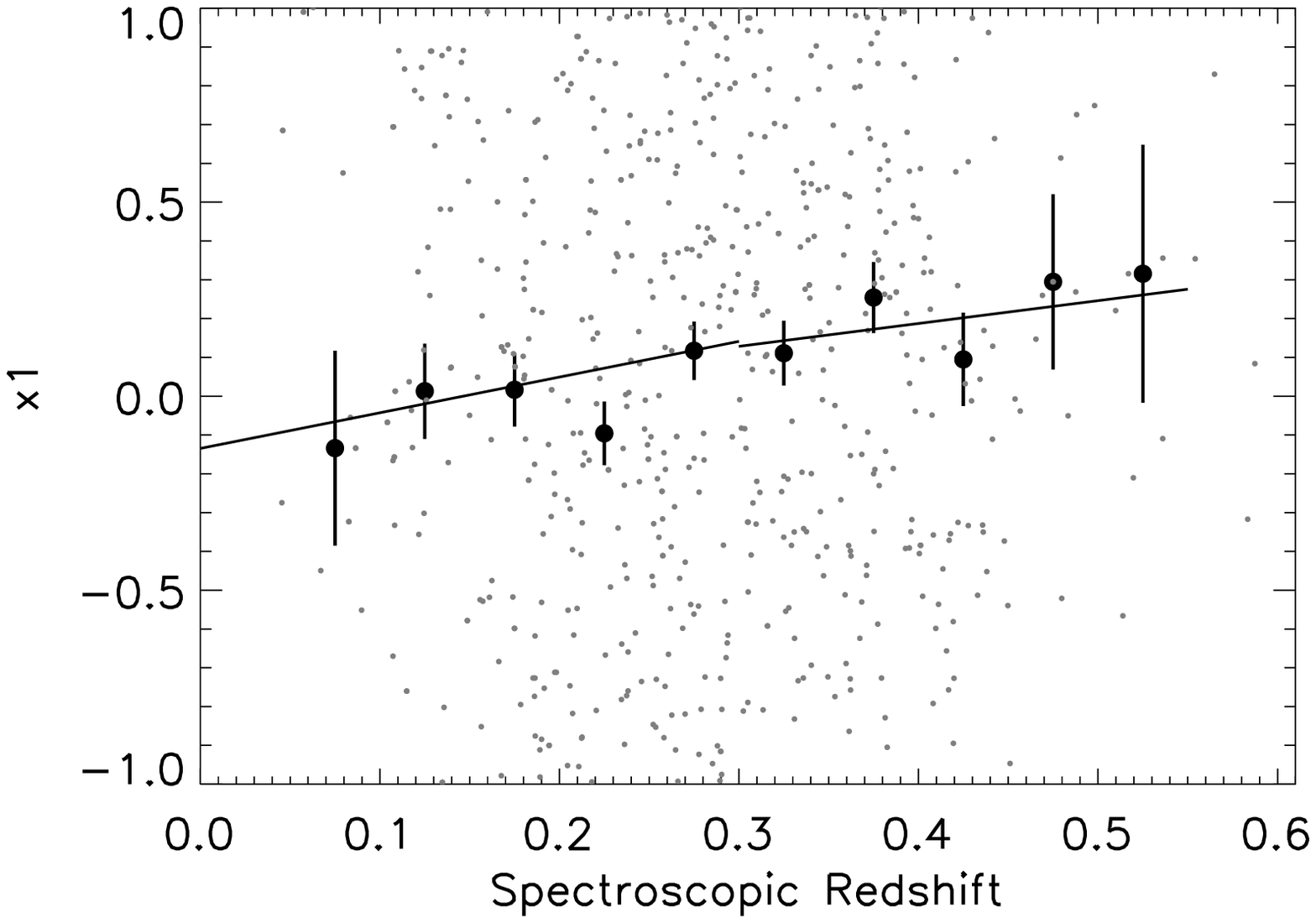}
\includegraphics[scale=0.5]{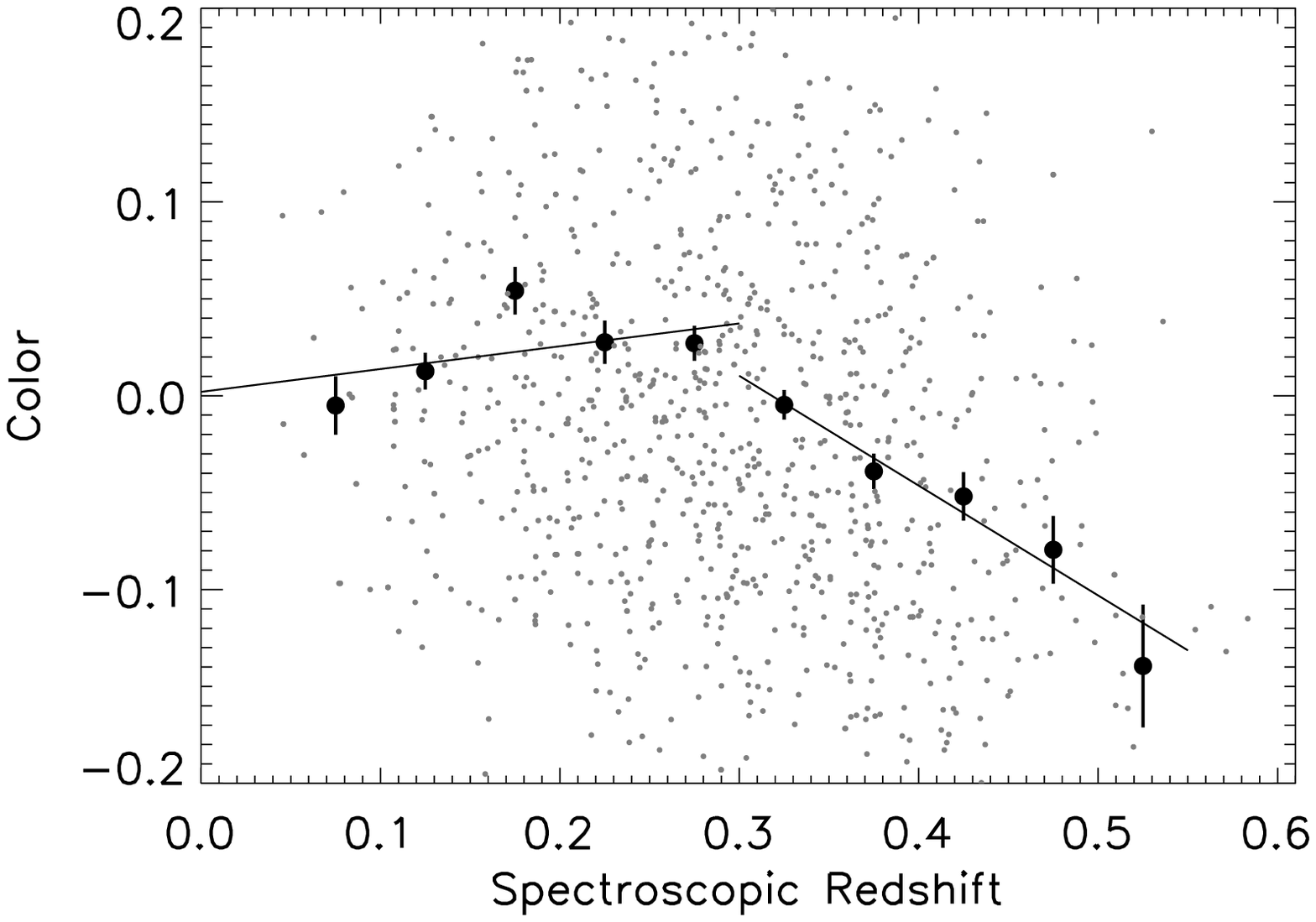}

\caption{ \label{fig:BOSS-best-fit-salt} TOP: Measured $x_1$ using the redshift prior for the entire BOSS and SDSS sample of SNe~Ia as a function of redshift.  The larger filled circles are the mean of the data points in a redshift bin of 0.05 with RMS error bands shown.  BOTTOM: Measured color when using redshift prior for entire BOSS and SDSS SNe~Ia sample as a function of spectroscopic redshift.  The larger filled circles are the mean of the data points in a bin in a redshift bin of 0.05. }  
\end{figure}

Figure~\ref{fig:BOSS-best-fit-salt} shows the SALT2 parameters $x_{1}$ and $c$ as a function of redshift for the entire photometric Z Ia sample.  Both the distribution of $x_1$ and $c$ evolve with increasing redshift.  At higher redshift, the sample tends towards higher values of $x_1$ and more negative values of $c$; these trends are to be expected due to the well-known Malmquist bias \citep{malmquist36a} in which only brighter SNe~Ia are identified at higher redshift.  This correlation begins roughly near $z=0.3$.  Assigning a turnover point at $z=0.3$, we split the sample into low redshift and high redshift SNe~Ia and fit linear models to demonstrate the effect of redshift incompleteness due to Malmquist bias.  For $x_1$ versus redshift, the slope of the low redshift and high redshift fit is consistent with 0.   For $c$, we see a slope at low redshift of $0.16\pm0.10$ and at high redshift we determine a slope of $-0.58 \pm 0.07$.  This analysis demonstrates the presence of Malmquist bias in the sample; a full correction must be applied from the cosmological perspective as done in C13.

\subsection{SN Light curve Properties With and Without Host Galaxy Redshifts}\label{subsec:bias}

To determine the impact of redshift on the light curve properties, we return to the sample of 1268 transients that were classified as either photometric Ia or photometric Z Ia.  This time, we fit a SALT2 light curve to each object twice, once allowing redshift as a free parameter, and once with the redshift constrained to that of the host galaxy.  We use the sample of SNe~Ia for which the SALT2 fits met the selection criteria both with and without the host galaxy redshift.  There are 888 SNe~Ia that meet this criteria.  We use the results to quantify the differences in the distributions of photometric redshift, $x_1$, and $c$.

\begin{figure}[htbp]
\includegraphics[scale=0.5]{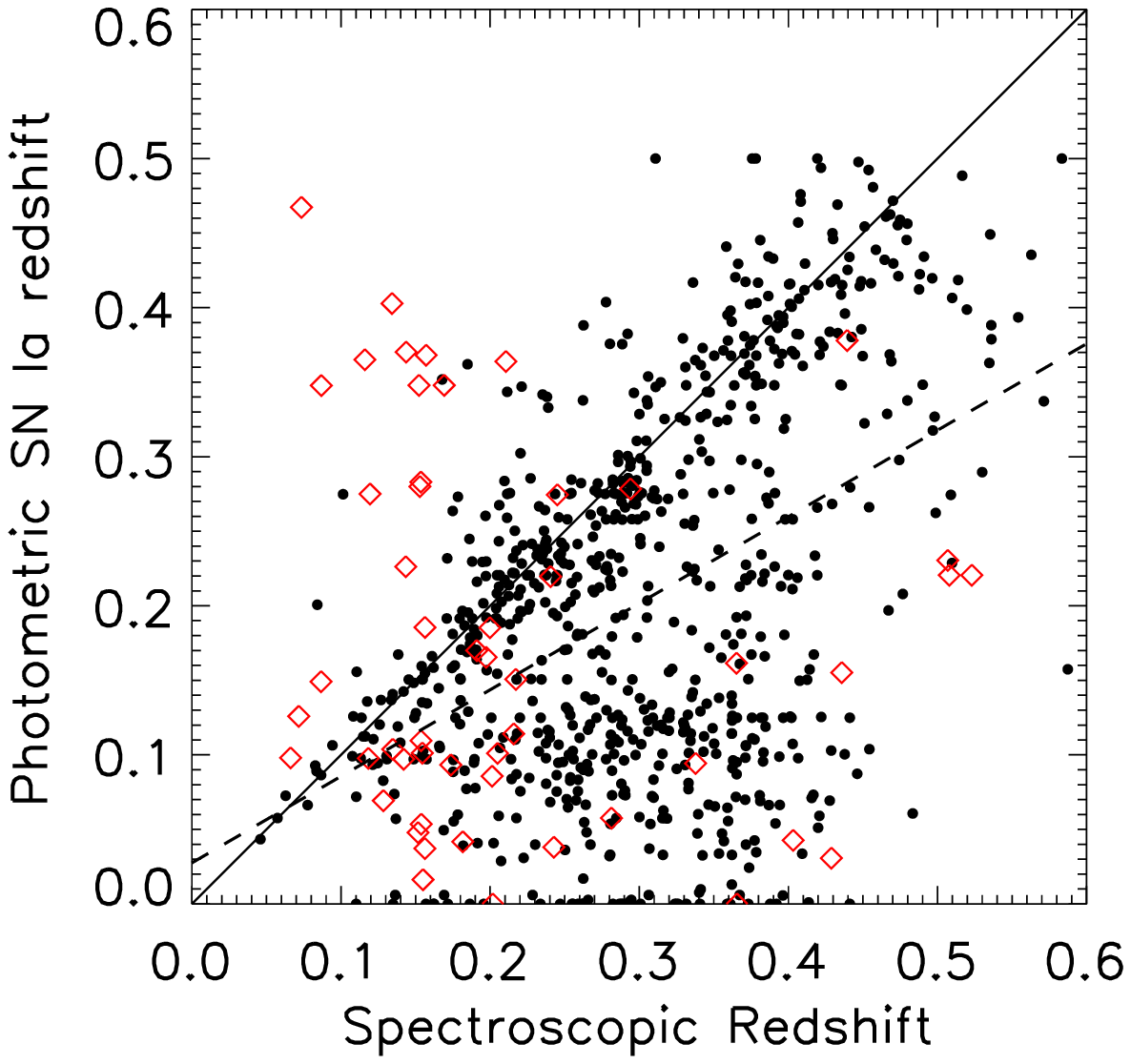}
\includegraphics[scale=0.5]{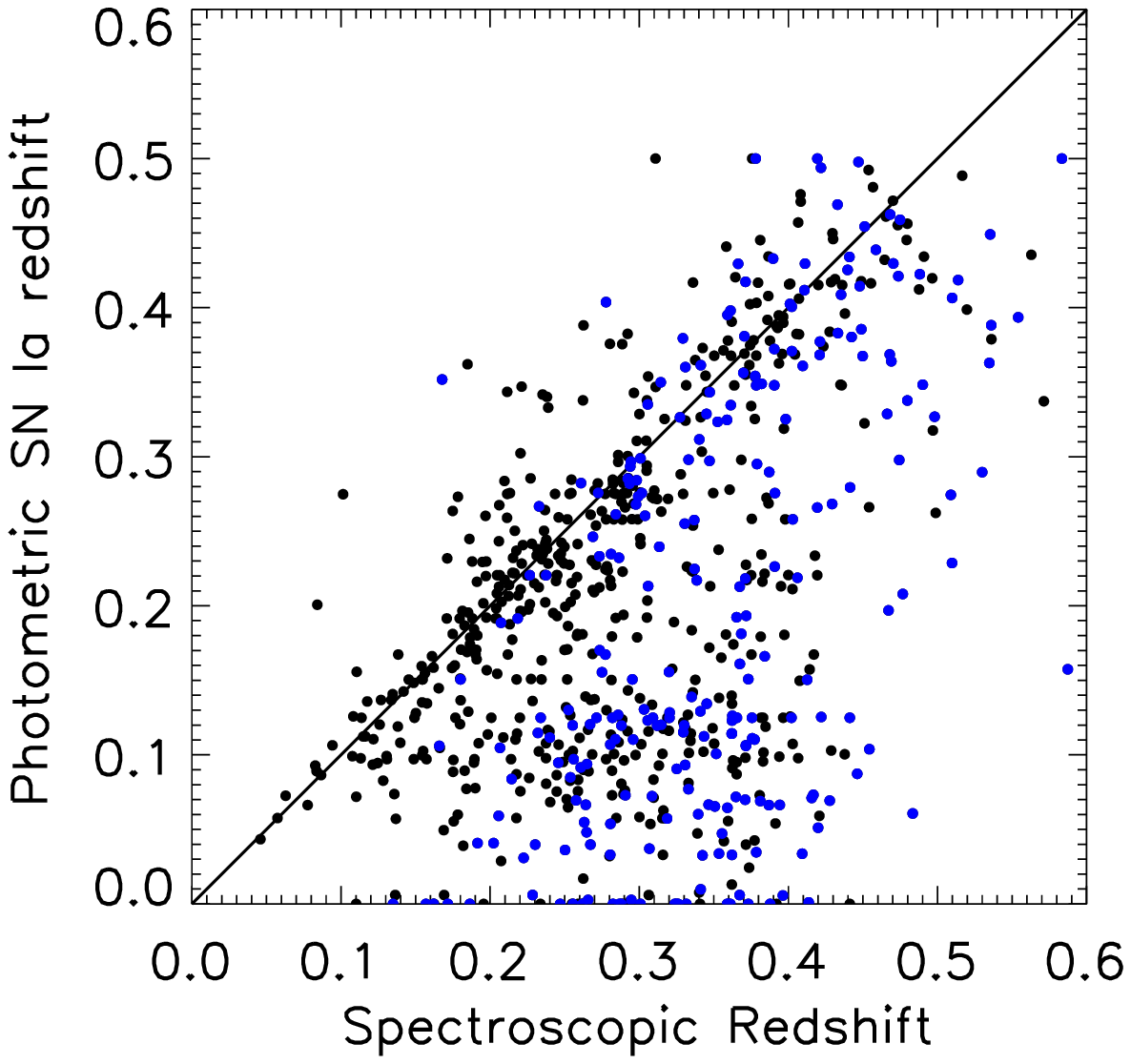}
\includegraphics[scale=0.5]{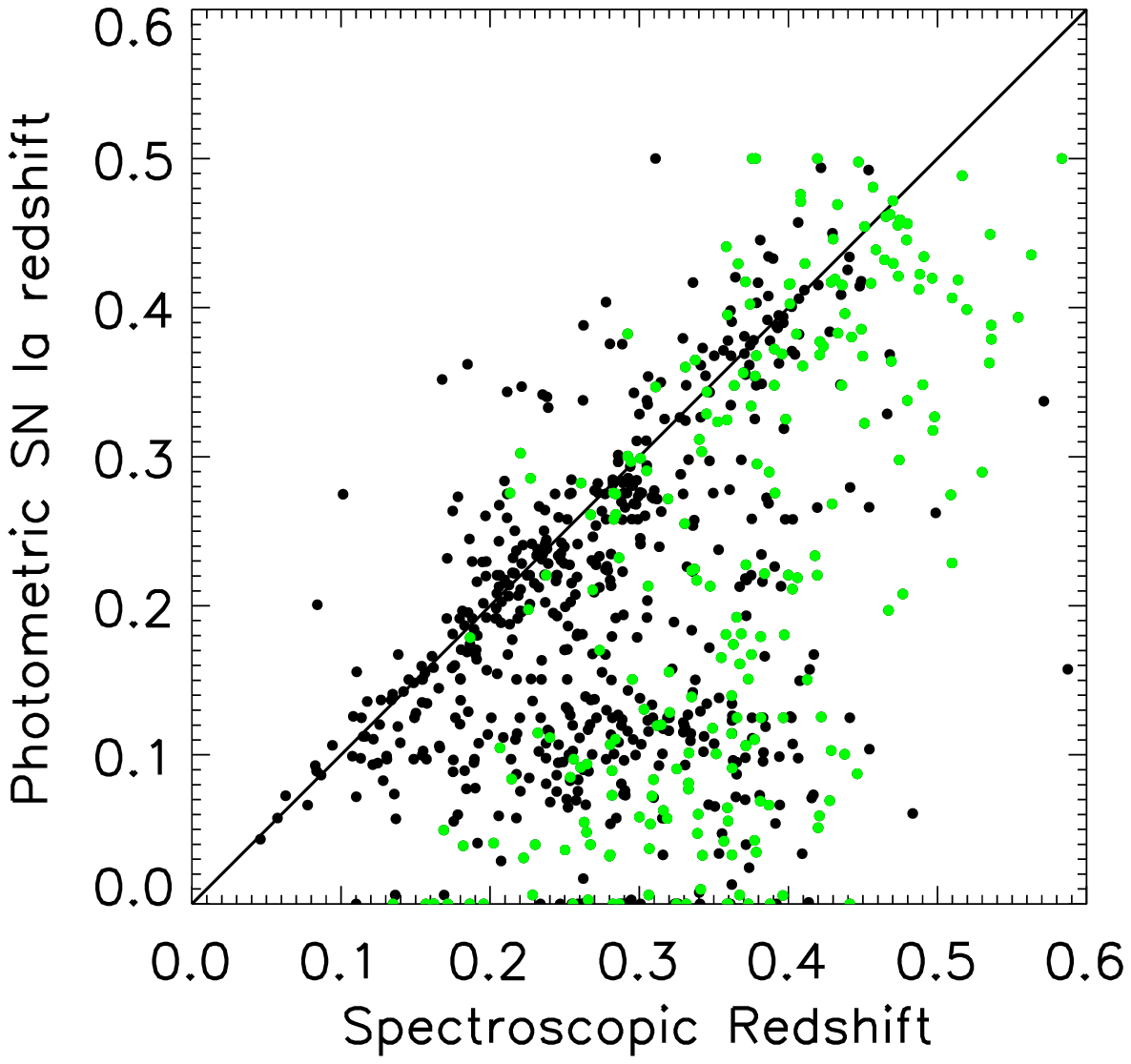}

\caption{ \label{fig:salt2-redshift}  The best fit SN~Ia photometric redshift versus the spectroscopic redshift.  TOP:  The black points represent candidates that are consistently classified as SNe~Ia with and without spectroscopic redshifts.  The red diamonds show the candidates identified as photometric Ia and not photometric Z Ia (likely contaminants).  The solid black line simply represents a slope of unity.  The dashed line is the line of best fit to the black data points.  MIDDLE:  The data with fewer than 8 epochs are shown in blue.  BOTTOM:  The data with S/N $\le 7$ are shown in green.  }  
\end{figure}

We first look at how the photometric redshift determined from the best-fit light curve compares to the spectroscopic redshift of the host galaxy (Figure~\ref{fig:salt2-redshift}). Instances in which a candidate is a photometric Ia and not a photometric Z Ia are highlighted independently; these are likely contaminants in the photometric Ia sample.  The redshift determined photometrically from the SN~Ia light curves is generally biased towards lower values.  The line of best fit has slope $0.58\pm0.39$, where one would expect a slope of unity for an unbiased photometric redshift.  As described below, the poor estimate of the redshift manifests itself in the derived color parameters.  The poor redshift estimate is likely due to low data quality in terms of a low number of epochs and small maximum S/N, as demonstrated in the two lower panels.

\begin{figure}[htbp]
\includegraphics[scale=0.5]{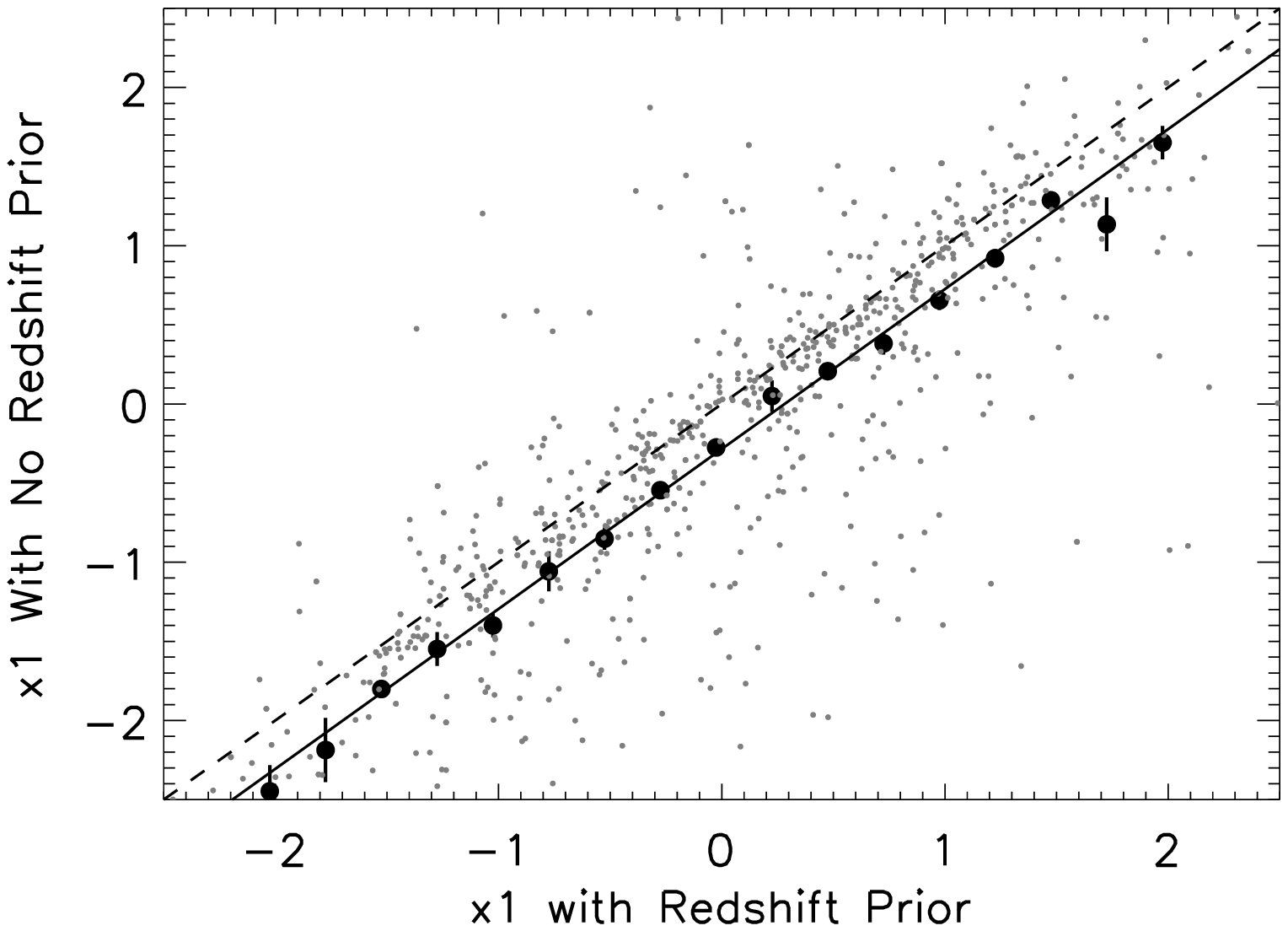}
\includegraphics[scale=0.5]{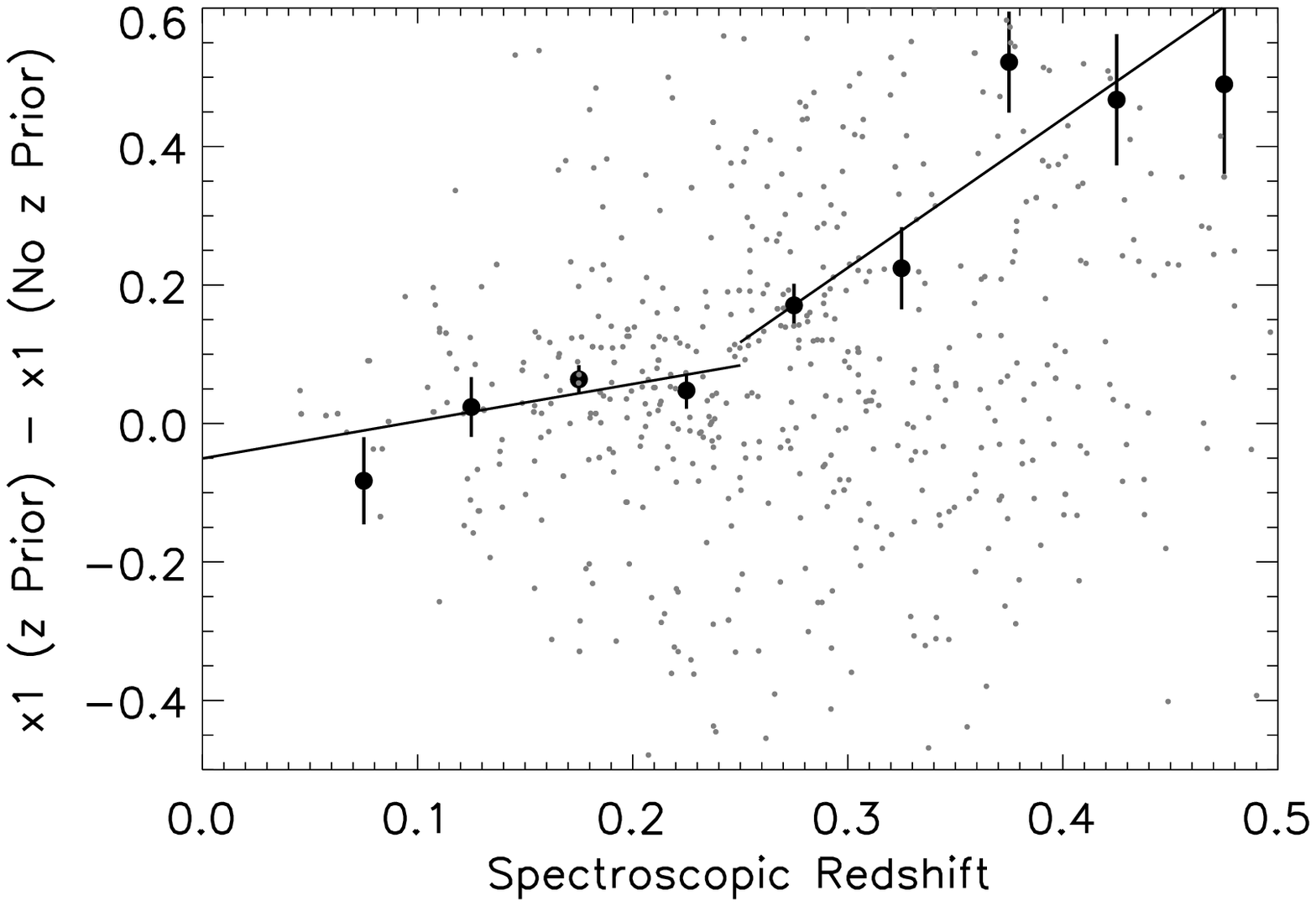}
\caption{ \label{fig:BOSS-best-fit-x1}  TOP:  Measured $x_1$ with flat redshift prior versus $x_1$ with redshift prior shown with a line of slope equal one.  BOTTOM: The difference between measured $x_1$ using spectroscopic redshift and measured $x_1$ assuming a flat redshift prior as a function of redshift. }  
\end{figure}

We examine the $x_1$ values determined by SALT2 with and without a spectroscopic redshift.  Figure~\ref{fig:BOSS-best-fit-x1} demonstrates the impact of redshift knowledge on estimates of $x_1$.  Fitting a simple linear model to the $x_1$ values derived with photometric redshift as a free parameter relative to $x_1$ values derived with the spectroscopic redshift, we find no bias in the slope and an offset.  The best-fit model has slope of $0.98 \pm 0.09$ and y-intercept $-0.28 \pm 0.09$.

\begin{figure}[htbp]
\includegraphics[scale=0.5]{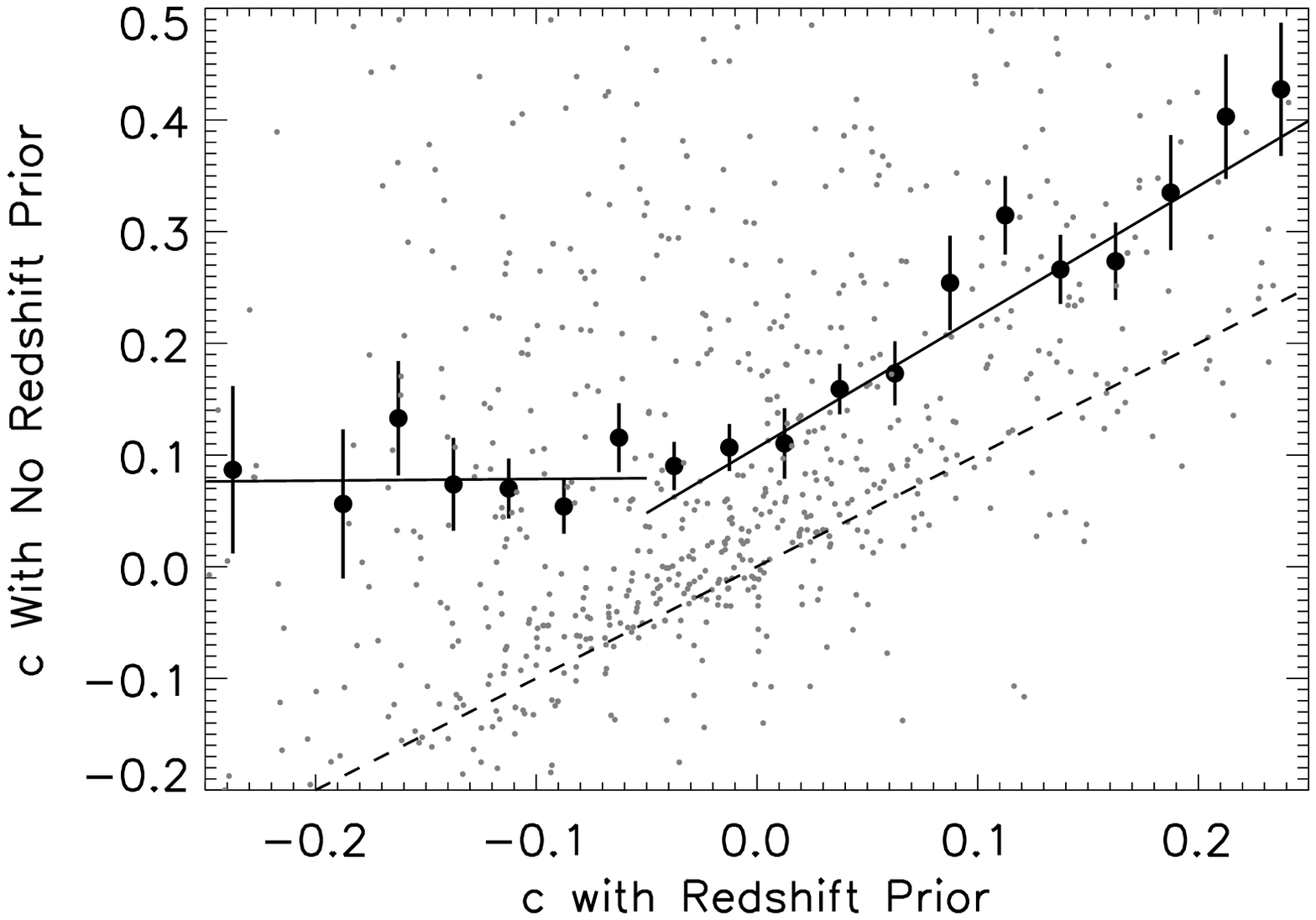}
\includegraphics[scale=0.5]{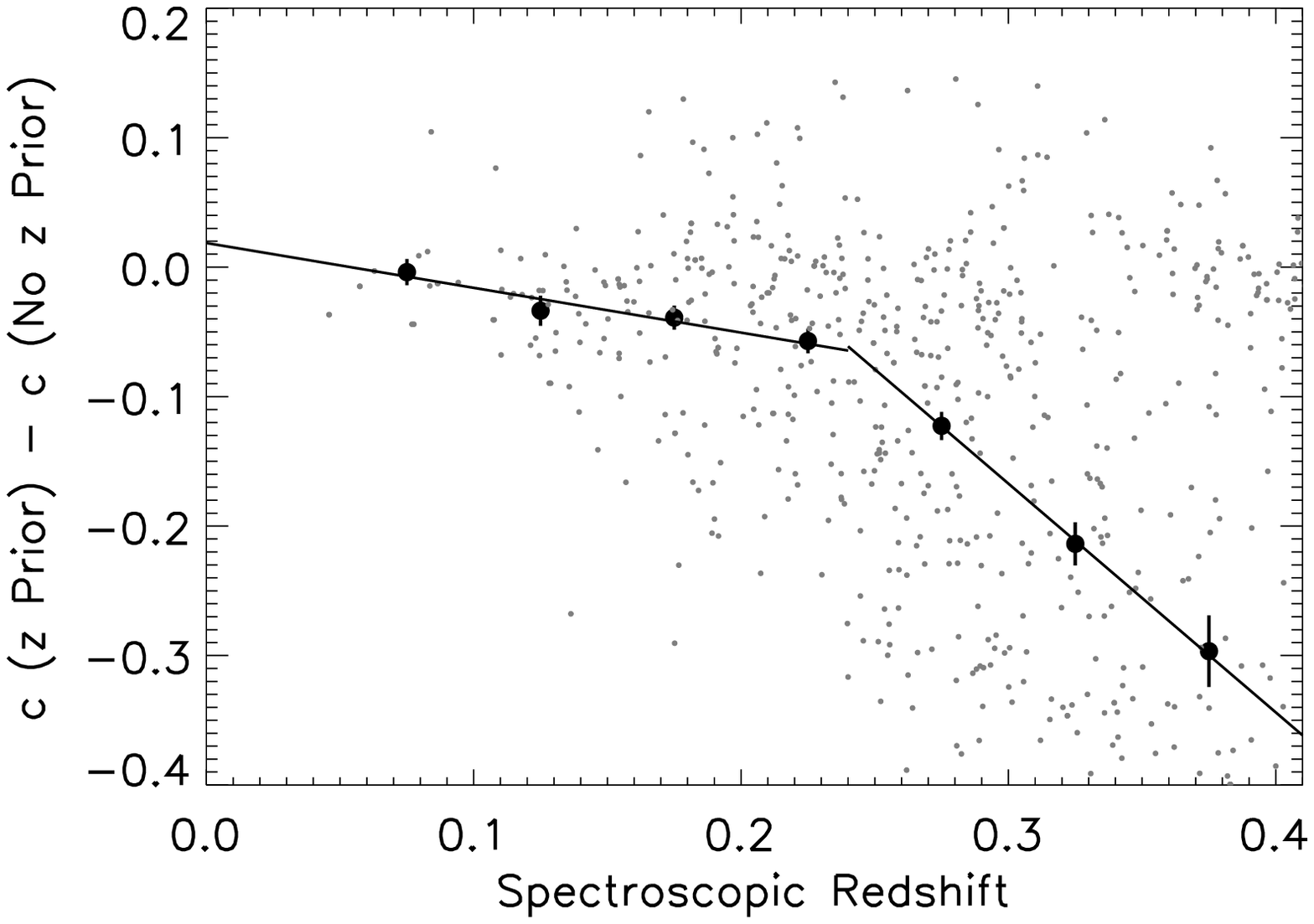}
\caption{ \label{fig:BOSS-best-fit-c}  TOP:  Measured color with flat redshift prior versus color with redshift prior shown with a dashed line of unity.  The solid line shows the line of best fit for the two redshift samples: $z <0.3$ and $z \ge 0.3$.  BOTTOM: The difference between measured $c$ using spectroscopic redshift and measured $c$ assuming a flat redshift prior as a function of redshift. }  
\end{figure}

We next examine the impact of redshifts on estimates of $c$.  The top panel of Figure~\ref{fig:BOSS-best-fit-c} compares estimates of $c$ with no spectroscopic redshift to estimates with a host galaxy redshift.  The SALT2 light curve fits clearly trend towards redder colors when redshift is included as a free parameter.  This bias toward red colors occurs over the full range of true SN color with a bias described by a linear fit with intercept of $0.14 \pm 0.04$ and slope of $0.8 \pm 0.4$.  We again divide the data, this time using the color estimates with redshift information.  We make the cut at $c_{\rm spec}= -0.05$.  We see a slope at low values of $c_{\rm spec}$ of $0.04\pm0.3$ and for high values of $c_{\rm spec}$ we determine a slope of $1.2 \pm 0.2$.  There is a significant change in estimates of $c$ between the two data samples.  The bottom panel shows that SNe at higher redshift produce more biased estimates of color when the precise redshift is not known.

In addition, we used simulations to examine the difference between the SALT2 parameters with and without a redshift prior.  The simulations show similar trends with the offsets found in  Figure \ref{fig:BOSS-best-fit-x1} and Figure \ref{fig:BOSS-best-fit-c} and that SALT2 creates biased estimates of $z$ and $c$ depending on whether a redshift prior is used.  

\begin{figure}[htbp]
\includegraphics[scale=0.5]{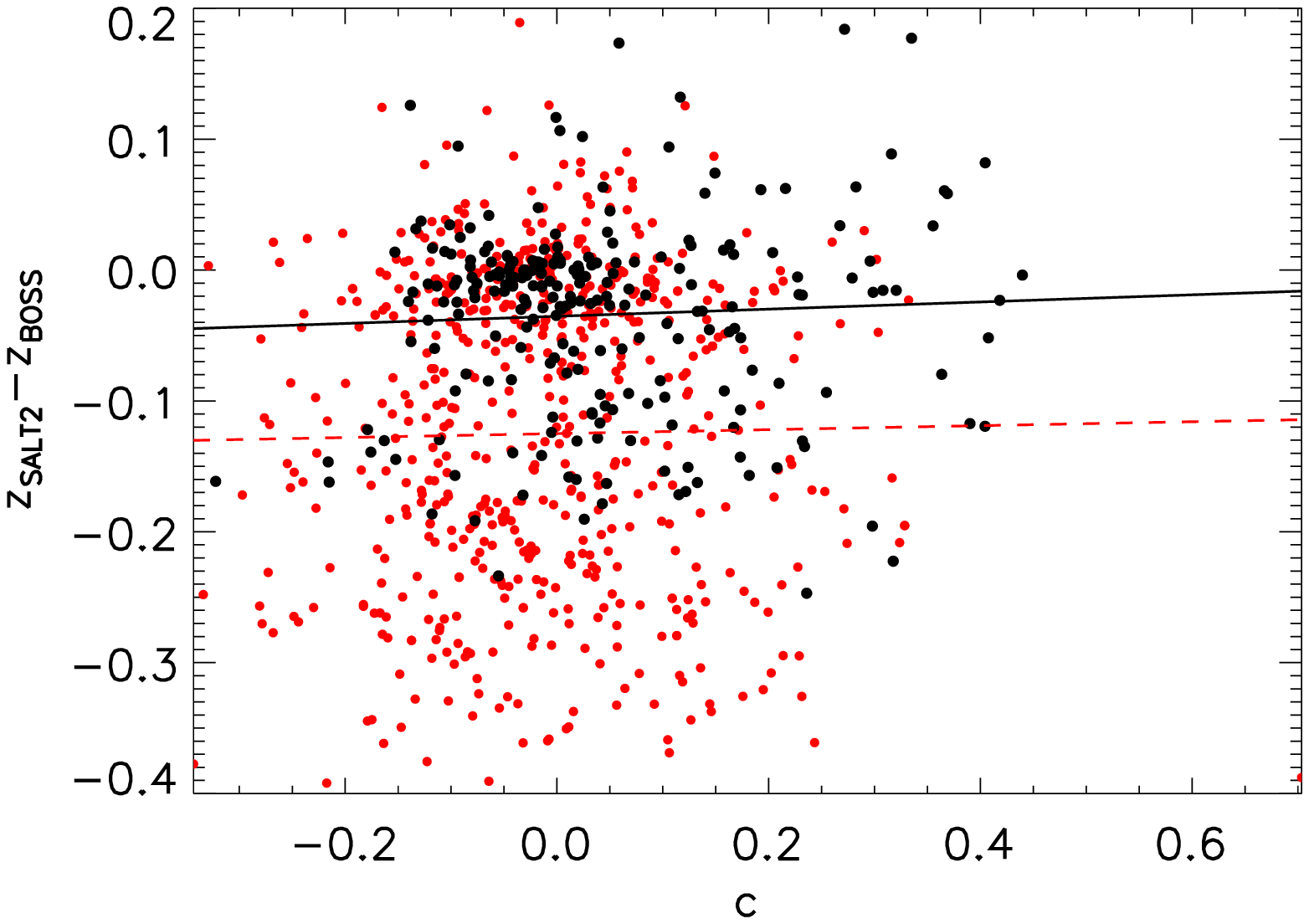}
\includegraphics[scale=0.5]{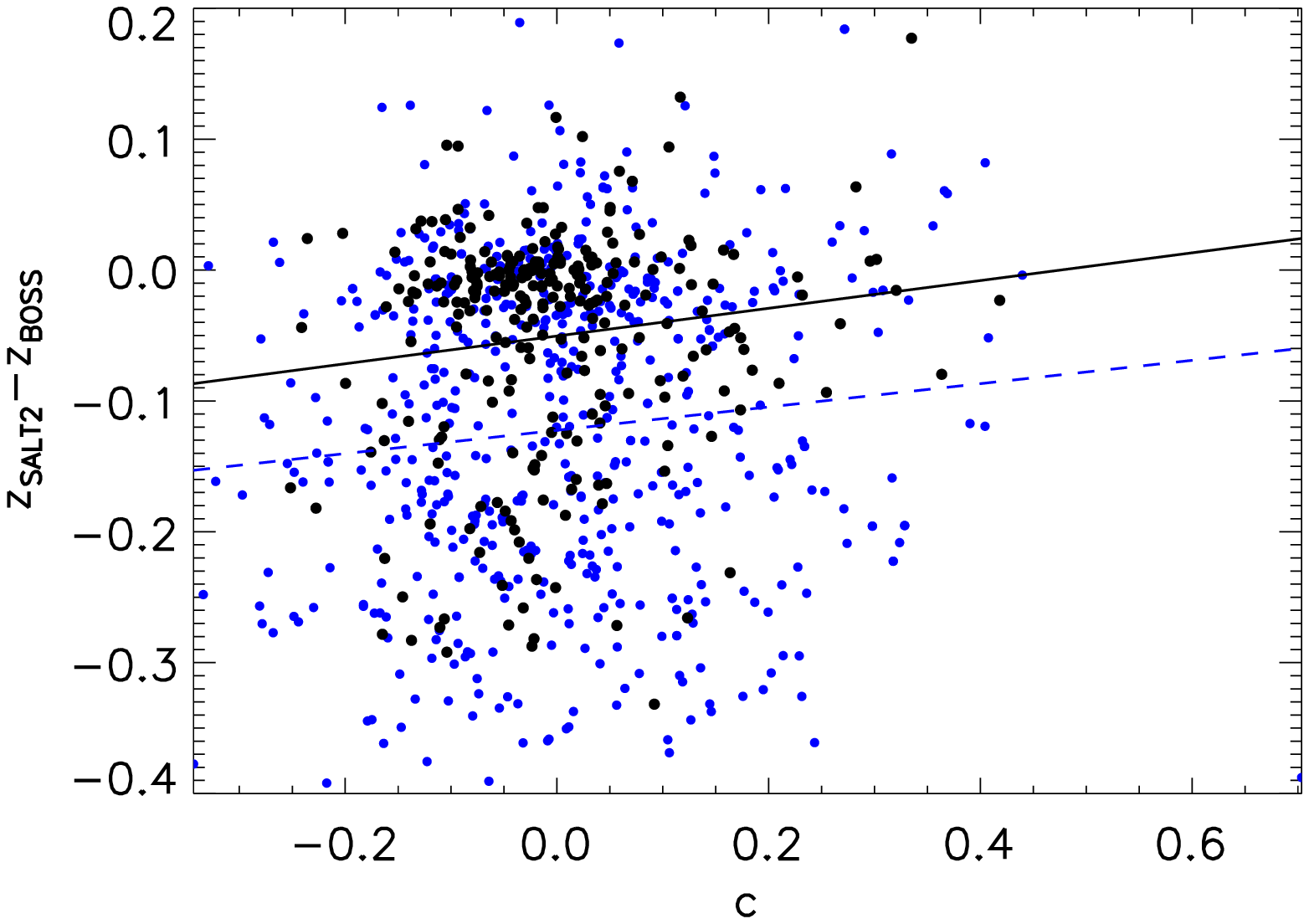}
\includegraphics[scale=0.5]{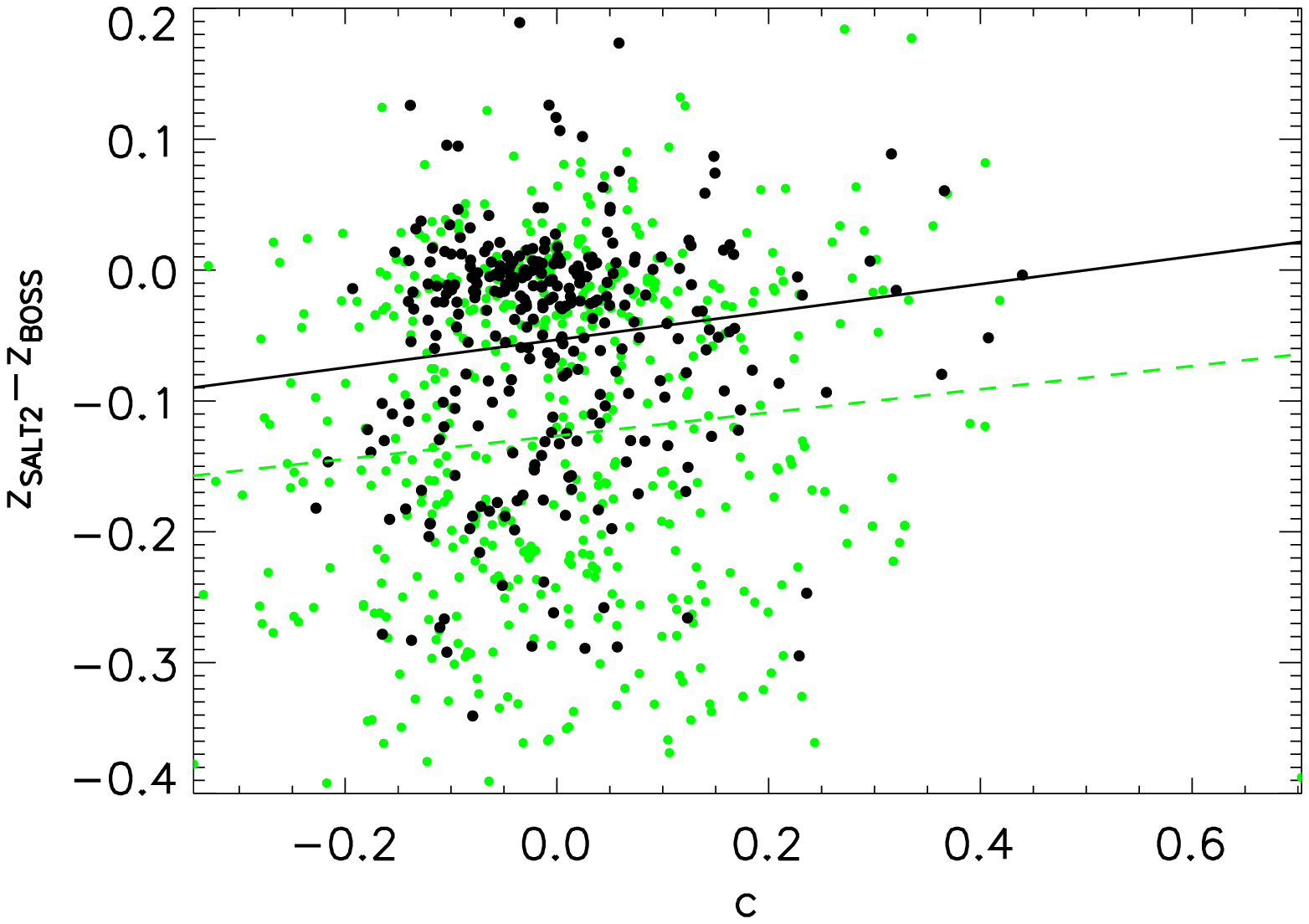}
\caption{ \label{fig:color-quality} The difference between the SALT2 redshift estimate and the host galaxy spectroscopic redshift as a function of color.  The solid, colored lines are the lines of best fit to the lower quality data.  TOP:  The red dots represent the sample with $z>0.25$.  MIDDLE:  The blue dots represent the sample with at most 9 epochs with S/N $>3$ in $r$-band.  BOTTOM:  The green dots represent candidates without at least one epoch of S/N $>10$ in $r$-band. }  
\end{figure}

The systematic underestimates in photometric redshift shown in Figure~\ref{fig:salt2-redshift} are likely responsible for the bias in the distribution of color estimates shown in Figure~\ref{fig:BOSS-best-fit-c}.  Figure~\ref{fig:color-quality} shows the relationship between bias in color estimates and bias in photometric redshift estimates.  The best-fit lines have positive slope, indicating that candidates for which the photometric redshift is underestimated are also interpreted to be redder.  As shown in the top panel, the effect is worse for high redshift SNe~Ia than for low redshift SNe~Ia.  A simple linear fit to the data with spectroscopic redshift $<0.25$ has a slope consistent with zero ($0.03 \pm 0.03$) and a y-intercept of $ -0.011 \pm 0.01 $.  The fit to the data at higher redshift has a slope of $0.20 \pm 0.03$ and a y-intercept of $-0.09 \pm 0.01$.  We attribute this trend to a general reduction in S/N and data quality at increasing redshift.

To assess the effect of data quality, we examine the trend in bias for samples cut on the number of epochs sampling the light curve in the middle panel and maximum S/N in the $r$-band in the bottom panel. As can be seen in the three panels, when including either low redshift SNe~Ia, SNe with at least 10 $r$-band epochs with S/N $>3$, or at least one $r$-band epoch with S/N $>$ 10, the bias in color and photometric redshift effectively disappears.

\begin{figure}[htbp]
\includegraphics[scale=0.5]{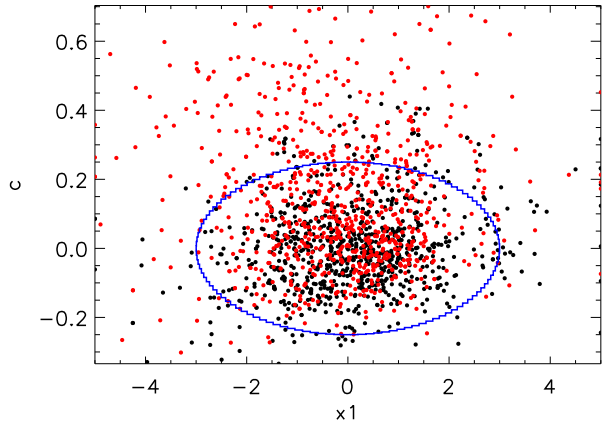}
\caption{\label{fig:salt2-ellipse} A plot of the SALT2 parameters $c$ vs $x_1$ including the ellipse found in C13.  The black points are determined using spectroscopic redshift as a prior and the red points are determined with a flat redshift prior.  }
\end{figure}

The bias in color estimates will lead to a different interpretation of the elliptical cuts in C13 if one does or does not use host galaxy redshifts in the SALT2 fits.  We reproduce the elliptical cuts described in Section~\ref{subsec:lightcurveparam} using the sample of 943 SNe~Ia with light curve parameters determined with and without host galaxy information.  These samples of photometric Ia and photometric Z Ia are shown with the elliptical cuts in $x_1-c$ parameter space in Figure~\ref{fig:salt2-ellipse}.  Only 79.8\% of the photometric Ia sample lie within the ellipse.  As a final test, we determine the number of photometric Z Ia inside the ellipse when using the SALT2 parameters determined with knowledge of the redshift.  Using the sample of 1074 photometric Z Ia with SALT2 light curve parameters described in Section~\ref{subsec:completeness}, we find 965 (89.9\%) SNe~Ia inside the ellipse.

\section{Conclusion}\label{sec:conclusion}

In this paper, we describe the impact of host galaxy spectra on the classification and interpretation of SN candidates from the SDSS-II SN data release.  We guide the reader through the SDSS and BOSS host galaxy spectra and the DR10 galaxy product.  We also provide analysis of the consistency in SN classification between a sample when only photometric data is available and a sample with well-constrained redshifts from host galaxy spectra.  Finally, we show significant differences in SALT2 light curve properties depending on whether or not the host galaxy redshift is used in the fit.

In total, 2998 spectra were obtained with BOSS and 2292 spectra were obtained with SDSS.  The redshift range for the combined sample is 0 $< z <$ 4.  The range of redshifts for host galaxies of the transients determined to be SNe~Ia with our most rigorous classification using host galaxy redshifts (``photometric Z Ia'') is 0.057 $< z <$ 0.688, and the range for the analogous core-collapse sample (photometric Z CC) is 0.041 $< z <$ 0.562.

Using the host galaxy spectroscopic redshift as a prior, we have photometrically typed these candidates using three different schemes.  In previous work \citep{kessler10a,bernstein12a}, simulations were performed to test completeness and purity of SNe samples with only photometric information; here we apply an independent test by evaluating the consistency of photometric classification with and without spectroscopic redshift.  We recommend using classifications derived from the nearest neighbor full photometric classification.   With the full photometric classification, there are 1120 SNe~Ia when using photometry alone and 1094 SNe~Ia when using host galaxy redshifts.  In total, 90 transients were classified as SNe~Ia only when host galaxy redshifts were used; 126 transients were classified as SNe~Ia when only photometric information was used.  Assuming the typing with redshifts to be more accurate, we interpret this result to imply that the contamination of our sample using only photometric information is 12.1\% and the completeness is 88.7\%.  We found that we can reduce the contamination from 12.1\% to 9.5\% by additionally requiring both early time coverage and late time coverage.  Slightly more than 50\% of the SDSS-II SNe~Ia meet this more stringent condition.  We find larger improvements by filtering on the light curve parameters derived from a photometric-only sample.  By including photometric SNe with SALT2 $x_1$ and $c$ parameter fits that lie within a simple ellipse defined in C13, we find a larger number of SNe~Ia with only 4.8\% contamination.

This sample of 522 photometric Ia with 4.8\% contamination appears to demonstrate the usefulness of photometric only samples.  However, the addition of redshift provides subtle, yet important, improvements on the sample of SNe~Ia.  First, when the spectroscopic redshift of the SN~Ia is not known, there is an additional degree of freedom in the SALT2 light curve fits, leading to a larger number of fits that fail selection cuts.  In addition, there is a bias in the estimate of photometric redshift and SN color that appears when the fits are performed without a redshift. These effects appear to be highly correlated with poor data quality and are significantly mitigated when the host galaxy spectroscopic redshifts are used. 

For future large SN surveys there will be far too many candidate transients for full spectroscopic confirmation to be feasible.   Obtaining the spectroscopic redshift of each host galaxy in multi-object spectroscopy is an efficient means to improve the science returns from these large samples.  The host galaxy redshift constrains the redshift axis of the Hubble diagram.  The addition of host galaxy spectra increases both the diversity of the host galaxy sample, allowing for fainter, more distant galaxies, and improves the observed host galaxy properties. Using the redshift removes a degree of freedom in light curve parameter space, thereby increasing the fraction of transients with reliable fits.  Finally, using the host galaxy redshifts mitigates an apparent bias in the SN light curve fits tending toward lower redshift estimates and redder colors.  The spectroscopic host galaxy properties can also be used to study the relationship between SN properties and the local environment.  Stellar mass, absolute luminosity, velocity dispersion, star formation rate, and metallicity can be derived from a sample of galaxies observed in imaging and spectroscopy to investigate trends in Hubble residuals and SN rates.  For these reasons, current SN projects such as DES and future projects such as LSST will benefit from a dedicated spectroscopic campaign to follow up host galaxies, even if the SN has faded.

Funding for the SDSS and SDSS-II has been provided by the Alfred P. Sloan Foundation, the Participating Institutions, the National Science Foundation, the U.S. Department of Energy, the National Aeronautics and Space Administration, the Japanese Monbukagakusho, the Max Planck Society, and the Higher Education Funding Council for England. The SDSS Web Site is http://www.sdss.org/. The SDSS is managed by the Astrophysical Research Consortium for the Participating Institutions. The Participating Institutions are the American Museum of Natural History, Astrophysical Institute Potsdam, University of Basel, Cambridge University, Case Western Reserve University, University of Chicago, Drexel University, Fermilab, the Institute for Advanced Study, the Japan Participation Group, Johns Hopkins University, the Joint Institute for Nuclear Astrophysics, the Kavli Institute for Particle Astrophysics and Cosmology, the Korean Scientist Group, the Chinese Academy of Sciences (LAMOST), Los Alamos National Laboratory, the Max-Planck-Institute for Astronomy (MPIA), the Max-Planck-Institute for Astrophysics (MPA), New Mexico State University, Ohio State University, University of Pittsburgh, University of Portsmouth, Princeton University, the United States Naval Observatory, and the University of Washington.

Funding for SDSS-III has been provided by the Alfred P. Sloan Foundation, the Participating Institutions, the National Science Foundation, and the U.S. Department of Energy Office of Science.  The SDSS-III web site is http://www.sdss3.org/.
SDSS-III is managed by the Astrophysical Research Consortium for the Participating Institutions of the SDSS-III Collaboration including the University of Arizona, the Brazilian Participation Group, Brookhaven National Laboratory, University of Cambridge, Carnegie Mellon University, University of Florida, the French Participation Group, the German Participation Group, Harvard University, the Instituto de Astrofisica de Canarias, the Michigan State/Notre Dame/JINA Participation Group, Johns Hopkins University, Lawrence Berkeley National Laboratory, Max Planck Institute for Astrophysics, , Max Planck Institute for Extraterrestrial Physics, New Mexico State University, New York University, Ohio State University, Pennsylvania State University, University of Portsmouth, Princeton University, the Spanish Participation Group, University of Tokyo, University of Utah, Vanderbilt University, University of Virginia, University of Washington, and Yale University.

The work of MO and KD was supported in part by the U.S. Department of Energy under Grant DE$-$SC0009959.    The support and resources from the Center for High Performance Computing at the University of Utah is gratefully acknowledged.  This work was partially supported by STFC grant ST/K00090X/1.  Please contact the author(s) to request access to research materials discussed in this paper.

\bibliographystyle{apj}
\bibliography{archive}

\end{document}

%% file: astromacro.tex
\usepackage{color}
\usepackage{graphicx}
 
\newif\ifdraftmodep
\draftmodepfalse

\newif\ifapjp
\apjpfalse

%% file: outlinemacro.tex
\usepackage{ifthen,color,graphicx}

\newcommand{\outlinestart}[1]
{
\ifthenelse{\boolean{#1}}{\begin{enumerate}}{}
}

\newcommand{\outline}[2]
{
\ifthenelse{\boolean{#1}}{\it \color {red} \item #2 \rm \color {black}}{}
}

\newcommand{\outlineend}[1]
{
\ifthenelse{\boolean{#1}}{\end{enumerate}}{}
}